\pdfoutput=1
\documentclass[12pt]{article}

\usepackage[margin=1in]{geometry}
\usepackage{graphicx}
\usepackage{subcaption}
\usepackage{caption}
\usepackage{setspace}
\usepackage{hyperref}
\expandafter\def\expandafter\UrlBreaks\expandafter{\UrlBreaks
  \do\a\do\b\do\c\do\d\do\e\do\f\do\g\do\h\do\i\do\j\do\k\do\l\do\m
  \do\n\do\o\do\p\do\q\do\r\do\s\do\t\do\u\do\v\do\w\do\x\do\y\do\z
  \do\A\do\B\do\C\do\D\do\E\do\F\do\G\do\H\do\I\do\J\do\K\do\L\do\M
  \do\N\do\O\do\P\do\Q\do\R\do\S\do\T\do\U\do\V\do\W\do\X\do\Y\do\Z
  \do\0\do\1\do\2\do\3\do\4\do\5\do\6\do\7\do\8\do\9\do\-}
\usepackage{booktabs}
\usepackage{array}
\usepackage{longtable}
\usepackage{natbib}
\usepackage{amsmath}

\graphicspath{{results_20may/}}

\title{Science under Threat? A Natural Experiment in Economics}

\author{Dominic Rohner, Oliver Vanden Eynde, and Philine Widmer\thanks{\scriptsize Rohner: University of Lausanne, CEPR (email: dominic.rohner@unil.ch); Vanden Eynde: Paris School of Economics, CNRS, and CEPR (email: o.vandeneynde@psemail.eu); Widmer: Paris School of Economics (email: philine.widmer@gmail.com). This research was supported by EUR grant ANR-17-EURE-0001 at PSE. Violaine D'Ortona provided excellent research assistance. We gratefully acknowledge helpful comments from various seminar participants and colleagues. All errors are of course ours.}}

\date{November 2026\\[3pt] {\small First version: February 2026. This version: November 2026.}}

\begin{document}

\maketitle
\thispagestyle{empty}

\begin{abstract}
Academic freedom is under strain worldwide. To study how scholars respond to (perceived) political pressure, we exploit a natural experiment in the United States: the publication in early 2025 of a ``stop list'' of words flagged by the government. We find that economists at universities relying more on federal funding reduce gender-, race-, and environment-related content with flagged words relative to other US and UK economists. Regarding mechanisms, our analyses suggest that self-censorship may explain this reduction better than direct grant cancellation. Estimates are similar when excluding papers acknowledging federal funding or controlling for universities' institutional measures and financial distress.

\smallskip\\
{\itshape Keywords:} Censorship, Science, Academic Freedom, Science Funding.
\end{abstract}

\clearpage

\numberwithin{equation}{section}
\renewcommand{\theequation}{\thesection.\arabic{equation}}
\setcounter{page}{1}

\section{Introduction}

Academic freedom has been challenged in many countries and by actors of various political orientations. The 2026 Academic Freedom Index \citep{kinzelbach2025academic} shows that over the 10-year period from 2015 to 2025, academic freedoms decreased in 50 countries (including the United States, India, Indonesia, Germany, Türkiye and Russia), while increasing in only 9 countries (most of which are relatively less populous).

Whether pressure on academic freedom actually translates into biases in the selection of research topics depends on how scholars respond. Do academics yield to political pressure, or are their choices of research topics guided by scientific interest alone, even when the institutions they work in have much to lose? This question matters because independent universities are an important safeguard against democratic backsliding. They have come under political pressure, as illustrated by the measured decline in academic freedom. Political actors may also contest established scientific evidence, for example on climate change and vaccines.

Studying scientific (self-)censorship is methodologically challenging, as there is rarely exogenous variation in academic freedom that can be easily measured. Furthermore, long publication delays make it difficult to study the timing of such changes, and it is difficult to disentangle scholars' behavior from potential biases in the publishing process. However, we can draw on a natural experiment that allows us to address these challenges.

In particular, we investigate how a ``stop list'' of words seen as undesired by the US government may have affected the choice of research topics in economics. In January 2025, the US government started screening federal research grants under several executive orders.\footnote{See Executive Order 14151 (``Ending Radical and Wasteful Government DEI Programs and Preferencing''), Executive Order 14168 (``Defending Women from Gender Ideology Extremism''), and Executive Order 14173 (``Ending Illegal Discrimination and Restoring Merit-Based Opportunity'') \citep{eo14151,eo14168,eo14173}.} In February 2025, it was reported that US federal agencies such as the National Science Foundation (NSF) and the Centers for Disease Control and Prevention (CDC) had started enforcing lists of ``flagged words'' -- including terms such as ``diversity,'' ``equity,'' ``inclusion,'' ``women,'' and ``climate crisis'' -- in research documents posted by these agencies and grant applications \citep{yourish2025words}.

The media started referring to a list of ``banned words,''\footnote{See, for example, \url{https://pen.org/banned-words-list/}. } even if the published list was not official (and possibly not exhaustive).\footnote{The keyword list is not merely inferred: a similar list appears in an official congressional document, Appendix~B of the Senate Commerce Committee report on DEI (diversity, equity and inclusion) grants at the National Science Foundation. The report, including the keyword appendix, was published on October~9, 2024; the accompanying database of flagged grants followed on February~11, 2025 (\url{https://www.commerce.senate.gov/2025/2/cruz-led-investigation-uncovers-2-billion-in-woke-dei-grants-at-nsf-releases-full-database}; the raw keyword files are also mirrored at \url{https://github.com/strubell/ted-cruz-dei-anal/tree/main/keywords}). The keyword appendix, thus, predates the November 2024 election. However, it was published by the committee minority and didn't imply any enforcement until the January 2025 executive orders and agency screening. The list NSF staff used to screen active grants in early February 2025 closely tracks this appendix, according to news reports \citep{garisto2025nsf, palmer2025hitlist}. Of the 197 base terms used in our analysis (Appendix Table~\ref{tab:censored_words}), 130 appear directly in the appendix and a further 19 overlap with an appendix entry as part of a longer phrase, in either direction (e.g., ``diversity'' within ``advancing diversity,'' or our term ``health equity'' containing ``equity''); the remainder are attested in cross-agency compilations circulated around the same period \citep{yourish2025words}.} Also, there was no strict ``ban,'' in the sense that the use of these words did not trigger automatic cancellations of projects and applications. However, this policy attracted considerable media attention, and was widely seen as a ``stop list'' of words to be avoided. By January 2026, 1,996 NSF grants had reportedly been canceled or suspended \citep{kozlov2026}.

Our analysis exploits the discontinuity in US government policy and the heterogeneity among scholars, more or less exposed to US government pressure, based on their institutions' reliance on federal research funding. The list is the visible and measurable part of this change in policy: it names the words, so we can track their use in research papers. The change itself was broader, from the executive orders to the screening and cancellation of grants. Our estimates capture exposed scholars' response to this change as a whole; the list is our measurable proxy, not the only part to which exposed scholars may have responded. Note that for our study purposes, it is not key whether the lists of flagged terms are binding. What matters for our question is that the ``stop list'' has affected institutions' and academics' \textit{perception} of political pressure (e.g., through the aforementioned media coverage). 

We can draw on the working paper culture in economics: unlike in most disciplines, it is standard practice to circulate working papers before they are peer-reviewed. Across science as a whole, preprints account for only about 4\% of research articles, and in many disciplines for less than 2\%; in economics, about one in five papers (20.8\% over 1991--2015) circulates as a working paper, a share exceeded only by physics \citep{xie2021preprint}. Working papers are posted before journal review. Accordingly, their release dates are close to the time of writing. This lets us date the response to the US policy change without the lags of refereeing and publication. Beyond providing a well-suited case study, economics is also a relevant setting because recent work has found that this discipline has a substantial impact on policymakers' decisions \citep{hjort2021research}.

In terms of identification strategy, we exploit heterogeneous exposure to the policy shock that unfolded in January/February 2025, allowing us to perform event-study and difference-in-difference analyses. 

Our results show that there was no statistically significant pre-trend within the US and that  (the perception of) scrutiny on research containing ``flagged words'' leads to a reduction in the use of these words in a sensitive context by economists working in universities that are more dependent on federal funding  (labeled ``High-Fed''), compared to those working in institutions that are less exposed to federal funding (referred to as ``Low-Fed''). In particular, comparing the post-treatment period to the pre-treatment period, we observe a 2.5 percentage point drop in the probability that a High-Fed paper uses any of the flagged words in the context of gender, race, or environment (GRE), relative to Low-Fed papers. An indicator for any flagged word, regardless of its context, shows no such drop. A language model classifies the context of each flagged word, and we validate this classification against the JEL codes of the NBER and CEPR papers: 35\% of the papers we classify as gender list the JEL code for the economics of gender, against 0.3\% of other papers, with similar gaps for race and environment (Section~\ref{sec_data}). A pooled difference-in-differences comparison of High-Fed against Low-Fed papers yields an estimate of similar magnitude ($-0.027$). Results are similar when High-Fed institutions are compared with UK institutions (instead of Low-Fed). At the author level, where author fixed effects absorb any time-invariant author characteristics, the within-US estimate is similar ($-0.027$) when authors are classified by whether the author team works at High-Fed institutions. When authors are classified only according to their own university, the effect at the author-level is smaller and is not statistically significant ($-0.012$).

When investigating the mechanisms, we first investigate heterogeneity across several dimensions, starting with departmental quality. Within the US, the point estimates are negative for both higher- and lower-ranked departments and larger at lower-ranked ones. Relative to the UK, the decline is concentrated in higher-ranked departments. We also examine the role of author attributes (gender, ethnicity, and seniority), and we find no strong evidence of heterogeneity along these dimensions.

Next, we investigate whether changes in research focus are semantic (i.e., just replacing some words with others in the abstract) or substantial (i.e., focusing on other research questions).\footnote{It is important to keep in mind that even changing terminology and moving content from the title page to the appendix is far from cosmetic. Past research has found that the placement of content matters \citep{Widmer2024} and that wording choices have substantial real-world effects \citep{djourelova2023persuasion}.} Two pieces of evidence speak to this, and both rest on the flagged vocabulary itself. First, within the US, the full-text estimates point in the same direction as the abstract-level ones for the papers that use the flagged words most densely, although most are imprecisely estimated: the response is not confined to abstract wording. Second, splitting the GRE indicator by whether a paper's core research question is about gender, race, or the environment shows that about half of the within-US decline comes from papers whose research question itself concerns these topics.\footnote{To see papers on these topics independently of the list, we also classify each abstract's topic with a separate call that does not use the list, and count a paper if it uses a flagged word in the context or discusses the topic. The list covers the vocabulary of gender and race well but that of the environment poorly. Before 2025, 85\% of the papers on gender or race use a flagged word in that context, against 25\% of the papers on the environment.}

Further, we study to what extent a ``mechanical'' effect of funding cuts could drive the results. We document that the reduction in word use is not confined to papers that acknowledge federal funding, which argues against a narrow direct-incentive effect. Indeed, the US government has no direct control over the content of working papers (even if research is federally funded). The ban is directly relevant only to existing and future federal grants. In this context, the impact of the word ban on research that does not acknowledge federal funding points toward self-censorship. 
Relatedly, the estimates are essentially unchanged when we control for university-level DEI policy changes (such as closing or renaming DEI offices) or for university-level financial distress signals. 

Finally, we find no High-Fed-specific decline in research output. US output falls relative to the UK after 2025, and more so for papers with GRE-related content, but these differences are imprecisely estimated. 

The current work relates to several strands of the literature. First, the pursuit of economic research is a topic of research in its own right \citep[e.g.,][]{Hamermesh2013,fourcade2015superiority,garg2025a}. A series of studies has investigated publication, citation, and implementation bias. The former, publication bias, refers to the fact that -- even if access to data is secured and the researcher has designed a credible empirical strategy -- significant results are more likely to be published, which may induce ``p-hacking'' by scholars \citep{brodeur2016star,brodeur2020methods,kasy2021forking}.\footnote{Retractions of published articles are relatively rare, but they can play an important role in correcting scientific knowledge when research findings are later identified as wrong \citep{Alabrese2022}.} Underpowered research designs could further skew evidence toward extreme findings \citep{ioannidis2017power}. Strategic behavior not only shapes which scientific findings are published but also influences citation practices. \cite{rubin2021systematic} show the importance of such citation bias, as research from the Journal of Business saw a decline in citations relative to other research after the journal's discontinuation. Another well-known type of bias is the so-called implementation bias, which refers to the fact that randomized control trials (RCTs) could be subject to external validity biases
\citep{chassang2012selective,banerjee2017decision}.

This existing work on scientific publications focuses on a set of biases related to the publication process, but it is not directly concerned with political pandering or reactions to political pressure. There is a growing literature on political censorship and propaganda in the media \citep[e.g.,][]{yanagizawadrott2014propaganda,guriev2019,Widmer2024}. However, the institutional structure and societal role of scientific research differ from those of the press, which has a more natural proximity to politics. In the broader educational sector, \citet{Cantoni2017} find that a politically motivated change in Chinese school curricula affected political attitudes. However, there is little evidence of political influence on the content of academic research. Still, experiences of scientific censorship appear to be relatively common in academia. A survey among US faculty in 2022 found that 4\% had been disciplined or threatened with discipline because of their research, academic talks, or non-academic publications \citep{honeycutt2023}.\footnote{In addition, researchers may adjust the way they communicate about science to the tastes of their audience \citep{Ratcliff2023}.} \cite{clark2023} argue that such scientific censorship is often driven by other researchers, who may be motivated by self-protection, benevolence toward peers, and prosocial concerns for the well-being of social groups. Biases in science funding, which may be at the root of self-censorship, are also understudied. Recent exceptions are by \citet{furnas2026}, who show that funders penalize US-Chinese scientific collaborations compared to US-German teams, and \cite{aghion2023does} who study the China Initiative, which made administrative procedures more complicated and funding less accessible for collaborative projects between Chinese and US researchers. They find that it has hampered the academic careers of Chinese scholars and Chinese-US collaborations. More broadly, \citet{iaria2018frontier} demonstrate that the output of researchers cut off from foreign frontier knowledge fell during and after World War I, when international scientific cooperation collapsed. Their work illustrates how political disruptions can reshape how science is produced.\footnote{\citet{waldinger2012peer} also exploits political interference in academia -- the dismissal of scientists in Nazi Germany -- to study peer effects in science.} Some studies also measure researchers' political biases. \citet{garg2025b} find that US academics who express themselves on Twitter diverge from general public opinion in topic focus, while \citet{AlabreseCapozzaGarg2024} show that political expression by scholars on social media may undermine their credibility.\footnote{This is also related to the literature on people's willingness to publicly express views when at risk of being sanctioned \citep{bursztyn2020extreme,braghieri2024political}.} 

This recent work does not speak to the extent to which scientific content itself is biased. On that question, \citet{jelveh2024political} and \citet{borjas2026ideological} find that economists' individual political attitudes influence their research. While there is evidence that institutional and political forces can change the behavior of institutional actors \citep[e.g.,][]{ash2025ideas,grosjean2023inflammatory,vandeneynde2018}, to the best of our knowledge, there is little evidence on how exogenous political shocks affect scholars' choices of research content, potentially biasing the production of scientific knowledge. Our paper addresses this gap.

The remainder of this paper is structured as follows. Section \ref{sec_data} presents the data and empirical framework, whereas Section \ref{sec_results} presents the findings. Section \ref{Sec_mechanisms} examines the mechanisms and channels of transmission. Finally, Section \ref{sec_conclusion} concludes. Further data description and additional empirical results are relegated to the Appendix.

\section{Data and Empirical Framework}\label{sec_data}

In what follows, we describe the data used and the empirical framework applied.

\subsection{Data}

\paragraph{Working papers.} We collect the universe of working papers published by the National Bureau of Economic Research (NBER) and the Centre for Economic Policy Research (CEPR) from January 2020 through June 2026, and complement them with economics working papers posted on arXiv over the same period.\footnote{\citet{garg2025a} also rely on a sample of CEPR and NBER working papers, studying economic research methodologies over time.} We obtain NBER papers via the NBER API, which provides structured metadata including title, authors (with profile URLs), abstract, publication date, and program affiliations. JEL codes are not part of that API metadata, so we extract them from the second page of each paper's PDF (Appendix~\ref{app_subsec_gre_core}).\footnote{The API endpoint is \texttt{https://www.nber.org/api/v1/working\_page\_listing/}.} We collect CEPR papers by scraping the CEPR discussion paper listing pages, extracting titles, authors, abstracts, dates, keywords, and JEL codes. For arXiv, we use the arXiv API to collect all papers listed in the three economics categories (econ.EM, econ.GN, and econ.TH). The API provides titles, authors, abstracts, submission dates, and subject categories.\footnote{The API endpoint is \texttt{https://export.arxiv.org/api/query}.} 
For all three series, we extract the paper text to obtain full-text term counts along with the abstract counts.

Since some papers are published in more than one series, we de-duplicate across the three series.\footnote{We match on titles (lowercased, with the CEPR discussion-paper prefix removed and non-alphanumeric characters collapsed; we focus on titles of at least 15 characters). We merge potential duplicate records only if they share at least one author surname.} Each paper enters as its first public release in any of the three series.
After de-duplication, our dataset contains 28,081 unique papers: 27.7\% from NBER, 22.6\% from CEPR, and 49.8\% from arXiv (Table~\ref{tab:summary_statistics}).\footnote{NBER and CEPR are independent, nonpartisan organizations run by academics for academics. While formally working papers are approved by program directors, \textit{de facto} the content of working papers is typically up to the affiliated researcher.}

\paragraph{Author affiliations and funding data.} We extract author affiliations from the second page of each PDF using OpenAI's open-weight model GPT-OSS, which identifies each author's university affiliation(s).
When an author lists multiple universities, we match the combined string to the institution found in the HERD funding reference list (see below). If both appear, we keep the affiliation mentioned first. \footnote{We extract affiliations separately for each paper, so an author who changes institutions lists different affiliations on different papers. The paper-level analysis and the team-majority assignment use the affiliations on each paper. The own-university assignment holds each author's affiliation fixed at their last paper before 2025 (Section~\ref{sec:methods}).}
We identify authors by name across papers and series: names that differ only in accents, case, punctuation, hyphens, or spacing are merged automatically. When variants are slightly different (e.g., John D. Doe vs. John Daniel Doe vs. John Doe), we send both names with their coauthors, affiliations, and paper titles to GPT-OSS. The model then decides whether they are the same person.%
With this procedure, we map 36,084 name strings to 34,541 authors.

Next, we match US universities to the Higher Education Research and Development (HERD) Survey (FY2024), which reports university-level R\&D expenditure by funding source. For UK universities, we rely on the Higher Education Statistics Agency (HESA),%
which provides analogous data on Research Council funding. To match the author-reported institution names from the papers to the standardized university names in the HERD and HESA lists, we use again the GPT-OSS model.
We extract an affiliation for 97.8\% of the author-paper observations.
Of the US affiliations, 86.5\% match a HERD university, and of the UK affiliations, 90.0\% match a HESA provider. The rest are mostly central banks, international organizations, government agencies, and firms.

Our key treatment variable, \textit{High-Fed}, indicates universities whose \emph{total-federal} funding share of total R\&D expenditure exceeds the median across all US universities in the HERD data. This is a broad measure of federal exposure: for US academic R\&D, the largest federal funders are the Department of Health and Human Services (which houses the National Institutes of Health), the Department of Defense, and the National Science Foundation (Table~\ref{tab:summary_statistics}). At the paper level, we classify papers by their author teams, resolving all ties toward treatment: a paper is a US paper if it has at least as many US-affiliated (HERD-matched) as UK-affiliated (HESA-matched) authors and at least one US-affiliated author, and a UK paper if UK-affiliated authors are in the strict majority. A US paper is High-Fed if at least half of all its authors are at High-Fed universities, counting UK-affiliated and unmatched authors as not High-Fed, and Low-Fed otherwise. At the author level, we report two complementary treatment assignments, described in Section~\ref{sec:methods}. For the UK, we construct an analogous indicator based on the Research Council (RC) funding share from the HESA data. Of the 28,081 papers, 15,022 are classified as US papers -- 7,621 High-Fed and 7,401 Low-Fed -- and 2,943 as UK papers. Paper-level summary statistics are in Table~\ref{tab:summary_statistics}; author-level summary statistics are in Appendix Table~\ref{tab:summary_statistics_author}.

\paragraph{Department rank.} As a measure of institutional research standing, we collect the RePEc\slash IDEAS ranking of economics departments.\footnote{\texttt{https://ideas.repec.org/top/top.econdept.html}. Institutions are ordered by a composite score: the harmonic mean of an institution's ranks across numerous bibliometric criteria, including publications, citations, and downloads. Lower scores denote \textit{higher} standing. We keep each institution's best-ranked entry.} We match these ranks to author affiliations by institution name.
For each paper, we average its departments' ranks and split papers at the median, distinguishing papers at higher- versus lower-ranked departments.

\paragraph{Targeted terms.} We use the list of terms flagged by US federal agencies following the executive orders of January 2025, as documented by \citet{yourish2025words}. This compilation contains 197 terms -- morphological variants included (e.g., ``barrier'' and ``barriers,'' ``bias'' and ``biased'') -- among them ``diversity,'' ``equity,'' ``inclusion,'' ``women,'' ``climate crisis,'' ``gender identity,'' and ``racial justice'' (see Appendix Table~\ref{tab:censored_words} for the complete list). We expand the list to 224 search patterns by adding UK spellings (e.g., ``marginalise''), hyphenation variants (e.g., ``anti-racism'' and ``antiracism''), and expansions of the list's combined forms.%

Several expressions in the list are nested. For instance, ``gender,'' ``gender-based,'' and ``gender-based violence'' are all separate entries. In such cases, we count the longest match, here ``gender-based violence'' instead of three matches.
We apply the same procedure to abstract-level and full-text counts. The expression matches whole words only and ignores case, so ``race'' does not match ``trace.'' On average, papers contain 0.91 targeted terms in the abstract and 60.6 in the full text (Table~\ref{tab:summary_statistics}).

A related official compilation is Appendix B of the Senate Commerce Committee minority-staff report on NSF grants, published in October 2024. The NSF's grant screening in February 2025 reportedly tracked this keyword list closely (see the Introduction). The appendix contains 696 keywords in five committee-assigned categories (gender, race, environmental justice, social justice, and status). It overlaps with our list (130 of our 197 base terms appear in it directly, and 133 of its keywords are identical to one of our 224 search patterns) but also contains vocabulary absent from the \citet{yourish2025words} compilation, such as environmental terms like ``climate change,'' ``net zero,'' and ``ESG.'' We use it in a robustness check.

\paragraph{Context classification.} Many targeted terms are ambiguous in economics -- ``equity'' may refer to financial equity, ``bias'' to statistical bias, and ``race'' to a competitive contest: Recall that abstracts contain 0.91 flagged terms on average. Hence, we need to distinguish substantive from incidental uses to capture the categories most closely associated with what the Trump administration characterizes as ``woke''.

To this end, we classify each distinct targeted term appearing in an abstract into one of five thematic categories (Gender, Race, Environment, Economic inequality, and Cannot say or other). We use OpenAI's open-weight GPT-OSS model for this classification.\footnote{Using the open-weight model as opposed to, say, ChatGPT, improves reproducibility and reduces costs.}
Appendix~\ref{app_subsec_classification} details the classification prompt. We validate the classification against the JEL codes assigned to NBER and CEPR papers. The authors assign these codes, and they are not derived from the abstract's vocabulary (Appendix~\ref{app_subsec_gre_core}, Table~\ref{tab:app_jel_gre}). Of the papers we classify as Gender, 35\% list J16 (Economics of Gender), against 0.3\% of all other papers. For Race and the codes J15 or J71 (Minorities, Race, and Migrants; Discrimination), the shares are 37 against 1.1\%; for Environment and the Q5 codes (Environmental Economics), 71 against 3.9\%. The codes with the highest odds of appearing on a GRE paper are J16, J71, and J15; the lowest are mostly finance and monetary codes.

Since each word is classified separately, papers with multiple targeted terms can appear in more than one category. Figure~\ref{fig:wordcloud_categories} displays odds-ratio word clouds for the four substantive categories.\footnote{Each word is scaled by its log-odds ratio of appearing in an abstract classified in a given category relative to abstracts not in that category, computed from unigram and bigram count vectorizations of all classified abstracts.} The Gender cloud is dominated by terms like ``women,'' ``female,'' ``gender gap,'' and ``gender differences''; Race by ``racial,'' ``racial disparities,'' ``black workers,'' and ``black americans''; Environment by ``pollution,'' ``air pollution,'' ``environmental quality,'' and ``wildfire smoke''; and Economic inequality by ``consumption inequality,'' ``wealth inequality,'' ``Gini coefficient,'' and ``income inequality.''
Our main analysis focuses on papers classified as Gender, Race, or Environment (GRE), which directly relate to the stated rationale behind the word ban \citep{yourish2025words}.

For robustness, we also build a measure that is independent of the GPT-OSS classifier: how densely targeted terms are used in the full paper text. We build an indicator for whether a paper's targeted-term mentions per page are at or above the median or the 90th percentile of the pre-2025 distribution. 

\paragraph{Core-question annotation.} Beyond classifying the context of individual terms, we annotate each paper containing at least one targeted term used in a Gender, Race, or Environment context (3,008 papers) by whether its \emph{core research question} is about gender, race, or the environment, or whether the GRE content is peripheral to a different question (for example, a side finding). Each abstract is judged by the GPT-OSS model in a separate, independent call, without any information about which terms were flagged. 62\% of GRE papers are annotated as core. The prompt, the protocol, and an external validation against JEL codes are in Appendix~\ref{app_subsec_gre_core}.

\paragraph{Topic classification.} The context classification and the core-question annotation are both based on the flagged words. A paper that discusses gender, race, or the environment without any flagged word would not be captured by these measures. Therefore, we also classify what each paper is about in the spirit of a supervised topic model: for each abstract, we send a call to the same large language model as before, GPT-OSS, to answer whether the abstract discusses gender, race or ethnicity, or the environment or climate. The prompt does not mention the list or point to any word in the abstract (Appendix~\ref{app_subsec_topics}). Before 2025, 9.2\% of papers discuss gender or race, and 7.9\% discuss the environment. This is a different measure, not necessarily a more comprehensive one: it reads the abstract as an external annotator would and draws its own boundaries. For example, it does not always count sexual orientation as a gender topic, whereas the word ban does contain LGBT-related terms. Against the JEL codes, the topic annotation performs at least as well as the context classification (Appendix~\ref{app_subsec_topics}).

\paragraph{NSF and SES mentions.} We want to test whether our effects are only driven by papers that acknowledge NSF funding directly, rather by a broader response from federally dependent institutions. Thus, we flag papers that mention ``NSF'' or ``SES'' (the Division of Social and Economic Sciences, the NSF division most relevant to economics) in their acknowledgment sections. We use these flags, and analogous flags for any federal agency, to conduct analyses that exclude papers acknowledging such funding (Figure~\ref{fig:nsf_exclude}).

\paragraph{University DEI actions.} To measure universities' own policy responses, we use the Chronicle of Higher Education's tracker of institutional DEI changes \citep{chronicle2026dei}. This tracker documents actions taken by US institutions since January 2023. Examples of such actions are diversity offices closed or renamed, jobs cut or retitled, training changed, diversity statements dropped, DEI language removed from websites and official statements, and other DEI-related activities. The Chronicle of Higher Education provides a public source for every entry (317 institutions). For each institution, we record whether it took any documented action, whether the action involved removing DEI language from websites or official statements, and whether the institution's state had passed legislation restricting DEI (using the tracker's state-legislation classification). We match the tracker institutions to the HERD universities in our sample.%
Of our 376 HERD-matched universities, 208 have at least one documented action, 116 of them in states that had passed no DEI legislation, and 97 with website- or language-removal actions.

\paragraph{University distress announcements and funding shocks.} We also date, for each university, its first public announcement of a hiring freeze, layoffs, or cuts to doctoral admissions attributed to federal funding cuts. Sources are the Ithaka S+R University Research Workforce Tracker \citep{ithaka2026tracker}, which collects publicly reported workforce actions that institutions themselves attribute to federal funding cuts, complemented by contemporaneous press reports (\textit{Inside Higher Ed}, \textit{STAT}, BestColleges, and \textit{University Business}); events are dated to the month of the first public announcement. This yields 53 sample universities with a dated distress announcement, spread from January 2025 to September 2025. Separately, we record the nine sample universities whose federal funding was publicly frozen or canceled at the level of the whole institution during our sample period (for example, the spring-2025 funding freezes at Columbia and Harvard), again dated to the month of the first action. These events capture publicly announced, university-wide actions only; individual grant terminations and changes in agency priorities are not observed. Matching to our sample uses the same deterministic name-matching procedure as for the DEI tracker.

\subsection{Empirical Framework}\label{sec:methods}

Our identification strategy exploits heterogeneous exposure to the policy shock that unfolded in January/February 2025. In what follows, we first show that the ``treatment group'' of more exposed scholars (i.e., those from US universities more dependent on federal funding) had a comparable pre-trend with the ``control group'' of less exposed scholars. Then, as the ``stop list'' of flagged words was publicized in February 2025, one might expect a different reaction in the treatment group than in the control group, which we investigate empirically.

Economics is also a discipline in which the direct financial stake in federal research funding is limited. Only 6.5\% of US papers in our sample acknowledge NSF funding (17.7\% any federal agency), and at the university level economics accounts for a median of 0.00\% of total federal R\&D expenditure. 53\% of in-sample universities come without any federally financed economics R\&D recorded  (Appendix Table~\ref{tab:app_fedfunding}). This leaves limited room for a direct financial channel. Since economics receives relatively little federal money, the direct loss of grant income alone is unlikely to explain a change in the content of economics research. We address this question in detail in Section~\ref{Sec_mechanisms}.

Note that we focus on the \textit{relative} reactions of more exposed versus less exposed scholars with respect to the policy change. Our setting allows us to detect to what extent academics react to political pressure, but we are not able to make general statements about the absolute levels of self-censorship, i.e., our data would not allow us to know if, during the pre-treatment period, scholars have self-censored themselves, and if yes, in what direction. Importantly, however, because our identification strategy is based on heterogeneous effects from a discrete policy change, we do not need such information on absolute levels. In line with our research question, we measure how much more the research content of more exposed academics changes in response to (perceived) political pressure.

\paragraph{Event study.} Our main specification is a yearly event study estimated as a linear probability model:
\begin{equation}\label{eq:eventstudy}
Y_{it} = \sum_{\substack{s=2020\\s\neq 2024}}^{2026} \beta_s \cdot \mathbf{1}[t=s] \times \text{HighFed}_i + \lambda \cdot \text{HighFed}_i + \theta_t \cdot \text{Cap}_i + \delta \cdot \text{Source}_i + \gamma_t + \varepsilon_{it}
\end{equation}
where $Y_{it}$ is a binary indicator equal to one if paper $i$ published in year $t$ contains any targeted term in a Gender, Race, or Environment context in its abstract, $\text{HighFed}_i$ indicates whether at least half of all the paper's authors are at above-median-federal-share universities, $\gamma_t$ are year fixed effects, and $\text{Source}_i$ is an NBER/CEPR/arXiv source indicator. The reference period is 2024, the last full year before the policy change. The coefficients of interest, $\beta_s$, trace the differential evolution of flagged-term usage between High-Fed and Low-Fed papers over time. As the exposure to federal funding could capture differences in the production of research that are unrelated to the funding environment, we control for time-varying effects of the economics research capacity of the paper's universities (interacted with time dummies). $\text{Cap}_i$ is the logarithm of one plus the number of papers in our sample dated before 2025 with at least one author at the university, averaged over the paper's US- and UK-affiliated authors. With year-specific coefficients $\theta_t$, the dynamic coefficients are net of differential trends by size; each university's capacity is measured before the policy change, so it cannot respond to treatment, and it enters every regression in the paper (interacted with Post in the difference-in-differences specifications).

At the paper level, we use heteroskedasticity-robust standard errors. At the author level, the unit of observation becomes the author-paper pair: a paper with $K$ coauthors contributes $K$ observations. We use two treatment assignments. Under the first, we estimate
\begin{equation}\label{eq:eventstudy_author}
Y_{ijt} = \sum_{\substack{s=2020\\s\neq 2024}}^{2026} \beta_s \cdot \mathbf{1}[t=s] \times \text{HighFed}_j + \theta_t \cdot \text{Cap}_j + \alpha_j + \gamma_t + \delta \cdot \text{Source}_i + \varepsilon_{ijt}
\end{equation}
where $j$ indexes authors, $i$ papers, and $t$ years. $\text{HighFed}_j$ is the status of author $j$'s own university, held fixed at the last pre-2025 affiliation (for authors whose first paper is dated 2025 or later, the affiliation on that first paper); it does not vary within author, so its main effect is absorbed by the author fixed effects $\alpha_j$, and $\text{Cap}_j$ is that university's capacity. Under the second assignment, the treatment is the classification of the paper and varies within author:
\begin{equation}\label{eq:eventstudy_teammaj}
Y_{ijt} = \sum_{\substack{s=2020\\s\neq 2024}}^{2026} \beta_s \cdot \mathbf{1}[t=s] \times \text{HighFed}_i + \lambda \cdot \text{HighFed}_i + \theta_t \cdot \text{Cap}_i + \alpha_j + \gamma_t + \delta \cdot \text{Source}_i + \varepsilon_{ijt}
\end{equation}
where $\text{HighFed}_i$ is the paper-level classification of equation~(\ref{eq:eventstudy}), that is, whether at least half of all the paper's authors are at above-median-federal-share universities. In both equations, $\alpha_j$ are author fixed effects, and standard errors are two-way clustered by university and paper, with US universities identified by their HERD match and UK universities by their HESA provider code (so that both the shared-institution and the shared-paper dependence across author-paper observations are accounted for).\footnote{As a robustness check on inference, we additionally recompute the paper-level estimates with standard errors clustered by the paper's modal university and with a wild cluster bootstrap; these results are presented alongside the other robustness checks. We also estimate pooled pre- and post-treatment averages and report the difference between them. }
As robustness checks, we estimate specifications with count-dependent variables (full-text word counts) using Poisson pseudo-maximum likelihood (PPML), controlling for the number of pages.

\paragraph{Reference group comparisons.} To assess robustness across different control groups, we estimate pooled difference-in-differences specifications:
\begin{equation}\label{eq:did}
Y_{it} = \beta \cdot \text{Post}_t \times \text{Treated}_i + \theta_1 \cdot \text{Post}_t \times \text{Cap}_i + \theta_2 \cdot \text{Cap}_i + \delta_1 \cdot \text{Post}_t + \delta_2 \cdot \text{Treated}_i + \delta_3 \cdot \text{Source}_i + \varepsilon_{it}
\end{equation}
where $\text{Post}_t = \mathbf{1}[\text{year} \geq 2025]$ and $\text{Treated}_i$ is a binary indicator that varies across comparisons, ordered by exposure to federal funding: US High-Fed vs.\ US Low-Fed, US High-Fed vs.\ UK, US Low-Fed vs.\ UK, and US (all) vs.\ UK. The same logic extends to all pairwise comparisons shown in Figure~\ref{fig:coefplot_refgroups}. Paper-level regressions use robust standard errors; author-level regressions replace $i$ with $ij$ subscripts and add author fixed effects $\alpha_j$ as in equations~(\ref{eq:eventstudy_author}) and~(\ref{eq:eventstudy_teammaj}), with standard errors two-way clustered by university and paper.

At the author level, we report two treatment assignments throughout. The first assigns each author the High-Fed (or US/UK) status of their own university. We hold this status fixed at the author's last affiliation before 2025. With author fixed effects, this specification captures how the same author changes under an exposure that is fixed in advance. The second assigns each author-paper observation the High-Fed (or US/UK) status of the paper itself, that is, of the majority of its team. The outcome is a property of the paper and shared by all of its coauthors. The team-majority assignment therefore matches the treatment to the paper. It asks whether the same author writes less sensitive content when working on a federally exposed team. This assignment is contemporaneous by construction, because authors choose their teams. We therefore read it as showing where the content response occurs, not as exposure fixed in advance.\footnote{In practice, however, team composition is stable across the policy change: within author, the probability of working on a majority-High-Fed team changes by only $-0.020$ ($p = 0.14$) after 2024 for authors from High-Fed relative to Low-Fed universities, so there is no sign of a general re-sorting of authors across teams after 2024. The two assignments coincide for 79\% of author-paper observations (79.2\% before 2025 and 80.1\% after): the team-majority measure largely captures own-affiliation exposure while additionally allowing for authors who publish across the High/Low boundary.}

\paragraph{Heterogeneity measures.} To study whether the self-censorship response varies across author subgroups, we construct two author-level characteristics. \textit{Gender} is classified as male or female from authors' first names using GPT-OSS (Appendix~\ref{app_subsec_gender}). \textit{Ethnicity} is predicted from the US Census 2010 surname data file, which contains approximately 162,000 surnames with associated race/ethnicity probabilities: each author's last name is looked up in the Census file, the highest-probability category is assigned, and we use the four largest groups -- White, Asian/Pacific Islander, Hispanic, and Black -- in the heterogeneity analysis; surnames not in the Census file (predominantly non-English names) remain unclassified. We additionally record seniority (years since PhD, collected via web searches; Appendix~\ref{app_subsec_phd}) and classify authors with more than 12 years since the PhD as senior (Table~\ref{tab:summary_statistics_author}).

\section{Results}\label{sec_results}

\begin{figure}[htbp]
\centering
\includegraphics[width=\textwidth]{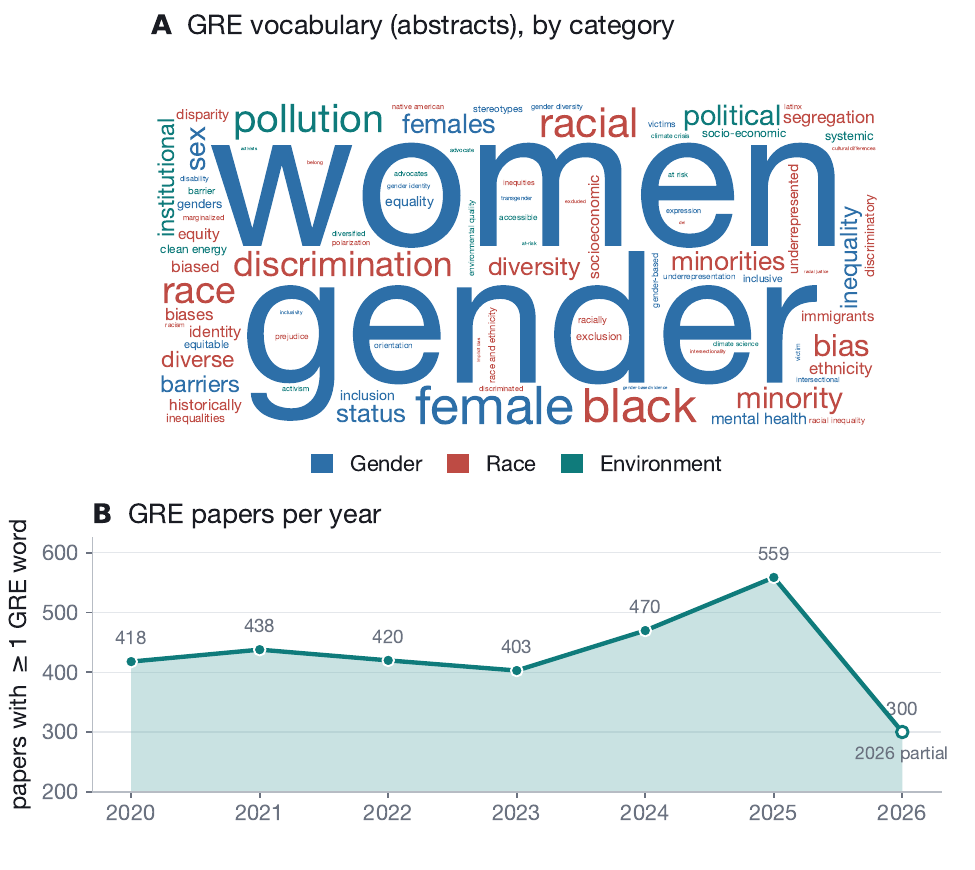}
\caption{Targeted Terms and Gender, Race, and Environment Content in Economics Working Papers}
\label{fig:barchart_gre}
\begin{minipage}{0.95\textwidth}
\footnotesize
\textit{Notes:} This figure describes the raw data. The word cloud displays the most frequent targeted terms used in a Gender, Race, or Environment (GRE) context in the abstracts of NBER, CEPR, and arXiv working papers (January 2020 through June 2026), with word size given by the number of abstracts in which the word is classified as Gender, Race, or Environment, and color by the most frequent of these three categories for the word (Economic inequality and Cannot say labels are not counted). The accompanying panel plots the yearly number of papers containing at least one targeted term used in a GRE context, as classified by GPT-OSS based on the surrounding text in the abstract; the 2026 count covers January through June. Targeted terms are the words flagged by US federal agencies following the executive orders of January 2025.
\end{minipage}
\end{figure}

In what follows, we first discuss key features of the raw data, then present the main econometric results. Figure~\ref{fig:barchart_gre} provides a first look at the data. The word cloud displays the most frequent targeted terms used in a Gender, Race, or Environment (GRE) context in paper abstracts; it is dominated by terms such as ``women,'' ``gender,'' ``female,'' ``black,'' and ``racial,'' alongside environment-related terms like ``pollution'' and ``clean energy.'' The accompanying panel plots the yearly number of papers containing at least one targeted term used in a GRE context -- the categories most closely associated with the stated rationale behind the word ban. The yearly number of such papers is roughly flat from 2020 to 2023, %
and rises in 2024 and 2025 to more than 500 per year.
We now examine how the prevalence of this sensitive content evolves differentially by institutions' exposure to federal funding.

\begin{figure}[htbp]
\centering
\includegraphics[width=\textwidth]{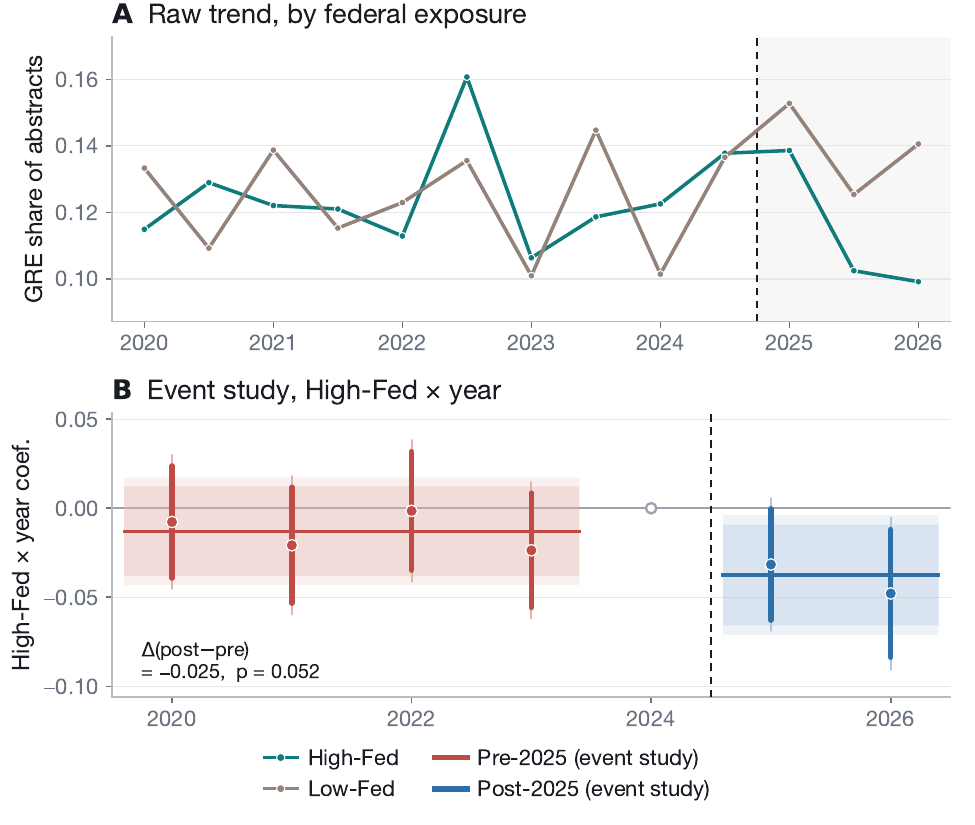}
\caption{Effect on Gender, Race, and Environment Content: Raw Trends and Event Study}
\label{fig:event_studies}
\begin{minipage}{0.95\textwidth}
\footnotesize
\textit{Notes:} Panel A plots the raw share of papers containing at least one targeted term in a Gender, Race, or Environment (GRE) context, separately for US High-Fed and Low-Fed institutions; the two groups follow parallel pre-trends and diverge after the policy change. Panel B plots yearly difference-in-differences coefficients comparing US High-Fed to US Low-Fed institutions, with 2024 (the last full year before the policy change) as the reference period. High-Fed universities are those whose total-federal share of R\&D expenditure (HERD FY2024, all fields) exceeds the cross-university median; Low-Fed universities are at or below the median. At the paper level, a US paper (at least as many US- as UK-affiliated matched authors, ties assigned to US) is classified as High-Fed if at least half of all its authors are at High-Fed universities (UK-affiliated and unmatched authors count as not High-Fed; ties assigned to High-Fed). The dependent variable is a binary indicator for whether the paper contains any targeted term used in a GRE context in its abstract (GPT-OSS classification). In Panel A, the gray area marks the post-2025 period; the year 2026 covers January through June. In Panel B, red shading indicates pre-treatment years and blue shading post-treatment years. Darker bands represent 90\% confidence intervals; lighter bands represent 95\% confidence intervals. Horizontal lines show pooled pre- and post-treatment averages. Paper-level estimates include source (NBER/CEPR/arXiv) fixed effects and use robust standard errors. All specifications control for economics research capacity interacted with year (Section~\ref{sec:methods}). Linear probability model estimates. See Appendix Figure~\ref{fig:raw_gre_author} for the author-level raw trend and Appendix Figure~\ref{fig:es_author} for the author-level event studies of equations~(\ref{eq:eventstudy_author}) and~(\ref{eq:eventstudy_teammaj}).
\end{minipage}
\end{figure}

\paragraph{Event study.} Figure~\ref{fig:event_studies} presents our main result. Panel A plots the raw share of papers containing at least one targeted term in a Gender, Race, or Environment context, separately for US High-Fed and Low-Fed institutions. The two groups follow parallel pre-trends -- the two series cross repeatedly, without a systematic level gap -- and diverge after the policy change: the High-Fed share drops while the Low-Fed share, if anything, rises. Panel B plots yearly event study coefficients from equation~(\ref{eq:eventstudy}), comparing the differential evolution of GRE-content prevalence between US High-Fed and Low-Fed institutions. The pre-treatment coefficients are stable from 2020 to 2023. They sit slightly below the 2024 reference year and show no trend toward the treatment effect. After the policy change, the paper-level coefficient drops %
sharply. Pooling the years, the difference between the post- and pre-period averages is $-0.025$ ($p = 0.052$), so the share of papers with GRE terms drops by 2.5 percentage points. At the paper level, the decline grows from 2025 to 2026, suggesting a gradual response to an evolving environment. The author-level event studies of equations~(\ref{eq:eventstudy_author}) and~(\ref{eq:eventstudy_teammaj}), which add author fixed effects that absorb time-invariant individual characteristics, are reported in Appendix Figure~\ref{fig:es_author}. Under the team-majority assignment the event study mirrors the paper-level pattern: no pre-treatment coefficient is statistically distinguishable from zero, and the post-treatment average lies $0.023$ below the pre-treatment average ($p = 0.090$). When we assign treatment by the author's own university, the within-US contrast is smaller and not statistically significant.

\begin{figure}[htbp]
\centering
\includegraphics[width=\textwidth]{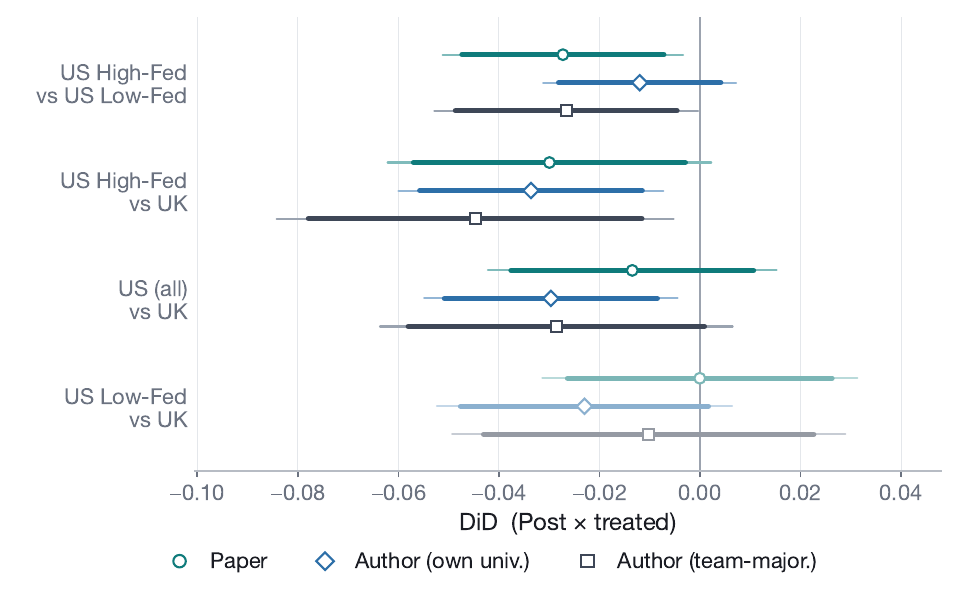}
\caption{Treatment Effects Across Reference Groups: Gender, Race, and Environment Content}
\label{fig:coefplot_refgroups}
\begin{minipage}{0.95\textwidth}
\footnotesize
\textit{Notes:} This figure displays pooled difference-in-differences coefficients (Post $\times$ Treated) for a range of treatment-control comparisons: US High-Fed vs.\ US Low-Fed, US High-Fed vs.\ UK, all US vs.\ UK, and US Low-Fed vs.\ UK. The US Low-Fed vs.\ UK comparison, in which neither group is highly exposed to federal funding, is drawn in lighter shades. Each row shows three estimates: the paper level (teal circles); the author level with treatment at the author's own basefixed university (blue diamonds); and the author level with treatment given by the paper's team classification (``team majority,'' slate squares). Post indicates papers dated 2025 or later. The dependent variable is a binary indicator for whether a paper contains any targeted term used in a Gender, Race, or Environment (GRE) context in its abstract (GPT-OSS classification). High-Fed universities are those whose total-federal share of R\&D expenditure (HERD FY2024, all fields) exceeds the cross-university median; Low-Fed universities are at or below the median. US papers have at least as many US-affiliated (HERD-matched) as UK-affiliated (HESA-matched) authors (ties assigned to US); UK papers have strictly more UK-affiliated authors. A US paper is High-Fed if at least half of all its authors are at High-Fed universities (UK-affiliated and unmatched authors count as not High-Fed; ties assigned to High-Fed). Thick bars represent 90\% confidence intervals; thin lines represent 95\% confidence intervals. Paper-level estimates use robust standard errors; author-level estimates include author fixed effects and cluster standard errors two-way by university and paper. All regressions control for economics research capacity interacted with Post (Section~\ref{sec:methods}). Linear probability model estimates. See Appendix Figures~\ref{fig:coefplot_2024pre}--\ref{fig:coefplot_anyword_fulltext}, \ref{fig:coefplot_redblue}, and~\ref{fig:coefplot_altinf} for robustness checks; a tabular version is in Table~\ref{tab:fig3}.
\end{minipage}
\end{figure}

\paragraph{Reference group comparisons.} Our main specification compares scholars in High- and Low-Fed US institutions, but our data allows us to construct a range of alternative treatment-control comparisons. Figure~\ref{fig:coefplot_refgroups} presents pooled difference-in-differences estimates from equation~(\ref{eq:did}) across such different treatment-control comparisons, at the paper level and under both author-level treatment assignments. At the paper level, the coefficient of our main specification is $-0.027$ ($p = 0.025$), comparing US High-Fed versus Low-Fed papers. We find similar reductions ($-0.030$, $p = 0.068$) for the comparison of US High-Fed versus UK papers. In contrast, US Low-Fed papers show no change against the UK -- confirming the importance of the exact funding environment for our findings. At the author level, we present results for two assignments that complement each other. When we assign treatment by the author's own university, both US-versus-UK comparisons are negative: $-0.034$ ($p = 0.012$) for US High-Fed versus the UK and $-0.030$ ($p = 0.021$) for all US versus the UK. However, the within-US High-Fed versus Low-Fed contrast is small ($-0.012$, $p = 0.221$), and the Low-Fed versus UK estimate is not statistically significant either ($-0.023$, $p = 0.12$). %
But, with the team-majority assignment, the author-level estimates mirror the paper level instead: the within-US contrast is $-0.027$ ($p = 0.047$) and High-Fed versus UK is $-0.045$ ($p = 0.026$), while Low-Fed versus UK is close to zero%
. Under the team-majority assignment, the same author thus writes less sensitive content when publishing with a majority-High-Fed team. In general, the findings of Figure~\ref{fig:coefplot_refgroups} show that the reduction in GRE content from universities that rely most on federal funding holds across a range of alternative treatment-control comparisons.

\paragraph{Robustness.} We check the robustness of the reference-group comparisons in several ways. The corresponding exhibits can be found in the Appendix.

\bigskip

\textit{Time window.} First, we shorten the pre-period to the two years before the policy change, so that 2023--2024 is compared with 2025--2026. In this specification, within-US estimates remain unchanged or become stronger (Appendix Figure~\ref{fig:coefplot_2024pre}). 
The comparisons with the UK shrink somewhat in this window. At the author level, US High-Fed versus UK remains statistically significant at the 10\% level under both assignments.

\bigskip

\textit{Continuous exposure.} Second, we replace the binary median split with the continuous total-federal share, standardized within the estimation sample (Appendix Figure~\ref{fig:coefplot_contdose} and Table~\ref{tab:app_contdose}). The paper-level decline holds: $-0.014$ per standard deviation ($p = 0.020$). The author-level estimates remain negative but miss the 10\% level, narrowly so under the team-majority assignment%
.

\bigskip

\textit{Full paper (term density).} Third, we move from abstracts to full paper texts with a measure that needs no context classification: an indicator for whether a paper's targeted-term mentions per page are at or above the median (0.67) or the 90th percentile (2.87) of the pre-2025 distribution (all papers, all three series pooled), estimated as a linear probability model (Appendix Figure~\ref{fig:app_dense}). Full-text measures are inherently noisier -- papers contain 60.6 targeted terms on average, compared with 0.9 in abstracts (Table~\ref{tab:summary_statistics}). 
At the median, we find no significant changes around the flagged word policy. 
In contrast, we find declines in the top decile (US High-Fed versus Low-Fed: $-0.020$, $p = 0.097$ at the paper level; $-0.017$, $p = 0.062$ and $-0.028$, $p = 0.031$ under the own-university and team-majority author assignments). The full-text response is thus among the papers that use the flagged words most densely.

\bigskip

\textit{Without context classification.} Fourth, we drop the context classification step entirely and use the raw targeted-term dictionary: an indicator for whether any targeted term appears anywhere in the abstract, and the raw count of targeted words in the full paper text, estimated by PPML with a page-count control (Appendix Figures~\ref{fig:coefplot_anyword} and~\ref{fig:coefplot_anyword_fulltext}). At the abstract margin, the raw indicator shows no differential change at any level%
; in the full text, the raw count declines at the own-university author level ($-0.08$, $p = 0.042$), while the paper-level and team-majority estimates are not statistically distinguishable from zero. As anticipated in Section~\ref{sec_data}, many targeted terms have incidental technical uses in economics, so the raw dictionary mixes uses in a gender, race, or environment context with technical and incidental ones; a top-decile indicator, which needs no classification, declines under all three assignments.

\bigskip

\textit{Broader vocabulary.} Fifth, we add the 563 Senate-appendix keywords that are not already among our search patterns (Section~\ref{sec_data}). They include 34 environmental terms, such as ``climate change,'' ``ESG,'' and ``net zero,'' which the \citet{yourish2025words} list lacks. The GRE estimates are similar or larger (Appendix Figure~\ref{fig:cruz_aug}): $-0.024$ ($p = 0.053$) within the US, against $-0.027$ with our vocabulary, and $-0.038$ ($p = 0.029$) for US High-Fed versus UK, against $-0.030$. The Environment component is $-0.005$ ($p = 0.46$) within the US, against $-0.010$, and $-0.019$ ($p = 0.077$) for US High-Fed versus UK, against $-0.014$.

\bigskip

\textit{Additional controls.} Sixth, adding an interaction between an indicator for universities in states that voted Republican in 2024 and the post-treatment dummy leaves all estimates essentially unchanged (Appendix Figure~\ref{fig:coefplot_redblue}).
Hence, our main findings do not appear to be driven by the broad state-level political climate.
Similarly, allowing for a differential post-2025 shift by departments' research standing -- the average RePEc/IDEAS rank of a paper's departments, interacted with Post -- leaves the US High-Fed versus Low-Fed estimate essentially unchanged; see Panel A of Appendix Figure~\ref{fig:rank}. So, the effect is not an artifact of department quality. Heterogeneity by department quality is discussed in Section~\ref{Sec_mechanisms} below.

\bigskip

\textit{Alternative inference.} Seventh, we recompute the paper-level estimates with standard errors clustered by the paper's modal university and with a wild cluster bootstrap (Rademacher weights, 99,999 replications, imposing the null; Appendix Figure~\ref{fig:coefplot_altinf} and Table~\ref{tab:app_altinf}). The within-US High-Fed versus Low-Fed estimate is unchanged (cluster-robust $p = 0.015$; wild bootstrap $p = 0.020$), as is US High-Fed versus UK ($p = 0.052$ and $p = 0.054$ respectively)%
. Author-level estimates are two-way clustered by university and paper unless noted otherwise.

\bigskip

\textit{Team composition.} Eighth, Low-Fed papers have more UK-affiliated and unmatched coauthors than High-Fed papers, since such coauthors count as not High-Fed in the paper rule (2.27 of 2.68 authors are US-matched for the average High-Fed paper, 1.88 of 3.33 for the average Low-Fed paper; Table~\ref{tab:summary_statistics}). We therefore re-estimate the within-US comparison on papers whose authors are all matched to HERD universities. On this subsample of 7,465 papers, the US High-Fed versus Low-Fed estimate is $-0.037$ ($p = 0.044$) at the paper level, against $-0.027$ in the full sample; the own-university and team-majority author estimates are $-0.015$ and $-0.029$, close to their full-sample values (Appendix Table~\ref{tab:app_teamcomp}).

\bigskip

\textit{Randomization inference.} Finally, we turn to randomization inference (Appendix Figure~\ref{fig:randinf}). For the within-US comparison we permute treatment: holding fixed the papers, authors, affiliations, outcomes, and dates, we randomly reassign High-Fed status across US universities, re-classify every paper and author from the reassigned labels, and re-estimate the same difference-in-differences on each of 100,000 draws. The observed paper-level estimate lies in the tail of the placebo distribution (two-sided permutation $p = 0.045$; $p = 0.040$ when High-Fed status is reassigned only among universities of similar size). The own-university author-level estimate, which is small and statistically insignificant in the baseline, is not distinguishable from the placebo draws ($p = 0.21$). Under the team-majority assignment, each author-paper observation takes the reassigned status of its paper. The team-majority estimate remains statistically significant at the 10\% level ($p = 0.067$; $p = 0.062$ with reassignment among universities of similar size). The comparisons with the UK do not lend themselves to a permutation of High-Fed status, because their contrast is between countries, which a reassignment among US universities does not randomize. For these comparisons we instead permute the outcome, randomly reassigning the binary GRE label across papers within each period (100,000 draws). The paper-level High-Fed versus UK estimate has a permutation $p$-value of $0.055$, and the own-university author-level High-Fed versus UK and US versus UK estimates remain statistically significant at the 10\% level ($p = 0.055$ and $p = 0.074$). So does the team-majority High-Fed versus UK estimate ($p = 0.050$).

\section{Mechanisms}\label{Sec_mechanisms}

The results so far show that researchers at federally dependent institutions reduced their use of targeted terms in gender, race, or environment contexts after the policy change. We now explore a series of dimensions of this response in order to understand the underlying mechanisms better.

\paragraph{Heterogeneity by department rank.} Universities differ in terms of prestige, and the academic standing of institutions could shape the response of researchers to political influence. Researchers at more prestigious institutions may feel more protected from political pressure. However, the higher visibility of their institutions may make them react more. In Appendix Figure~\ref{fig:rank}, we assess the role of institutional research standing, measured by the RePEc\slash IDEAS ranking of economics departments. The first exercise we conduct is essentially a robustness check of the main result. We confirm that the US High-Fed versus Low-Fed effect is essentially unchanged when we allow for a differential post-2025 shift by department rank%
, so the effect is not an artifact of department quality. We then present the proper heterogeneity analysis. Estimating the effect separately by rank half, the decline is, if anything, larger at lower-ranked departments (paper level: $-0.030$, $p = 0.091$, versus $-0.020$, $p = 0.234$ at higher-ranked ones), but these differences are not significant. The US-versus-UK comparison, in turn, is clearly concentrated in higher-quality institutions, a pattern that holds whether quality is split across the pooled sample or within each country. Taken together, these patterns do not paint a consistent picture about the role of the universities' prestige.

\paragraph{Heterogeneity by university size.} One could hypothesize that an institution's size affects the sensitivity to political pressure through the same channels as academic prestige. High-Fed universities may simply be the ones with more federal money at stake. If so, our effect would capture the amount of funding exposed, not the dependence on it. Appendix Table~\ref{tab:app_size} separates the two. We first add a control for the university's total federal R\&D dollars (in logs) interacted with Post. The High-Fed effect is, if anything, larger: $-0.032$ with and $-0.027$ without the control at the paper level, $-0.036$ and $-0.027$ under the team-majority assignment. We then enter the federal share and federal dollars jointly as continuous measures. The share predicts the decline ($-0.021$ per standard deviation, $p = 0.005$, at the paper level; $-0.020$, $p = 0.004$, team-majority). Dollars enter with a positive sign. Finally, we conduct a heterogeneity analysis and split universities at the median of federal dollars. The point estimates are negative at both smaller and larger recipients (paper level: $-0.017$ and $-0.027$; team-majority: $-0.031$ and $-0.026$), none of them is statistically significant on its own, and neither is the difference. The High-Fed gradient thus reflects dependence on federal funding; the amount of federal money at stake does not explain it.

\begin{figure}[htbp]
\centering
\includegraphics[width=\textwidth]{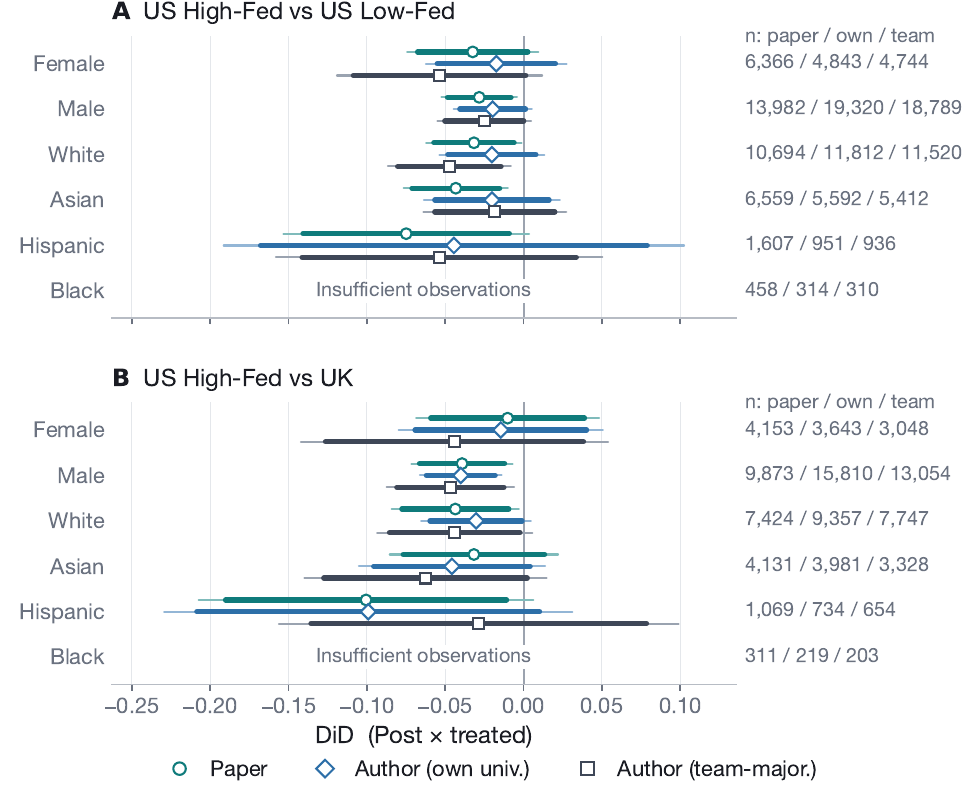}
\caption{Treatment Effects by Author Characteristics}
\label{fig:heterogeneity}
\begin{minipage}{0.95\textwidth}
\footnotesize
\textit{Notes:} This figure reports difference-in-differences coefficients (Post $\times$ Treated) estimated separately for subsamples of authors defined by predicted gender and Census-predicted ethnicity, for the US High-Fed versus US Low-Fed comparison (Panel A) and the US High-Fed versus UK comparison (Panel B). Teal circles are paper-level estimates (subsample: papers with at least one author of that group); blue diamonds are author-level estimates with treatment at the author's own basefixed university (subsample: authors of that group); slate squares assign the same authors the treatment of the paper's team classification (``team majority,'' as in Figure~\ref{fig:coefplot_refgroups}). Gender is classified from authors' first names; ethnicity is predicted from the US Census 2010 surname file (highest-probability category). Groups with fewer than 600 papers are labeled rather than plotted. The dependent variable is a binary indicator for whether a paper contains any targeted term used in a Gender, Race, or Environment (GRE) context in its abstract (GPT-OSS classification). High-Fed universities are those whose total-federal share of R\&D expenditure (HERD FY2024, all fields) exceeds the cross-university median. US papers have at least as many US-affiliated (HERD-matched) as UK-affiliated (HESA-matched) authors (ties assigned to US); UK papers have strictly more UK-affiliated authors. A US paper is High-Fed if at least half of all its authors are at High-Fed universities (UK-affiliated and unmatched authors count as not High-Fed; ties assigned to High-Fed). Thick bars represent 90\% confidence intervals; thin lines represent 95\% confidence intervals. Paper-level estimates use robust standard errors; author-level estimates include author fixed effects and cluster standard errors two-way by university and paper. All regressions control for economics research capacity interacted with Post (Section~\ref{sec:methods}). Linear probability model estimates. A tabular version is in Table~\ref{tab:fig7}.
\end{minipage}
\end{figure}

\paragraph{Heterogeneity by authors.} We might expect the response to political pressure to differ with the characteristics of the authors. Figure~\ref{fig:heterogeneity} reports the difference-in-differences coefficient (Post $\times$ Treated) estimated separately for subsamples of authors defined by predicted gender and Census-predicted ethnicity, for the US High-Fed versus Low-Fed and US High-Fed versus UK comparisons, at the paper level and under both author-level treatment assignments. The team-majority estimates can be read as follows: Does the same author produce less sensitive content when working on a High-Fed project after the policy change, and does that response differ between, say, men and women? For the US High-Fed versus Low-Fed comparison, the coefficient is negative and of comparable magnitude across gender (paper level: female $-0.032$, male $-0.028$) and across the predicted-White ($-0.032$) and predicted-Asian ($-0.043$) surname groups. Wald tests on the group interaction in pooled specifications, at both author levels, find no pair of the groups shown statistically distinguishable (all $p \geq 0.23$). These effects could be surprising at first sight, as we might have expected female scholars and scholars from minority ethnic backgrounds to be more vulnerable and, therefore, more responsive to political pressure. However, these scholars could also have a stronger intrinsic motivation to study GRE topics. The lack of heterogeneity along the gender and ethnicity dimensions could be a combination of these counteracting mechanisms.

We also find no evidence that the response differs by seniority, where senior authors are those with more than 12 years since the PhD. For the US High-Fed versus Low-Fed comparison, the team-majority estimate is $-0.033$ for senior and $-0.026$ for junior authors. The own-university estimates are $-0.018$ and $-0.024$ (tests of equality: $p = 0.95$ and $p = 0.56$).\footnote{Standard errors are between 0.014 and 0.019. Seniority is known for 89\% of the observations in the within-US estimation samples.
Authors without a recorded PhD year are excluded, and current PhD students are classified as junior. We collected PhD years for authors at US universities, so the split is limited to the within-US comparison.}
This result may appear counterintuitive, as one might expect junior scholars to be more vulnerable. However, like scholars from underrepresented groups, they could also be more motivated to work on GRE topics. In addition, senior scholars more often lead research groups and could weigh threats to funding more heavily. On balance, the combination of these mechanisms could explain the lack of heterogeneity.

In general, we find no evidence that the effect is concentrated in a particular demographic group or in a particular career stage.

\begin{figure}[htbp]
\centering
\includegraphics[width=\textwidth]{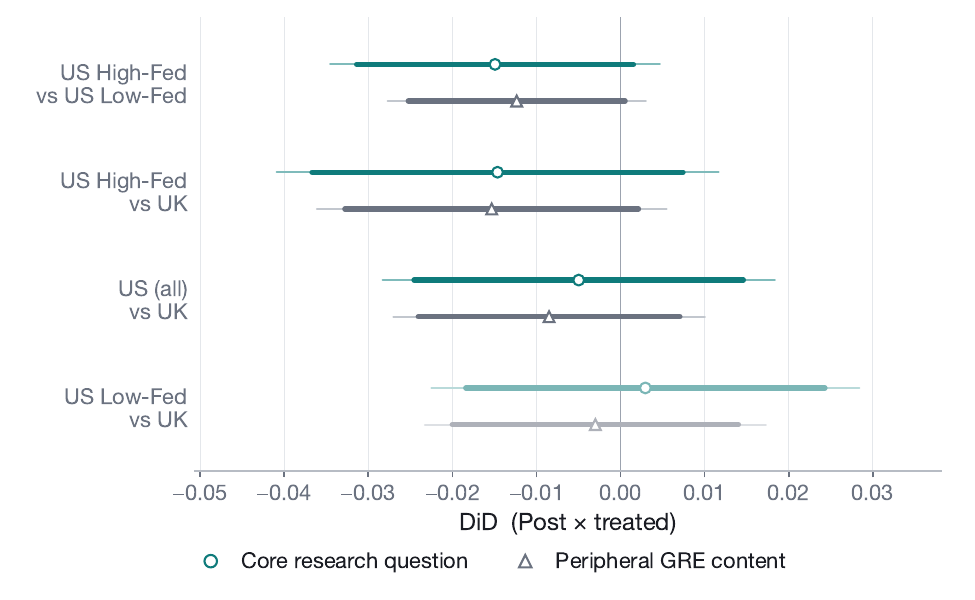}
\caption{Treatment Effects by Core versus Peripheral Research Question}
\label{fig:gre_core}
\begin{minipage}{0.95\textwidth}
\footnotesize
\textit{Notes:} This figure displays difference-in-differences coefficients (Post $\times$ Treated) for the binary GRE indicator split by the paper's core research question: papers whose core research question is about gender, race, or the environment (teal circles) versus papers whose GRE content is peripheral to a different research question (gray triangles). Each GRE paper's abstract is annotated by GPT-OSS, blind to which targeted terms were flagged (Appendix~\ref{app_subsec_gre_core}); 62\% of GRE papers are annotated as core. The two indicators partition the GRE outcome: each indicator is defined on the full sample, with papers of the other component coded zero rather than dropped, so within each comparison the two coefficients sum to the aggregate paper-level estimate in Figure~\ref{fig:coefplot_refgroups}. Rows are the four treatment-control comparisons; the US Low-Fed vs.\ UK comparison, in which neither group is highly exposed to federal funding, is drawn in lighter shades. All estimates are at the paper level. High-Fed universities are those whose total-federal share of R\&D expenditure (HERD FY2024, all fields) exceeds the cross-university median. US papers have at least as many US-affiliated (HERD-matched) as UK-affiliated (HESA-matched) authors (ties assigned to US); UK papers have strictly more UK-affiliated authors. A US paper is High-Fed if at least half of all its authors are at High-Fed universities (UK-affiliated and unmatched authors count as not High-Fed; ties assigned to High-Fed). Thick bars represent 90\% confidence intervals; thin lines represent 95\% confidence intervals. Paper-level estimates use robust standard errors. All regressions control for economics research capacity interacted with Post (Section~\ref{sec:methods}). Linear probability model estimates.
\end{minipage}
\end{figure}

\begin{figure}[htbp]
\centering
\includegraphics[width=\textwidth]{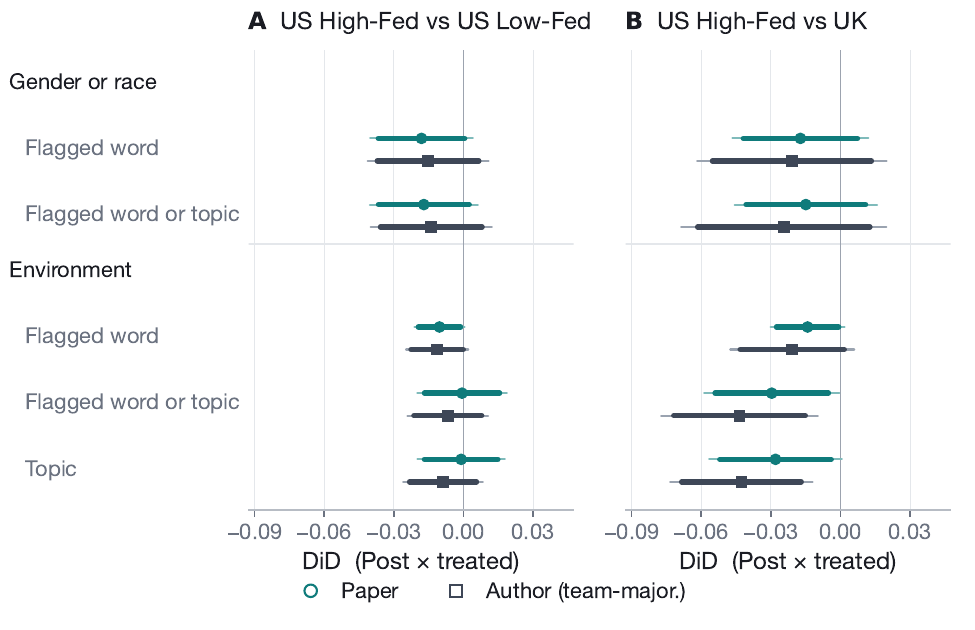}
\caption{Treatment Effects by Content: Flagged Words and Topics}
\label{fig:topics}
\begin{minipage}{0.95\textwidth}
\footnotesize
\textit{Notes:} This figure displays difference-in-differences coefficients (Post $\times$ Treated) for content on gender or race (combined) and on the environment. ``Flagged word'' means that the abstract uses a targeted term in this context, as assigned by GPT-OSS based on the surrounding text. Gender, Race, and Environment are the subcomponents of the combined GRE measure used in the main analysis -- the content areas most closely associated with the stated rationale behind the word ban. ``Topic'' means that a separate GPT-OSS call, which does not use the list of targeted terms, says that the abstract discusses the topic (Appendix~\ref{app_subsec_topics}); ``Flagged word or topic'' is either. For gender and race the topic row is omitted, as 85\% of the papers on these topics use a flagged word in that context (it is in Table~\ref{tab:content_union}). A paper can appear in several rows. Panel A compares US High-Fed with US Low-Fed papers, and Panel B compares US High-Fed with UK papers. The remaining comparisons (all US vs.\ UK and US Low-Fed vs.\ UK) are in Appendix Figure~\ref{fig:topics_addcomps}. Each row shows paper-level estimates (teal circles) and team-majority author-level estimates (slate squares). High-Fed universities are those whose total-federal share of R\&D expenditure (HERD FY2024, all fields) exceeds the cross-university median. US papers have at least as many US-affiliated (HERD-matched) as UK-affiliated (HESA-matched) authors (ties assigned to US); UK papers have strictly more UK-affiliated authors. A US paper is High-Fed if at least half of all its authors are at High-Fed universities (UK-affiliated and unmatched authors count as not High-Fed; ties assigned to High-Fed). Thick bars represent 90\% confidence intervals; thin lines represent 95\% confidence intervals. Paper-level estimates use robust standard errors; author-level estimates include author fixed effects and cluster standard errors two-way by university and paper. All regressions control for economics research capacity interacted with Post (Section~\ref{sec:methods}). Linear probability model estimates. All rows are tabulated, with all three assignments, in Table~\ref{tab:content_union}.
\end{minipage}
\end{figure}

\paragraph{Core versus peripheral research questions.} To what extent is the drop in GRE words purely cosmetic? In principle, the decline in GRE content could mean two things. Fewer papers may ask research questions about gender, race, or the environment. Or, the topics may be unchanged and described with fewer targeted terms, a purely lexical adjustment. To locate the decline, we split the GRE indicator by the core-question annotation of Section~\ref{sec_data}: papers whose core research question is about gender, race, or the environment, versus papers in which the GRE content is peripheral to a different question. In our classification, every GRE paper is either core or peripheral, so the two indicators add up to the GRE indicator. We estimate both on the full sample of Figure~\ref{fig:coefplot_refgroups}, with the same controls, and papers without GRE content count as zero in both. The two coefficients therefore sum exactly to the aggregate paper-level estimate. Figure~\ref{fig:gre_core} reports the decomposition at the paper level for all four reference groups. Within the US, the High-Fed versus Low-Fed coefficient is $-0.015$ ($p = 0.14$) for core papers, about half of the aggregate effect, against $-0.012$ ($p = 0.11$) for peripheral ones; the two components, however, are not statistically distinguishable from each other ($p = 0.85$). The High-Fed versus UK comparison splits more evenly ($-0.015$ and $-0.015$, neither significant on its own), and US Low-Fed versus UK shows no change in either component. What the decomposition does show is that about half of the point estimate comes from papers whose research question itself is about gender, race, or the environment. For these papers, the targeted vocabulary names the subject of the study rather than an incidental detail. The presence of the effect among core papers thus suggests that the response is not confined to incidental uses of the flagged words.\footnote{This claim rests on the core annotation and on the flagged vocabulary. The core indicator requires a flagged term, so a core paper that drops every flagged term from its abstract leaves the GRE set altogether. The decomposition therefore locates the decline within the flagged vocabulary; it does not count papers on these topics independently of it. The topic classification of Figure~\ref{fig:topics} does. We report the decomposition at the paper level. Appendix~\ref{app_subsec_gre_core} validates the annotation against JEL codes.}

\paragraph{Content heterogeneity.} To what extent are the results driven by a specific type of content? Figure~\ref{fig:topics} shows the estimates separately for gender or race (combined) and for the environment. The first row of each block is the flagged-word outcome of the main analysis, restricted to that context. All of these coefficients lie to the left of zero, in both panels and under both assignments. Hence, the point estimates are negative for both gender or race and the environment, although each category alone rests on few papers: from 2025, 193 High-Fed papers use a flagged word in a gender or race context and 44 in an environment context. For US High-Fed versus Low-Fed (Panel A), the paper-level coefficient is $-0.018$ ($p = 0.109$) for gender or race and $-0.010$ ($p = 0.061$) for the environment (Table~\ref{tab:content_union}). In the remaining comparisons (Appendix Figure~\ref{fig:topics_addcomps}), neither flagged-word estimate changes significantly.

The other rows add the topic classification of Section~\ref{sec_data}, which reads each abstract without the list: a paper counts if it uses a flagged word in the context or if the call says that the abstract discusses the topic. The topic call is a different measure, and what it adds depends on how much of a topic's vocabulary the list covers. For gender and race it adds little, because 85\% of the papers on these topics already use a flagged word in that context: the combined row moves with the flagged-word row ($-0.017$, $p = 0.15$, at the paper level; $-0.014$ under the team-majority assignment), and the two cannot be told apart. For the environment the list covers only 25\% of the papers on the topic, and the rows separate. Within the US, the flagged-word measure falls while the combined measure 
and the topic alone 
stay near zero, although these estimates are imprecise; the pattern is consistent with High-Fed authors dropping the listed environment terms and keeping the topic. Against the UK (Panel B), the combined and topic rows for the environment fall ($-0.030$, $p = 0.046$, and $-0.028$, $p = 0.055$, at the paper level), and they fall by about as much for US Low-Fed papers (Appendix Figure~\ref{fig:topics_addcomps}%
), so this is a US-wide movement relative to the UK rather than a High-Fed one. The own-university estimates are in Table~\ref{tab:content_union}.

Overall, the flagged-word estimates point to a decline for both gender or race and the environment. Each category alone rests on few papers and is imprecisely estimated. For the environment, the High-Fed decline appears only in the flagged words, not in the topic.

\begin{figure}[htbp]
\centering
\includegraphics[width=\textwidth]{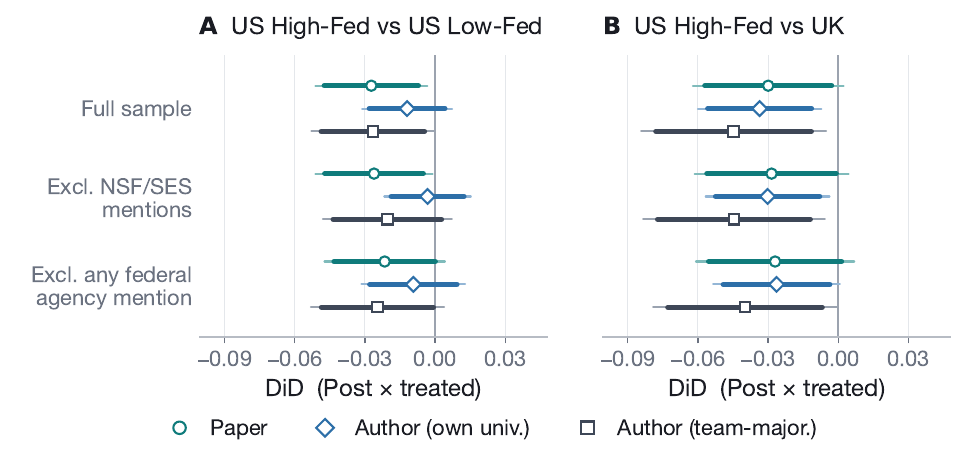}
\caption{Treatment Effects Excluding Federal-Agency Mentions}
\label{fig:nsf_exclude}
\begin{minipage}{0.95\textwidth}
\footnotesize
\textit{Notes:} This figure displays difference-in-differences coefficients (Post $\times$ Treated) for the full sample and for subsamples that progressively exclude papers acknowledging federal-agency funding in their acknowledgment sections: first papers mentioning the NSF or its Division of Social and Economic Sciences (SES), then papers mentioning any federal agency. The two panels show the US High-Fed vs.\ US Low-Fed and US High-Fed vs.\ UK comparisons; the remaining comparisons (all US vs.\ UK and US Low-Fed vs.\ UK) are in Appendix Figure~\ref{fig:nsf_exclude_addcomps}. Within each panel, rows are the three samples, with paper-level estimates (teal circles), own-university author-level estimates (blue diamonds), and team-majority author-level estimates (slate squares). The x-axis is identical to Figure~\ref{fig:coefplot_refgroups} for direct comparability. Because the word ban applies directly only to federal grant documents and agency communications -- not to working papers -- the persistence of the effect after excluding papers that acknowledge federal funding points to self-censorship rather than a mechanical response to individual grant terms. The dependent variable is a binary indicator for whether a paper contains any targeted term used in a Gender, Race, or Environment (GRE) context in its abstract (GPT-OSS classification). High-Fed universities are those whose total-federal share of R\&D expenditure (HERD FY2024, all fields) exceeds the cross-university median. US papers have at least as many US-affiliated (HERD-matched) as UK-affiliated (HESA-matched) authors (ties assigned to US); UK papers have strictly more UK-affiliated authors. A US paper is High-Fed if at least half of all its authors are at High-Fed universities (UK-affiliated and unmatched authors count as not High-Fed; ties assigned to High-Fed). Thick bars represent 90\% confidence intervals; thin lines represent 95\% confidence intervals. Paper-level estimates use robust standard errors; author-level estimates include author fixed effects and cluster standard errors two-way by university and paper. All regressions control for economics research capacity interacted with Post (Section~\ref{sec:methods}). Linear probability model estimates. A tabular version is in Table~\ref{tab:fig5}.
\end{minipage}
\end{figure}

\paragraph{Direct effects of federal funding.} To what extent do our findings capture the direct impact of federal funding policies on federally funded research? Our treatment variable captures university-level dependence on federal funding rather than whether a specific paper or author received a federal grant.
Figure~\ref{fig:nsf_exclude} drops papers that acknowledge federal funding: first those that mention the NSF or its Division of Social and Economic Sciences (SES), then those that mention any federal agency. The treatment effect persists in both samples.
Appendix Figure~\ref{fig:nsf_exclude_addcomps} reports the remaining comparisons.

As discussed in Section~\ref{sec:methods}, direct federal funding of economics research is low. Since the word ban applies \textit{directly} only to federal grant documents and agency communications -- not to working papers -- the reduced use of targeted terms in research that does not acknowledge federal funding points to self-censorship rather than a mechanical response to individual grant terms. These results are also consistent with researchers responding to real or perceived expectations from their university administrations. It is plausible that such pressure operates across disciplines in universities that rely heavily on federal funding. This would explain why fields with limited exposure to federal funding themselves, like economics, are still affected.

\begin{figure}[htbp]
\centering
\includegraphics[width=\textwidth]{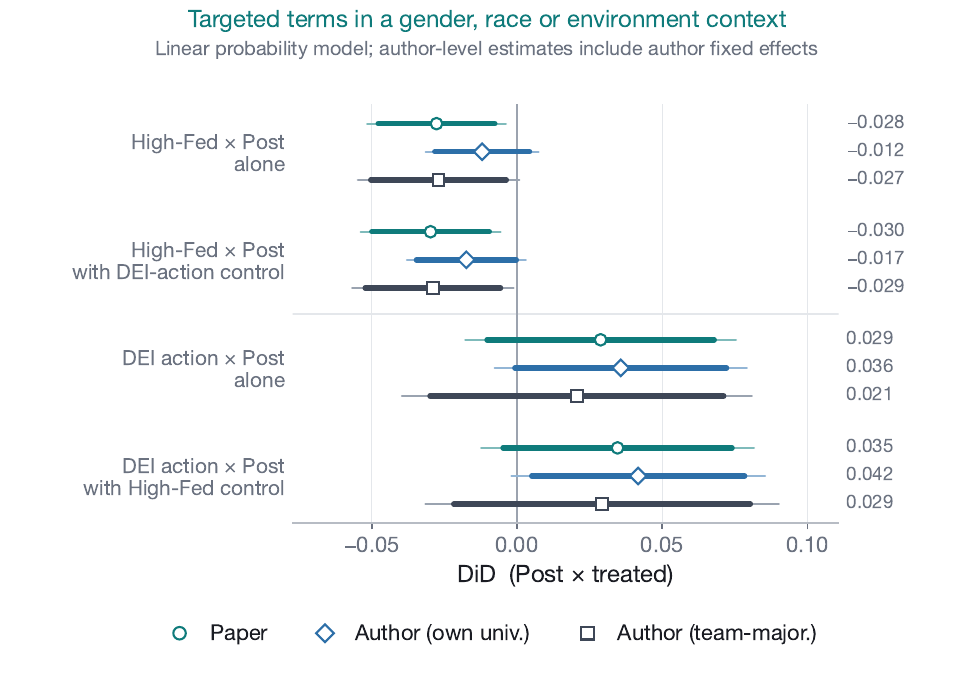}
\caption{Treatment Effects and Universities' Own DEI Actions}
\label{fig:mech_dei}
\begin{minipage}{0.95\textwidth}
\footnotesize
\textit{Notes:} This figure displays difference-in-differences coefficients (Post $\times$ Treated) for the High-Fed and DEI-action interactions, each estimated alone and with the other included in the same regression. DEI action is an indicator for universities with at least one documented action in the Chronicle of Higher Education's DEI tracker, matched to the HERD universities by a deterministic name crosswalk (Section~\ref{sec_data}; 208 of 376 universities). Rows are the four specifications; within each row, paper-level estimates (teal circles), own-university author-level estimates (blue diamonds), and team-majority author-level estimates (slate squares). The dependent variable is a binary indicator for whether a paper contains any targeted term used in a Gender, Race, or Environment (GRE) context in its abstract (GPT-OSS classification). The sample is restricted to US papers (paper level: the 15,011 papers whose matched authors have a FY2024 federal share, against 15,022 in Figure~\ref{fig:coefplot_refgroups}) and US authors (author level). High-Fed universities are those whose total-federal share of R\&D expenditure (HERD FY2024, all fields) exceeds the cross-university median; a US paper is High-Fed if at least half of all its authors are at High-Fed universities (UK-affiliated and unmatched authors count as not High-Fed; ties assigned to High-Fed). Thick bars represent 90\% confidence intervals; thin lines represent 95\% confidence intervals. Paper-level estimates use robust standard errors; own-university author-level estimates include author fixed effects and cluster standard errors by university; team-majority author-level estimates include author fixed effects and cluster standard errors two-way by university and paper. All regressions control for economics research capacity interacted with Post (Section~\ref{sec:methods}). Linear probability model estimates.
\end{minipage}
\end{figure}

\paragraph{University DEI actions.} Universities themselves changed policies during this period: many closed or renamed diversity offices, removed DEI language from websites and official statements, or dropped diversity statements (Section~\ref{sec_data}). Authors at such universities might therefore adjust their vocabulary in response to institutional guidance rather than to the federal word list itself. Exposure to this channel is widespread and, if anything, larger in the treated group: 94.5\% of High-Fed and 79.9\% of Low-Fed author-paper observations are at universities with at least one documented action. Figure~\ref{fig:mech_dei} reports the High-Fed $\times$ Post and DEI-action $\times$ Post interactions, each estimated alone and jointly in the same regression. The High-Fed coefficient is essentially unchanged when the DEI-action interaction is added%
, and the DEI-action interaction itself is positive and statistically insignificant at the 5\% level throughout (for example, $+0.029$, $p = 0.23$, at the paper level).\footnote{The High-Fed coefficient is essentially unchanged when the DEI indicator is restricted to website- and language-removal actions -- the action closest to the vocabulary we measure, and nearly balanced across the two groups (53.4\% of High-Fed and 57.6\% of Low-Fed author-papers) -- and when actions are split by whether the university's state had passed anti-DEI legislation (equality of the two coefficients: $p = 0.97$ at the author level, $p = 0.44$ at the paper level). These results are available upon request.} In a nutshell, documented institutional DEI activity thus does not predict the change in GRE content and does not absorb the High-Fed effect. Two limitations apply: identification comes from the minority of universities without documented DEI actions, which are disproportionately smaller and Low-Fed; and the tracker records public administrative actions, which may not be reflected in internal guidance issued to faculty.

\paragraph{University-level financial shocks.} We next use the dates of universities' own funding-related events (Section~\ref{sec_data}). If the response were transmitted through the author's own institution, GRE content at event universities would fall after that institution's announcement or funding shock, relative to universities without such an event. We therefore ask whether anything happens after the university's own event.

\begin{figure}[htbp]
\centering
\includegraphics[width=\textwidth]{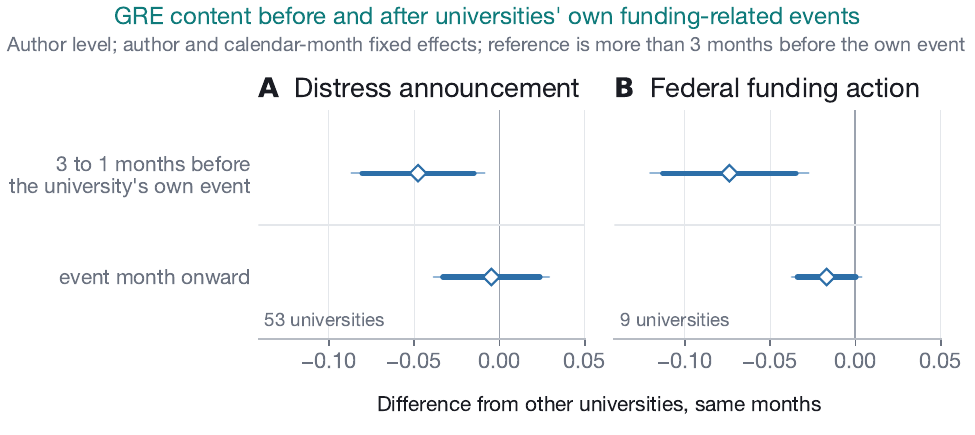}
\caption{GRE Content Before and After Universities' Own Funding-Related Events}
\label{fig:mech_timing}
\begin{minipage}{0.95\textwidth}
\footnotesize
\textit{Notes:} This figure displays event-time coefficients for the binary GRE indicator around each university's own event. Panel A dates each university's first public announcement of a hiring freeze, layoffs, or reductions in doctoral admissions attributed to federal funding cuts (53 universities, January 2025 to September 2025); Panel B dates the first publicly announced federal funding freeze or cancellation affecting the university as a whole (9 universities); sources and construction in Section~\ref{sec_data}. Rows are two event-time windows relative to the university's own event: the three months immediately before it, and the period from the event month onward. Coefficients are measured relative to the omitted category, four or more months before the event; authors at universities with no such event serve as controls. All specifications include author fixed effects, calendar-month fixed effects, and source indicators, with standard errors clustered by university; the sample is author-paper observations at HERD-matched universities with base-fixed affiliations. All regressions control for economics research capacity interacted with Post (Section~\ref{sec:methods}). Estimates are author-level and drawn as blue open diamonds accordingly. The coefficients are not cumulative, and calendar-month fixed effects absorb movements common to all universities. Equality of the two coefficients is rejected with $p = 0.032$ in Panel A and $p = 0.024$ in Panel B. Thick bars represent 90\% confidence intervals; thin lines represent 95\% confidence intervals. Linear probability model estimates.
\end{minipage}
\end{figure}

Figure~\ref{fig:mech_timing} reports author-level event-time estimates around the university's first public distress announcement -- hiring freezes, layoffs, or doctoral-admission cuts attributed to federal funding cuts (Panel A; 53 universities, with announcement months spread from January 2025 to September 2025) -- and around the university's first publicly announced federal funding freeze or cancellation (Panel B; 9 universities). The specifications include author and calendar-month fixed effects, so the movement common to all US authors in any given month is absorbed; authors at universities with no such event serve as controls. We distinguish two event-time windows: the three months immediately before the university's own event, and the period from the event month onward. Each coefficient measures how papers written at event universities in that window differ from papers at other universities in the same calendar months, relative to the same universities more than three months before their event.\footnote{For an author-paper observation of author $a$ posted in month $t$ at (base-fixed) university $u$, we estimate $y_{aut} = \sum_{k} \beta_k D^{k}_{u,t} + \alpha_a + \lambda_t + \varphi_{s} + \theta \, \text{Post}_t \times \text{Cap}_u + \varepsilon_{aut}$, where the $D^{k}_{u,t}$ indicate that university $u$ has an event and that $t$ falls in event-time window $k \in \{-3$ to $-1$, $0+\}$ months relative to the event month; the omitted category is four or more months before the event, together with all observations at universities without an event. $\alpha_a$ are author fixed effects, $\lambda_t$ calendar-month fixed effects, $\varphi_s$ source indicators, and $\text{Cap}_u$ the economics research capacity of university $u$ (Section~\ref{sec:methods}); the outcome is the binary GRE indicator and standard errors are clustered by university. Identification of event time separately from calendar time comes from universities being hit at different dates.}

In both panels, the negative coefficient sits in the three months before the university's own event ($-0.048$, $p = 0.017$, in Panel A; $-0.074$, $p = 0.002$, in Panel B). For most event universities this window falls between October 2024 and March 2025, around the national policy change of January and February 2025, so it does not date the response relative to the policy. The informative part is what follows. From the event month onward, the coefficient is $-0.005$ ($p = 0.78$) in Panel A and $-0.017$ ($p = 0.11$) in Panel B: no additional decline after the university's own announcement, and a further decline after a federal funding freeze that is not statistically significant. We reject that the pre-event and post-event coefficients are equal ($p = 0.032$ in Panel A; $p = 0.024$ in Panel B). 
Two features matter for reading these estimates. They are not cumulative, and calendar-month fixed effects absorb the aggregate movement: near-zero post-event coefficients mean that event universities no longer move differently from other universities, not that content recovered. As noted in Section~\ref{sec_data}, these events capture publicly announced, university-wide actions only; individual grant terminations and changes in agency priorities are not observed.\footnote{Such terminations would reach most directly the papers that acknowledge federal funding;  as discussed above, Figure~\ref{fig:nsf_exclude} confirms that the results are robust to excluding those papers.} In sum, a university's own distress announcement adds no further decline. A federal funding freeze may add one, but the High-Fed interaction remains, so these events do not drive our results.

\paragraph{Research output.} Finally, we study whether the documented changes in research content reflect shifts in research productivity. In line with our findings, political influence could reduce academic output -- a deterrent effect on research activity. In contrast, if research output increases, the reduction in the use of GRE words may be a decline relative to non-GRE papers and hence mechanical. Appendix Figure~\ref{fig:paperlength} reports PPML estimates for paper length (number of pages) across the reference-group comparisons: the within-US paper-level estimate is $-0.030$ ($p = 0.10$), about 3\% of pages; the other paper-level coefficients are close to zero, and the author-level cross-country point estimates are reductions of at most about 4\% of pages. We find little evidence that authors shortened their papers. Appendix Figure~\ref{fig:papercounts} estimates the effect on the average number of papers per author per year: the within-US High-Fed versus Low-Fed estimate is not statistically significant, and the negative cross-country estimates are, if anything, larger for US Low-Fed than for US High-Fed authors, providing no evidence of a High-Fed-specific reduction in output.

Splitting research output into GRE and non-GRE papers, the larger post-2025 declines are for GRE papers (Appendix Figure~\ref{fig:papercounts_split}): for US versus UK, the GRE-paper estimate is $-0.23$ versus $-0.03$ for non-GRE papers, although the difference of $-0.21$ is imprecisely estimated under university-level clustering (cluster-bootstrap $p = 0.056$). For the within-US High-Fed versus Low-Fed comparison, the difference between GRE and non-GRE papers is not statistically significant. These output results also speak to the mechanical channel behind the content
estimates. Because the GRE indicator is a share of papers, an author who added
non-sensitive papers while keeping sensitive work unchanged would mechanically
dilute the GRE share without altering any paper's content. Dilution, however,
requires output to expand. In the cross-country comparisons output, if anything,
falls (Appendix Figure~\ref{fig:papercounts}), and the point estimates of the decline are concentrated in GRE papers rather than non-GRE papers (Appendix Figure~\ref{fig:papercounts_split}). A falling GRE share combined with flat-to-falling total output implies a genuine reduction in the level of GRE research. For the within-US comparison, the change in relative output is not statistically significant ($+0.040$, $p = 0.41$).

\section{Conclusion}\label{sec_conclusion}

Self-censorship in science is particularly hard to measure. In early 2025, the US administration began scrutinizing federally funded research based on words to be avoided. One might have expected researchers to ignore this change and work solely according to the principles of academic freedom. Even if researchers reacted, one might have expected the impact to be limited to the content of new federal grant applications. Instead, we find that economists at universities more dependent on federal funding reduced the use of newly flagged words in sensitive contexts. This impact is not confined to research that acknowledges federal funding, and it is not absorbed by controls for university-level DEI measures or financial distress. Instead, the changing funding environment appears to have triggered a broader effect on the content of economics research: less of the flagged vocabulary in gender, race, and environment contexts in abstracts and, within the US, fewer full texts that use the flagged words densely, with part of the decline coming from papers whose core question concerns these topics. 

As academic freedom has declined around the world in recent years, self-censorship responses have implications for the resilience of science. Understanding better how scholars react to political pressure in various contexts and countries is important. Recent research has shown that, on average, populist governments are followed after 15 years by a GDP per capita that is 10\% lower than a plausible non-populist counterfactual \citep{funke2023populist}. As such an economic record could be electorally costly, populist leaders have incentives not only to weaken democratic institutions, but also to curtail independent scrutiny by the media and science. To defend civil and political rights and to support evidence-based policymaking, strengthening academic freedom is essential.

\clearpage
\bibliographystyle{plainnat}
\bibliography{ref}

\begin{thebibliography}{45}
\providecommand{\natexlab}[1]{#1}
\providecommand{\url}[1]{\texttt{#1}}
\expandafter\ifx\csname urlstyle\endcsname\relax
  \providecommand{\doi}[1]{doi: #1}\else
  \providecommand{\doi}{doi: \begingroup \urlstyle{rm}\Url}\fi

\bibitem[Aghion et~al.(2023)Aghion, Antonin, Paluskiewicz, Stromberg, Wargon, Westin, and Sun]{aghion2023does}
Philippe Aghion, Celine Antonin, Luc Paluskiewicz, David Stromberg, Raphael Wargon, Karolina Westin, and Xueping Sun.
\newblock Does chinese research hinge on us co-authors? evidence from the china initiative.
\newblock 2023.

\bibitem[Alabrese(2022)]{Alabrese2022}
Eleonora Alabrese.
\newblock {Bad Science: Retractions and Media Coverage}.
\newblock Available at SSRN: \url{https://ssrn.com/abstract=4324218}, 2022.
\newblock URL \url{https://ssrn.com/abstract=4324218}.

\bibitem[Alabrese et~al.(2024)Alabrese, Capozza, and Garg]{AlabreseCapozzaGarg2024}
Eleonora Alabrese, Francesco Capozza, and Prashant Garg.
\newblock {Politicized Scientists: Credibility Cost of Political Expression on Twitter}.
\newblock Available at SSRN: \url{https://ssrn.com/abstract=4922427}, 2024.
\newblock URL \url{https://ssrn.com/abstract=4922427}.

\bibitem[Ash et~al.(2025)Ash, Chen, and Naidu]{ash2025ideas}
Elliott Ash, Daniel~L. Chen, and Suresh Naidu.
\newblock {Ideas Have Consequences: The Impact of Law and Economics on American Justice}.
\newblock \emph{The Quarterly Journal of Economics}, 2025.
\newblock \doi{10.1093/qje/qjaf042}.
\newblock Advance access.

\bibitem[Banerjee et~al.(2017)Banerjee, Chassang, and Snowberg]{banerjee2017decision}
Abhijit~V Banerjee, Sylvain Chassang, and Erik Snowberg.
\newblock {Decision Theoretic Approaches to Experiment Design and External Validity}.
\newblock In \emph{Handbook of economic field experiments}, volume~1, pages 141--174. Elsevier, 2017.

\bibitem[Borjas and Breznau(2026)]{borjas2026ideological}
George~J Borjas and Nate Breznau.
\newblock {Ideological Bias in the Production of Research Findings}.
\newblock \emph{Science Advances}, 12\penalty0 (1):\penalty0 eadz7173, 2026.

\bibitem[Braghieri(2024)]{braghieri2024political}
Luca Braghieri.
\newblock Political correctness, social image, and information transmission.
\newblock \emph{American Economic Review}, 114\penalty0 (12):\penalty0 3877--3904, 2024.

\bibitem[Brodeur et~al.(2016)Brodeur, L{\'e}, Sangnier, and Zylberberg]{brodeur2016star}
Abel Brodeur, Mathias L{\'e}, Marc Sangnier, and Yanos Zylberberg.
\newblock {Star Wars: The Empirics Strike Back}.
\newblock \emph{American Economic Journal: Applied Economics}, 8\penalty0 (1):\penalty0 1--32, 2016.

\bibitem[Brodeur et~al.(2020)Brodeur, Cook, and Heyes]{brodeur2020methods}
Abel Brodeur, Nikolai Cook, and Anthony Heyes.
\newblock {Methods Matter: P-hacking and Publication Bias in Causal Analysis in Economics}.
\newblock \emph{American Economic Review}, 110\penalty0 (11):\penalty0 3634--3660, 2020.

\bibitem[Bursztyn et~al.(2020)Bursztyn, Egorov, and Fiorin]{bursztyn2020extreme}
Leonardo Bursztyn, Georgy Egorov, and Stefano Fiorin.
\newblock From extreme to mainstream: The erosion of social norms.
\newblock \emph{American economic review}, 110\penalty0 (11):\penalty0 3522--3548, 2020.

\bibitem[Cantoni et~al.(2017)Cantoni, Chen, Yang, Yuchtman, and Zhang]{Cantoni2017}
Davide Cantoni, Yuyu Chen, David~Y. Yang, Noam Yuchtman, and Y.~Jane Zhang.
\newblock Curriculum and ideology.
\newblock \emph{Journal of Political Economy}, 125\penalty0 (2):\penalty0 338--392, 2017.
\newblock \doi{10.1086/690951}.
\newblock URL \url{https://doi.org/10.1086/690951}.

\bibitem[Chassang et~al.(2012)Chassang, Padr{\'o}~i Miquel, and Snowberg]{chassang2012selective}
Sylvain Chassang, Gerard Padr{\'o}~i Miquel, and Erik Snowberg.
\newblock {Selective Trials: A Principal-Agent Approach to Randomized Controlled Experiments}.
\newblock \emph{American Economic Review}, 102\penalty0 (4):\penalty0 1279--1309, 2012.

\bibitem[Clark et~al.(2023)Clark, Jussim, Frey, Stevens, al~Gharbi, Aquino, Bailey, Barbaro, Baumeister, Bleske-Rechek, Buss, Ceci, Del~Giudice, Ditto, Forgas, Geary, Geher, Haider, Honeycutt, Joshi, Krylov, Loftus, Loury, Lu, Macy, Martin, McWhorter, Miller, Paresky, Pinker, Reilly, Salmon, Stewart-Williams, Tetlock, Williams, Wilson, Winegard, Yancey, and von Hippel]{clark2023}
Cory~J. Clark, Lee Jussim, Komi Frey, Sean~T. Stevens, Musa al~Gharbi, Karl Aquino, J.~Michael Bailey, Nicole Barbaro, Roy~F. Baumeister, April Bleske-Rechek, David Buss, Stephen Ceci, Marco Del~Giudice, Peter~H. Ditto, Joseph~P. Forgas, David~C. Geary, Glenn Geher, Sarah Haider, Nathan Honeycutt, Hrishikesh Joshi, Anna~I. Krylov, Elizabeth Loftus, Glenn Loury, Louise Lu, Michael Macy, Chris~C. Martin, John McWhorter, Geoffrey Miller, Pamela Paresky, Steven Pinker, Wilfred Reilly, Catherine Salmon, Steve Stewart-Williams, Philip~E. Tetlock, Wendy~M. Williams, Anne~E. Wilson, Bo~M. Winegard, George Yancey, and William von Hippel.
\newblock {Prosocial Motives Underlie Scientific Censorship by Scientists: A Perspective and Research Agenda}.
\newblock \emph{Proceedings of the National Academy of Sciences}, 120\penalty0 (48):\penalty0 e2301642120, 2023.
\newblock \doi{10.1073/pnas.2301642120}.
\newblock URL \url{https://www.pnas.org/doi/abs/10.1073/pnas.2301642120}.

\bibitem[Cutler et~al.(2026)Cutler, Baiocchi, Gretzinger, Hicks, Dutton, and Smith]{chronicle2026dei}
Sonel Cutler, Aisha Baiocchi, Erin Gretzinger, Maggie Hicks, Christa Dutton, and Jasper Smith.
\newblock {Tracking Higher Ed's Dismantling of DEI}, 2026.
\newblock URL \url{https://www.chronicle.com/article/tracking-higher-eds-dismantling-of-dei}.
\newblock Interactive tracker, accessed: 2026-08-27.

\bibitem[Djourelova(2023)]{djourelova2023persuasion}
Milena Djourelova.
\newblock Persuasion through slanted language: Evidence from the media coverage of immigration.
\newblock \emph{American Economic Review}, 113\penalty0 (3):\penalty0 800--835, 2023.

\bibitem[{Executive Office of the President}(2025{\natexlab{a}})]{eo14151}
{Executive Office of the President}.
\newblock Executive order 14151: Ending radical and wasteful government {DEI} programs and preferencing.
\newblock Federal Register, Vol.~90, January 2025{\natexlab{a}}.
\newblock URL \url{https://www.federalregister.gov/executive-order/14151}.

\bibitem[{Executive Office of the President}(2025{\natexlab{b}})]{eo14168}
{Executive Office of the President}.
\newblock Executive order 14168: Defending women from gender ideology extremism and restoring biological truth to the federal government.
\newblock Federal Register, Vol.~90, January 2025{\natexlab{b}}.
\newblock URL \url{https://www.federalregister.gov/executive-order/14168}.

\bibitem[{Executive Office of the President}(2025{\natexlab{c}})]{eo14173}
{Executive Office of the President}.
\newblock Executive order 14173: Ending illegal discrimination and restoring merit-based opportunity.
\newblock Federal Register, Vol.~90, January 2025{\natexlab{c}}.
\newblock URL \url{https://www.federalregister.gov/executive-order/14173}.

\bibitem[Fourcade et~al.(2015)Fourcade, Ollion, and Algan]{fourcade2015superiority}
Marion Fourcade, Etienne Ollion, and Yann Algan.
\newblock {The Superiority of Economists}.
\newblock \emph{Journal of economic perspectives}, 29\penalty0 (1):\penalty0 89--114, 2015.

\bibitem[Funke et~al.(2023)Funke, Schularick, and Trebesch]{funke2023populist}
Manuel Funke, Moritz Schularick, and Christoph Trebesch.
\newblock {Populist Leaders and the Economy}.
\newblock \emph{American Economic Review}, 113\penalty0 (12):\penalty0 3249--3288, 2023.

\bibitem[Furnas et~al.(2026)Furnas, Jia, Roberts, and Wang]{furnas2026}
Alexander~C Furnas, Ruixue Jia, Margaret~E Roberts, and Dashun Wang.
\newblock {Geopolitics in the Evaluation of International Scientific Collaboration}.
\newblock Working Paper 34789, National Bureau of Economic Research, February 2026.
\newblock URL \url{http://www.nber.org/papers/w34789}.

\bibitem[Garg and Fetzer(2025{\natexlab{a}})]{garg2025a}
Prashant Garg and Thiemo Fetzer.
\newblock {Causal Claims in Economics}.
\newblock \emph{Working Paper}, 2025{\natexlab{a}}.
\newblock URL \url{https://arxiv.org/abs/2501.06873}.

\bibitem[Garg and Fetzer(2025{\natexlab{b}})]{garg2025b}
Prashant Garg and Thiemo Fetzer.
\newblock {Political Expression of Academics on Twitter}.
\newblock \emph{Nature Human Behaviour}, page 1815, 09 2025{\natexlab{b}}.

\bibitem[Garisto and Kozlov(2025)]{garisto2025nsf}
Dan Garisto and Max Kozlov.
\newblock Exclusive: how {NSF} is scouring research grants for violations of {Trump}'s orders.
\newblock \emph{Nature}, February 2025.
\newblock \doi{10.1038/d41586-025-00365-z}.
\newblock URL \url{https://www.nature.com/articles/d41586-025-00365-z}.

\bibitem[Grosjean et~al.(2023)Grosjean, Masera, and Yousaf]{grosjean2023inflammatory}
Pauline Grosjean, Federico Masera, and Hasin Yousaf.
\newblock {Inflammatory Political Campaigns and Racial Bias in Policing}.
\newblock \emph{The Quarterly Journal of Economics}, 138\penalty0 (1):\penalty0 413--463, 2023.

\bibitem[Guriev and Treisman(2019)]{guriev2019}
Sergei Guriev and Daniel Treisman.
\newblock {Informational Autocrats}.
\newblock \emph{Journal of Economic Perspectives}, 33\penalty0 (4):\penalty0 100–127, November 2019.
\newblock \doi{10.1257/jep.33.4.100}.
\newblock URL \url{https://www.aeaweb.org/articles?id=10.1257/jep.33.4.100}.

\bibitem[Hamermesh(2013)]{Hamermesh2013}
Daniel~S. Hamermesh.
\newblock {Six Decades of Top Economics Publishing: Who and How?}
\newblock \emph{Journal of Economic Literature}, 51\penalty0 (1):\penalty0 162–72, March 2013.
\newblock \doi{10.1257/jel.51.1.162}.
\newblock URL \url{https://www.aeaweb.org/articles?id=10.1257/jel.51.1.162}.

\bibitem[Hjort et~al.(2021)Hjort, Moreira, Rao, and Santini]{hjort2021research}
Jonas Hjort, Diana Moreira, Gautam Rao, and Juan~Francisco Santini.
\newblock {How Research Affects Policy: Experimental Evidence from 2,150 Brazilian Municipalities}.
\newblock \emph{American Economic Review}, 111\penalty0 (5):\penalty0 1442--1480, 2021.

\bibitem[Honeycutt et~al.(2023)Honeycutt, Stevens, and Kaufmann]{honeycutt2023}
Nathan Honeycutt, Sean~T Stevens, and Eric Kaufmann.
\newblock {The Academic Mind in 2022: What Faculty Think About Free Expression and Academic Freedom on Campus}.
\newblock Report, The Foundation for Individual Rights and Expression, 2023.
\newblock URL \url{https://www.thefire.org/research-learn/academic-mind-2022-what-faculty-think-about-free-expression-and-academic-freedom}.

\bibitem[Iaria et~al.(2018)Iaria, Schwarz, and Waldinger]{iaria2018frontier}
Alessandro Iaria, Carlo Schwarz, and Fabian Waldinger.
\newblock {Frontier Knowledge and Scientific Production: Evidence from the Collapse of International Science}.
\newblock \emph{The Quarterly Journal of Economics}, 133\penalty0 (2):\penalty0 927--991, 2018.

\bibitem[Ioannidis et~al.(2017)Ioannidis, Stanley, and Doucouliagos]{ioannidis2017power}
John~PA Ioannidis, Tom~D Stanley, and Hristos Doucouliagos.
\newblock {The Power of Bias in Economics Research}, 2017.

\bibitem[{Ithaka S+R}(2026)]{ithaka2026tracker}
{Ithaka S+R}.
\newblock {University Research Workforce Tracker}, 2026.
\newblock URL \url{https://sr.ithaka.org/our-work/the-research-enterprise/university-research-workforce-tracker/}.
\newblock Public tracker, accessed: 2026-08-27.

\bibitem[Jelveh et~al.(2024)Jelveh, Kogut, and Naidu]{jelveh2024political}
Zubin Jelveh, Bruce Kogut, and Suresh Naidu.
\newblock {Political Language in Economics}.
\newblock \emph{The Economic Journal}, 134\penalty0 (662):\penalty0 2439--2469, 2024.

\bibitem[Kasy(2021)]{kasy2021forking}
Maximilian Kasy.
\newblock {Of Forking Paths and Tied Hands: Selective Publication of Findings, and what Economists should do about it}.
\newblock \emph{Journal of Economic Perspectives}, 35\penalty0 (3):\penalty0 175--192, 2021.

\bibitem[Kinzelbach et~al.(2026)Kinzelbach, Lindberg, Lott, and Panaro]{kinzelbach2025academic}
Katrin Kinzelbach, Staffan~I Lindberg, Lars Lott, and Angelo~V Panaro.
\newblock Academic freedom index update 2026.
\newblock \emph{Available at SSRN 6430978}, 2026.

\bibitem[Kozlov et~al.(2026)Kozlov, Tollefson, and Garisto]{kozlov2026}
Max Kozlov, Jeff Tollefson, and Dan Garisto.
\newblock {US Science after a Year of Trump: What Has Been Lost and What Remains}, January 2026.
\newblock URL \url{https://www.nature.com/immersive/d41586-026-00088-9/index.html}.
\newblock Nature.com, accessed: 2026-01-30.

\bibitem[Palmer(2025)]{palmer2025hitlist}
Kathryn Palmer.
\newblock The {NSF}'s higher ed research `hit list'.
\newblock Inside Higher Ed, February 2025.
\newblock URL \url{https://www.insidehighered.com/news/government/science-research-policy/2025/02/26/nsfs-higher-ed-research-hit-list}.

\bibitem[Ratcliff et~al.(2023)Ratcliff, Harvill, and Wicke]{Ratcliff2023}
Chelsea~L. Ratcliff, Blue Harvill, and Rebekah Wicke.
\newblock {Understanding Public Preferences for Learning about Uncertain Science: Measurement and Individual Difference Correlates}.
\newblock \emph{Frontiers in Communication}, Volume 8 - 2023, 2023.
\newblock ISSN 2297-900X.
\newblock \doi{10.3389/fcomm.2023.1245786}.
\newblock URL \url{https://www.frontiersin.org/journals/communication/articles/10.3389/fcomm.2023.1245786}.

\bibitem[Rubin and Rubin(2021)]{rubin2021systematic}
Amir Rubin and Eran Rubin.
\newblock {Systematic Bias in the Progress of Research}.
\newblock \emph{Journal of Political Economy}, 129\penalty0 (9):\penalty0 2666--2719, 2021.

\bibitem[Vanden~Eynde et~al.(2018)Vanden~Eynde, Kuhn, and Moradi]{vandeneynde2018}
Oliver Vanden~Eynde, Patrick~M. Kuhn, and Alexander Moradi.
\newblock {Trickle-Down Ethnic Politics: Drunk and Absent in the Kenya Police Force (1957-1970)}.
\newblock \emph{American Economic Journal: Economic Policy}, 10\penalty0 (3):\penalty0 388–417, August 2018.
\newblock \doi{10.1257/pol.20160384}.
\newblock URL \url{https://www.aeaweb.org/articles?id=10.1257/pol.20160384}.

\bibitem[Waldinger(2012)]{waldinger2012peer}
Fabian Waldinger.
\newblock {Peer Effects in Science: Evidence from the Dismissal of Scientists in Nazi Germany}.
\newblock \emph{The Review of Economic Studies}, 79\penalty0 (2):\penalty0 838--861, 2012.

\bibitem[Widmer(2024)]{Widmer2024}
Philine Widmer.
\newblock {Is Propaganda Front-Page News?}
\newblock \emph{Working Paper}, January 2024.
\newblock \doi{10.2139/ssrn.4686681}.
\newblock URL \url{https://ssrn.com/abstract=4686681}.

\bibitem[Xie et~al.(2021)Xie, Shen, and Wang]{xie2021preprint}
Boya Xie, Zhihong Shen, and Kuansan Wang.
\newblock Is preprint the future of science? a thirty year journey of online preprint services.
\newblock \emph{arXiv preprint arXiv:2102.09066}, 2021.

\bibitem[Yanagizawa-Drott(2014)]{yanagizawadrott2014propaganda}
David Yanagizawa-Drott.
\newblock {Propaganda and Conflict: Evidence from the Rwandan Genocide}.
\newblock \emph{The Quarterly Journal of Economics}, 129\penalty0 (4):\penalty0 1947--1994, 2014.

\bibitem[Yourish et~al.(2025)Yourish, Daniel, Datar, White, and Gamio]{yourish2025words}
Karen Yourish, Annie Daniel, Saurabh Datar, Isaac White, and Lazaro Gamio.
\newblock {These Words Are Disappearing in the New Trump Administration}.
\newblock The New York Times, March 2025.
\newblock 7 March 2025.

\end{thebibliography}

\appendix
\setcounter{figure}{0}
\setcounter{table}{0}
\renewcommand{\thefigure}{A\arabic{figure}}
\renewcommand{\thetable}{A\arabic{table}}

\clearpage
\section{Data Appendix}

\subsection{Context Classification and Full List of Flagged Words}\label{app_subsec_classification}

In this Appendix subsection we will start by discussing the prompt used for the classification of contexts and then include below a list of all flagged terms (Table \ref{tab:censored_words}).

For each word-paper pair (i.e., each distinct targeted term appearing in an abstract; repeated occurrences of the same term in one abstract are classified once), we classify the context in which the word is used. The following prompt is sent to GPT-OSS (the open-weight GPT-OSS-120B model; temperature 0, low reasoning effort, one call per word-paper pair), with \texttt{\{word\}} and \texttt{\{abstract\}} filled for each observation:

\begin{quote}
\small
\textit{The following word is on a 2025 compilation of terms that US federal agencies were instructed to avoid or limit: \{word\}}

\smallskip
\textit{This word appears in the economics research abstract below. In which context is this word used?}

\smallskip
\textit{Contexts:}\\
\textit{- Economic inequality}\\
\textit{- Race}\\
\textit{- Gender}\\
\textit{- Environment}\\
\textit{- Cannot say or other}

\smallskip
\textit{If multiple contexts could apply, choose the dominant one.}

\smallskip
\textit{Abstract: ``\{abstract\}''}

\smallskip
\textit{Return ONLY the context name.}
\end{quote}

The full-text measures do not use this classification: a paper's term density is the number of targeted-term occurrences in its full text divided by its number of pages, and the dense indicators flag papers at or above the median (0.67 mentions per page) or the 90th percentile (2.87) of this density among all papers dated before 2025 (all three series pooled); full paper texts are never sent to the model.

\subsection{Core-Question Annotation}\label{app_subsec_gre_core}

The mechanisms analysis (Figure~\ref{fig:gre_core}) splits GRE papers by whether the paper's core research question is about gender, race, or the environment. For each of the 3,008 papers with at least one GRE-classified targeted term, the abstract (first 1,500 characters) is sent to GPT-OSS (\texttt{gpt-oss-120b}) in a separate, independent call, once per paper, with temperature 0, low reasoning effort, and at most 2,000 output tokens; the annotator sees only the abstract, not which terms were flagged. The following prompt is used, with \texttt{\{abstract\}} filled for each paper:

\begin{quote}
\small
\textit{The following is an economics research abstract.}

\smallskip
\textit{Is the paper's core research question about gender, race, or the environment?}

\smallskip
\textit{Answer ``core'' if at least one of these topics is a central subject of the study.}\\
\textit{Answer ``peripheral'' otherwise.}

\smallskip
\textit{Abstract: ``\{abstract\}''}

\smallskip
\textit{Think about what the paper's core research question is. Then, on the last line of your reply, return exactly one word: core or peripheral.}
\end{quote}

The final word of the reply is the label; 62.4\% of GRE papers are annotated as core. The framing is neutral -- no mention of censorship or the word ban -- and the final instruction lets the model reason briefly before answering.

To validate the annotation externally, we compare it with JEL codes, which are available for NBER and CEPR papers but not for arXiv. CEPR JEL codes come from the CEPR discussion-paper listing pages; NBER JEL codes are extracted from the second page of each paper PDF, as the NBER API metadata does not include them. Of the 2,067 NBER and CEPR GRE papers, 1,940 (94\%) carry JEL codes (1,286 annotated core, 654 peripheral). Table~\ref{tab:gre_core_validation_jel} reports, for every JEL code attached to at least 50 GRE papers, the number and share of core and peripheral papers listing the code and the odds ratio of the code appearing on a core versus a peripheral paper (with the Haldane--Anscombe small-sample correction). The ordering supports the annotation: the gender and race codes concentrate at the core-leaning end -- Economics of Gender (J16, odds ratio 7.0), Discrimination (J71, 2.6), and Minorities, Race, and Migrants (J15, 2.9), with Climate (Q54) at 2.0 and Air Pollution (Q53) at 1.0 -- while the peripheral-leaning end consists of fields that use GRE vocabulary without being about it, such as Health Behavior (I12, 0.42) and Health Policy (I18, 0.48).

\begin{table}[htbp]
\centering
\caption{JEL Codes on Core versus Peripheral GRE Papers}
\label{tab:gre_core_validation_jel}
\small
\setlength{\tabcolsep}{6pt}
\begin{tabular}{llccc}
\toprule
 JEL & Field & Odds ratio & Core $n$ (\%) & Peripheral $n$ (\%) \\
\midrule
\addlinespace[2pt]
  J16 & Economics of Gender & 7.01 & 336 (26.1) & 31 (4.7) \\
  J15 & Minorities, Race, Migrants & 2.91 & 202 (15.7) & 39 (6.0) \\
  J71 & Discrimination & 2.63 & 67 (5.2) & 13 (2.0) \\
  Q58 & Environmental Policy & 1.98 & 51 (4.0) & 13 (2.0) \\
  Q54 & Climate & 1.97 & 73 (5.7) & 19 (2.9) \\
  Z13 & Social Norms, Identity & 1.66 & 43 (3.3) & 13 (2.0) \\
  O12 & Development: Household & 1.62 & 45 (3.5) & 14 (2.1) \\
  J21 & Labor Force & 1.55 & 49 (3.8) & 16 (2.4) \\
  J22 & Labor Supply & 1.52 & 74 (5.8) & 25 (3.8) \\
  I23 & Higher Education & 1.35 & 48 (3.7) & 18 (2.8) \\
  N32 & US Labor History 1913-- & 1.24 & 37 (2.9) & 15 (2.3) \\
  H23 & Externalities, Pigouvian Taxes & 1.17 & 35 (2.7) & 15 (2.3) \\
  D72 & Political Processes & 1.10 & 50 (3.9) & 23 (3.5) \\
  E24 & Employment, Wages (Macro) & 1.07 & 34 (2.6) & 16 (2.4) \\
  J12 & Marriage, Family Structure & 1.05 & 52 (4.0) & 25 (3.8) \\
  Q53 & Air Pollution, Waste & 1.04 & 68 (5.3) & 33 (5.0) \\
  J24 & Human Capital, Skills & 1.04 & 100 (7.8) & 49 (7.5) \\
  I21 & Analysis of Education & 1.01 & 48 (3.7) & 24 (3.7) \\
  J31 & Wage Structure & 0.95 & 53 (4.1) & 28 (4.3) \\
  D91 & Behavioral Micro & 0.94 & 41 (3.2) & 22 (3.4) \\
  I14 & Health and Inequality & 0.88 & 72 (5.6) & 41 (6.3) \\
  I24 & Education and Inequality & 0.84 & 64 (5.0) & 38 (5.8) \\
  J13 & Fertility, Children & 0.77 & 103 (8.0) & 66 (10.1) \\
  O15 & Human Resources, Development & 0.71 & 34 (2.6) & 24 (3.7) \\
  I1 & Health: General & 0.66 & 33 (2.6) & 25 (3.8) \\
  I18 & Health Policy & 0.48 & 48 (3.7) & 49 (7.5) \\
  I12 & Health Behavior & 0.42 & 43 (3.3) & 50 (7.6) \\
\bottomrule
\end{tabular}

\begin{minipage}{0.95\textwidth}
\vspace{4pt}
\footnotesize
\textit{Notes:} External validation of the core-question annotation against JEL codes, for the 1,940 NBER and CEPR GRE papers with JEL codes (1,286 annotated core, 654 peripheral; arXiv papers carry no JEL codes). CEPR codes are from the CEPR listing pages; NBER codes are extracted from the second page of each paper PDF. Each code is counted once per paper, and the table keeps codes attached to at least 50 GRE papers. Core $n$ and Peripheral $n$ are the numbers of core and peripheral papers listing the code, with the within-group percentage in parentheses. The odds ratio is the odds of the code appearing on a core relative to a peripheral paper (Haldane--Anscombe correction); rows are sorted by the odds ratio.
\end{minipage}
\end{table}

\paragraph{JEL codes and the context classification.} The same JEL codes also validate the context classification that defines GRE papers in the first place, before any core annotation. Of the 14,098 NBER and CEPR papers in our sample, 13,111 carry JEL codes (99\% of NBER and 86\% of CEPR papers): 1,940 GRE papers and 11,171 non-GRE papers. Table~\ref{tab:app_jel_gre} compares the two groups. Panel A matches each classified category to its JEL field: 35\% of Gender papers list J16 against 0.3\% of other papers (odds ratio 182); 37\% of Race papers list J15 or J71 against 1.1\% (odds ratio 51); 71\% of Environment papers list a Q5 code against 3.9\% (odds ratio 60). Panel B ranks every code attached to at least 50 papers by the odds of appearing on a GRE rather than a non-GRE paper: J16 (91), J71 (33), and J15 (21) lead, followed by Air Pollution (Q53, 13), while the bottom of the ranking consists mostly of finance and monetary codes (F31, D43, E58, F34, F30, E50) that essentially never appear on a GRE paper.

\begin{table}[htbp]
\centering
\caption{Validation of the Context Classification against JEL Codes}
\label{tab:app_jel_gre}
\small
\setlength{\tabcolsep}{4.5pt}
\begin{tabular}{llccc}
\toprule
\multicolumn{5}{l}{\textit{Panel A: Category-specific codes}} \\
\addlinespace[2pt]
Classified category & Target JEL codes & Classified (\%) & Other papers (\%) & Odds ratio \\
\midrule
  Gender & J16 & 34.9 & 0.3 & 181.88 \\
  Race & J15 or J71 & 36.8 & 1.1 & 50.62 \\
  Environment & Q5 (Q50--Q59) & 71.0 & 3.9 & 60.24 \\
  Any GRE & J16, J15, J71, or Q5 & 45.7 & 5.3 & 15.15 \\
\midrule
\addlinespace[2pt]
\multicolumn{5}{l}{\textit{Panel B: JEL codes with the 12 highest and 6 lowest odds ratios}} \\
\addlinespace[2pt]
JEL code & Field & GRE $n$ (\%) & Non-GRE $n$ (\%) & Odds ratio \\
\midrule
  J16 & Economics of Gender & 367 (18.9) & 28 (0.3) & 91.32 \\
  J71 & Discrimination & 80 (4.1) & 14 (0.1) & 33.29 \\
  J15 & Minorities, Race, Migrants & 241 (12.4) & 74 (0.7) & 21.17 \\
  Q53 & Air Pollution, Waste & 101 (5.2) & 46 (0.4) & 13.20 \\
  Q52 & Pollution Control Costs & 42 (2.2) & 27 (0.2) & 9.07 \\
  J12 & Marriage, Family Structure & 77 (4.0) & 51 (0.5) & 8.98 \\
  D13 & Household Production & 40 (2.1) & 27 (0.2) & 8.64 \\
  N32 & US Labor History 1913- & 52 (2.7) & 40 (0.4) & 7.64 \\
  N31 & US Labor History pre-1913 & 29 (1.5) & 25 (0.2) & 6.75 \\
  J13 & Fertility, Children & 169 (8.7) & 167 (1.5) & 6.29 \\
  I14 & Health and Inequality & 113 (5.8) & 114 (1.0) & 6.00 \\
  Q50 &  & 28 (1.4) & 30 (0.3) & 5.44 \\
  \multicolumn{5}{c}{$\cdots$} \\
  E50 &  & 0 (0.0) & 99 (0.9) & 0.03 \\
  F30 &  & 0 (0.0) & 101 (0.9) & 0.03 \\
  F34 &  & 0 (0.0) & 106 (0.9) & 0.03 \\
  E58 &  & 1 (0.1) & 314 (2.8) & 0.03 \\
  D43 &  & 0 (0.0) & 143 (1.3) & 0.02 \\
  F31 &  & 0 (0.0) & 169 (1.5) & 0.02 \\
\bottomrule
\end{tabular}
\begin{minipage}{0.95\textwidth}
\vspace{4pt}
\footnotesize
\textit{Notes:} Sample: the 13,111 NBER and CEPR papers of our sample with non-empty JEL codes (99\% of NBER and 86\% of CEPR papers; arXiv papers carry no JEL codes). CEPR codes come from the discussion-paper listing pages, NBER codes from the second page of each paper PDF. A paper is GRE if at least one targeted term in its abstract is classified as used in a Gender, Race, or Environment context (1,940 GRE papers, 11,171 non-GRE papers); category flags are defined likewise. Panel A reports, for each classified category, the share of papers listing the target code(s) among papers assigned to the category and among all other papers, and the odds ratio. Panel B reports, for every JEL code attached to at least 50 papers, the number and share of GRE and non-GRE papers listing the code and the odds ratio of the code appearing on a GRE versus a non-GRE paper; only the 12 highest and 6 lowest of the 290 codes are shown. Odds ratios use the Haldane--Anscombe correction. Codes are counted once per paper.
\end{minipage}
\end{table}

\subsection{Topic Classification}\label{app_subsec_topics}

The content analysis (Figure~\ref{fig:topics}) adds a measure that does not depend on the list of targeted terms. Each of the 28,081 abstracts in the sample is sent to GPT-OSS (\texttt{gpt-oss-120b}) in a separate call, once per paper, with temperature 0, low reasoning effort, and at most 2,000 output tokens, with the following prompt and \texttt{\{abstract\}} filled for each paper:

\begin{quote}
\small
\textit{Does the economics research abstract below discuss any of the following topics?}

\smallskip
\textit{- Gender}\\
\textit{- Race or ethnicity}\\
\textit{- Environment or climate}

\smallskip
\textit{Abstract: ``\{abstract\}''}

\smallskip
\textit{If it does, return ONLY the topic (or topics, separated by commas).}\\
\textit{If it does not, return ONLY: Topics not present.}
\end{quote}

A topic named in the reply gives the paper that topic; ``Topics not present'' gives none. Every reply was parseable. The prompt does not mention the list of targeted terms and does not point to any word in the abstract. Across all papers, 6.0\% discuss gender, 3.9\% race or ethnicity, and 8.3\% the environment or climate (8.9\% gender or race).

The topic call is a different measure from the context classification, not necessarily a more comprehensive one: it draws its own boundaries. For example, it does not always count sexual orientation as a gender topic, whereas our context classification labels terms such as LGBT, transgender, and gender identity as gender. Table~\ref{tab:app_topic_validation} validates it against the JEL codes that the authors assign (Appendix~\ref{app_subsec_gre_core}): the call labels 92\% of the papers filed under J16 as gender, 67\% of those under J15 or J71 as race, and 90\% of those under a Q5 code as environment, and its papers carry the corresponding code far more often than other papers (33 against 0.3\% for gender). The flagged-word measure catches the gender and race papers about as often (91 and 64\%) but only 34\% of the environment papers: the list covers the vocabulary of gender and race well and that of the environment poorly, which is why the topic call adds little for gender and race and matters for the environment (Section~\ref{Sec_mechanisms}).

\begin{table}[htbp]
\centering
\caption{Validation of the Topic Classification against JEL Codes}
\label{tab:app_topic_validation}
\small
\setlength{\tabcolsep}{4.5pt}
\begin{tabular}{lccccc}
\toprule
 & & \multicolumn{2}{c}{Share of code papers caught (\%)} & \multicolumn{2}{c}{Share listing the code (\%)} \\
\cmidrule(lr){3-4} \cmidrule(lr){5-6}
JEL code & Papers & Topic call & Flagged word & Topic-call papers & Other papers \\
\midrule
  J16 (gender) & 395 & 92 & 91 & 32.8 & 0.3 \\
  J15 or J71 (race) & 392 & 67 & 64 & 33.0 & 1.0 \\
  Q5 (environment) & 753 & 90 & 34 & 64.5 & 0.6 \\
\bottomrule
\end{tabular}
\begin{minipage}{0.95\textwidth}
\vspace{4pt}
\footnotesize
\textit{Notes:} For the NBER and CEPR papers with JEL codes (13,111; arXiv papers carry none), the share of papers listing the code (J16 Economics of Gender; J15 Minorities, Race, and Migrants, or J71 Discrimination; Q5 Environmental Economics) that the topic call labels with the corresponding topic, and the share that uses a targeted term in that context (the flagged-word measure); then the share listing the code among papers the topic call labels with the topic and among all other papers. Codes are counted once per paper.
\end{minipage}
\end{table}

\subsection{Gender Classification}\label{app_subsec_gender}

Using GPT-OSS (model \texttt{gpt-oss-120b}), we classify author gender from first names. The following prompt is used:

\begin{quote}
\small
\textit{Classify the likely gender of this person based on their name.}\\
\textit{Respond with exactly one of: male, mostly male, mostly female, female, impossible to know.}

\smallskip
\textit{Name: \{name\}}
\end{quote}

\noindent Responses are collapsed into a binary variable: ``male'' and ``mostly male'' map to male; ``female'' and ``mostly female'' map to female; ``impossible to know'' is treated as missing.

\subsection{Ethnicity Prediction}\label{app_subsec_ethnicity}

Predicted ethnicity is based on the US Census Bureau's 2010 Frequently Occurring Surnames file, which lists approximately 162,000 surnames that together cover most of the US population. For each surname, the file provides the percentage distribution across six racial/ethnic categories: White, Black, Asian/Pacific Islander, Hispanic, American Indian/Alaska Native, and Two or More Races. We extract each author's last name, normalize it to uppercase ASCII, and look it up in the Census file. The predicted ethnicity is the category with the highest probability for that surname. Authors whose surnames do not appear in the file are classified as ``Unknown.''

\subsection{PhD Graduation Years}\label{app_subsec_phd}

PhD graduation years are collected via automated web searches using Claude Haiku agents, which query university faculty pages, CVs, Wikipedia, RePEc profiles, and Google Scholar for each author. Seniority (years since PhD) is computed as the difference between the paper's publication year and the PhD graduation year and is therefore time-varying. Authors with more than 12 years since the PhD are classified as senior. Current PhD students are classified as junior; authors for whom no PhD year is found are excluded from the seniority analysis.

\begin{table}[htbp]
\centering
\caption{List of Targeted Terms}
\label{tab:censored_words}
\small
\begin{tabular}{p{0.93\textwidth}}
\toprule
\\[-6pt]
accessible, activism, activists, advocacy, advocate, advocates, affirming care, all-inclusive, allyship, anti-racism, antiracist, assigned at birth, assigned female at birth, assigned male at birth, at risk, barrier, barriers, belong, bias, biased, biased toward, biases, biases towards, biologically female, biologically male, BIPOC, Black, breastfeed + people, breastfeed + person, chestfeed + people, chestfeed + person, clean energy, climate crisis, climate science, commercial sex worker, community diversity, community equity, confirmation bias, cultural competence, cultural differences, cultural heritage, cultural sensitivity, culturally appropriate, culturally responsive, DEI, DEIA, DEIAB, DEIJ, disabilities, disability, discriminated, discrimination, discriminatory, disparity, diverse, diverse backgrounds, diverse communities, diverse community, diverse group, diverse groups, diversified, diversify, diversifying, diversity, enhance the diversity, enhancing diversity, environmental quality, equal opportunity, equality, equitable, equitableness, equity, ethnicity, excluded, exclusion, expression, female, females, feminism, fostering inclusivity, GBV, gender, gender based, gender based violence, gender diversity, gender identity, gender ideology, gender-affirming care, genders, Gulf of Mexico, hate speech, health disparity, health equity, hispanic minority, historically, identity, immigrants, implicit bias, implicit biases, inclusion, inclusive, inclusive leadership, inclusiveness, inclusivity, increase diversity, increase the diversity, indigenous community, inequalities, inequality, inequitable, inequities, inequity, injustice, institutional, intersectional, intersectionality, key groups, key people, key populations, Latinx, LGBT, LGBTQ, marginalize, marginalized, men who have sex with men, mental health, minorities, minority, most risk, MSM, multicultural, Mx, Native American, non-binary, nonbinary, oppression, oppressive, orientation, people + uterus, people-centered care, person-centered, person-centered care, polarization, political, pollution, pregnant people, pregnant person, pregnant persons, prejudice, privilege, privileges, promote diversity, promoting diversity, pronoun, pronouns, prostitute, race, race and ethnicity, racial, racial diversity, racial identity, racial inequality, racial justice, racially, racism, segregation, sense of belonging, sex, sexual preferences, sexuality, social justice, sociocultural, socioeconomic, status, stereotype, stereotypes, systemic, systemically, they/them, trans, transgender, transsexual, trauma, traumatic, tribal, unconscious bias, underappreciated, underprivileged, underrepresentation, underrepresented, underserved, undervalued, victim, victims, vulnerable populations, women, women and underrepresented \\[2pt]
\bottomrule
\multicolumn{1}{p{0.93\textwidth}}{\footnotesize \textit{Notes:} The 197 terms reported in \citet{yourish2025words}, morphological variants included (e.g., ``barrier'' and ``barriers''). The ``+'' notation denotes word combinations (e.g., ``breastfeed + people'' matches ``breastfeeding people''). For text matching, we expand the list to include UK spellings (e.g., marginalise, polarisation) and hyphenation variants (e.g., anti-racism, antiracism), yielding 224 search patterns in total.} \\
\end{tabular}
\end{table}

\clearpage

\begin{table}[htbp]
\centering
\caption{Summary Statistics: Paper Level}
\label{tab:summary_statistics}
\small
\begin{tabular}{l*{3}{>{\centering\arraybackslash}p{2.35cm}}}
\toprule
  & \multicolumn{1}{c}{All} & \multicolumn{1}{c}{US High-Fed} & \multicolumn{1}{c}{US Low-Fed} \\
\midrule
\addlinespace[6pt]
\textit{Panel A: Paper characteristics} \\
\addlinespace[3pt]
  Number of papers & 28,081 & 7,621 & 7,401 \\
  Targeted word count (abstract) & 0.911 & 0.949 & 1.063 \\
  & (1.989) & (2.101) & (2.180) \\
  Targeted word count (full text) & 60.63 & 68.15 & 73.94 \\
  & (99.51) & (107.75) & (115.68) \\
  Number of pages & 46.6 & 50.9 & 51.8 \\
  & (26.6) & (26.4) & (25.9) \\
  Number of authors & 2.81 & 2.68 & 3.33 \\
  & (1.74) & (1.36) & (2.14) \\
\addlinespace[6pt]
\textit{Panel B: Context classification} \\
\addlinespace[3pt]
  Gender, race, or environment & 0.107 & 0.121 & 0.128 \\
  Gender & 0.056 & 0.055 & 0.069 \\
  Race & 0.034 & 0.052 & 0.048 \\
  Environment & 0.024 & 0.025 & 0.023 \\
\addlinespace[6pt]
\textit{Panel C: Funding and affiliation} \\
\addlinespace[3pt]
  Number of US-affiliated authors & 1.12 & 2.27 & 1.88 \\
  Number of UK-affiliated authors & 0.22 & 0.05 & 0.11 \\
  Federal share of R\&D (\%) & 22.1 & 51.9 & 29.9 \\
  \quad Health \& Human Services (NIH) & 60.2 & 63.5 & 56.8 \\
  \quad Defense & 11.2 & 11.7 & 10.6 \\
  \quad National Science Foundation & 12.8 & 11.4 & 14.2 \\
  Share NBER & 0.277 & 0.508 & 0.471 \\
  Share CEPR & 0.226 & 0.111 & 0.180 \\
  Share arXiv & 0.498 & 0.381 & 0.350 \\
  Any author at red-state university & 0.217 & 0.149 & 0.284 \\
\bottomrule
\end{tabular}
\begin{minipage}{0.95\textwidth}
\vspace{4pt}
\footnotesize
\textit{Notes:} Standard deviations in parentheses. US papers have at least as many US- as UK-affiliated matched authors (ties assigned to US). US High-Fed = papers where at least half of all authors are from universities with above-median total-federal funding share (UK-affiliated and unmatched authors count as not High-Fed; ties assigned to High-Fed); US Low-Fed = the remaining US papers. The federal split uses HERD FY2024 data and the FY2024 cross-university median. The paper-level federal share averages the FY2024 university values over all authors of the paper, UK-affiliated and unmatched authors counted as zero. The three agency rows report each agency's share of total federal R\&D (the three largest federal funders of academic R\&D), averaged over papers with at least one HERD-matched author. Context shares (GPT-OSS classification of flagged words) are computed over all papers (including those with zero targeted words). Full-text term counts and page counts cover all three series; for arXiv papers, full texts come from the arXiv source files and page counts from the arXiv PDFs. Red-state = university located in a state that voted Republican in the 2024 presidential election.
\end{minipage}
\end{table}

\clearpage

\begin{table}[htbp]
\centering
\caption{Summary Statistics: Author Level}
\label{tab:summary_statistics_author}
\small
\begin{tabular}{l*{3}{>{\centering\arraybackslash}p{2.35cm}}}
\toprule
  & \multicolumn{1}{c}{All US} & \multicolumn{1}{c}{US High-Fed} & \multicolumn{1}{c}{US Low-Fed} \\
\midrule
\addlinespace[6pt]
\textit{Panel A: Paper characteristics} \\
\addlinespace[3pt]
  Number of author-paper pairs & 31,455 & 19,173 & 12,282 \\
  Unique authors & 11,454 & 7,021 & 4,905 \\
  Gender, race, or environment & 0.136 & 0.131 & 0.145 \\
  Targeted word count (abstract) & 1.055 & 0.996 & 1.147 \\
  & (2.210) & (2.152) & (2.294) \\
\addlinespace[6pt]
\textit{Panel B: Author characteristics} \\
\addlinespace[3pt]
  Female & 0.219 & 0.210 & 0.232 \\
  White & 0.457 & 0.445 & 0.476 \\
  Minority & 0.295 & 0.295 & 0.296 \\
  Surname not in Census & 0.248 & 0.261 & 0.228 \\
  Years since PhD & 15.9 & 16.0 & 15.8 \\
  & (12.9) & (12.8) & (13.0) \\
  Senior ($>$12 years since PhD) & 0.497 & 0.497 & 0.497 \\
  Red state & 0.247 & 0.177 & 0.358 \\
\addlinespace[6pt]
\textit{Panel C: Federal funding composition} \\
\addlinespace[3pt]
  Federal share of R\&D (\%) & 55.1 & 62.2 & 44.0 \\
  \quad Health \& Human Services (NIH) & 58.6 & 62.8 & 52.0 \\
  \quad Defense & 11.2 & 12.2 & 9.5 \\
  \quad National Science Foundation & 13.8 & 11.7 & 17.0 \\
\bottomrule
\end{tabular}
\begin{minipage}{0.95\textwidth}
\vspace{4pt}
\footnotesize
\textit{Notes:} Unique-author counts in the exposure-group columns are non-exclusive: authors affiliated with both High-Fed and Low-Fed universities across papers are counted in each group. Unit of observation is the author-paper pair, restricted to US-affiliated authors matched to HERD funding data. Standard deviations in parentheses. US High-Fed = author at a US university with above-median total-federal funding share; US Low-Fed = US-affiliated author not classified as High-Fed. The federal split uses HERD FY2024 data and the FY2024 cross-university median. The three agency rows report each agency's share of total federal R\&D, averaged over HERD-matched author-paper pairs. The context classification of flagged words (GPT-OSS) is measured at the paper level. Ethnicity predicted from US Census 2010 surname data. Years since PhD and seniority are computed for author-paper pairs with a recorded PhD year or a current PhD student (77\% of pairs); senior = more than 12 years since PhD; current PhD students count as junior. Red state = university in a state that voted Republican in the 2024 presidential election.
\end{minipage}
\end{table}

\clearpage

\section{Results Appendix}

\begin{figure}[htbp]
\centering
\includegraphics[width=\textwidth]{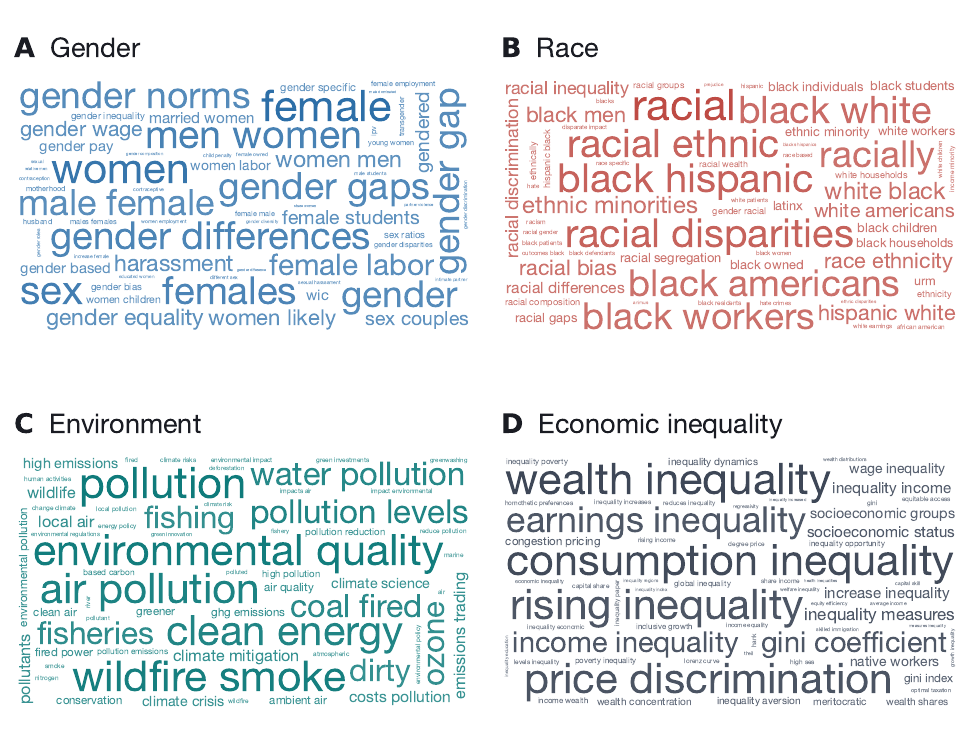}
\caption{Odds-Ratio Word Clouds by Context Category}
\label{fig:wordcloud_categories}
\begin{minipage}{0.95\textwidth}
\footnotesize
\textit{Notes:} Each panel displays the most distinctive unigrams and bigrams for abstracts classified in the given category, with word size proportional to the log odds ratio of the term appearing in abstracts of that category relative to abstracts not in that category. Abstracts are tokenized into unigrams and bigrams (minimum document frequency of 5, English stop words excluded), based on the classified abstracts of NBER, CEPR, and arXiv working papers (January 2020--June 2026). Classification by GPT-OSS based on the surrounding text in the abstract.
\end{minipage}
\end{figure}

\clearpage

\IfFileExists{results_20may/fig_app_raw_gre_author.pdf}{%
\begin{figure}[htbp]
\centering
\includegraphics[width=\textwidth]{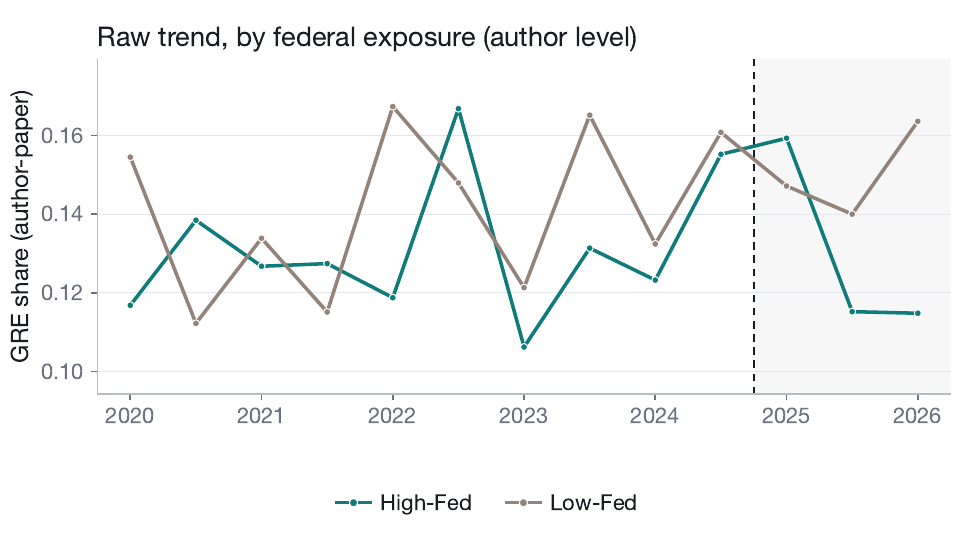}
\caption{Raw GRE Content Shares by Federal Exposure: Author Level}
\label{fig:raw_gre_author}
\begin{minipage}{0.95\textwidth}
\footnotesize
\textit{Notes:} This figure repeats the raw descriptive comparison of Figure~\ref{fig:event_studies} (Panel A) at the author level. It plots the share of author-paper observations containing at least one targeted term used in a Gender, Race, or Environment (GRE) context, separately for authors at US High-Fed and Low-Fed institutions. High-Fed universities are those whose total-federal share of R\&D expenditure (HERD FY2024, all fields) exceeds the cross-university median; the classification is based on each author's own university. The dashed vertical line indicates the policy change (January 2025). Classification by GPT-OSS based on the surrounding text in the abstract. US authors only.
\end{minipage}
\end{figure}
\clearpage
}{}
\IfFileExists{results_20may/fig_app_es_author.pdf}{%
\begin{figure}[htbp]
\centering
\includegraphics[width=\textwidth]{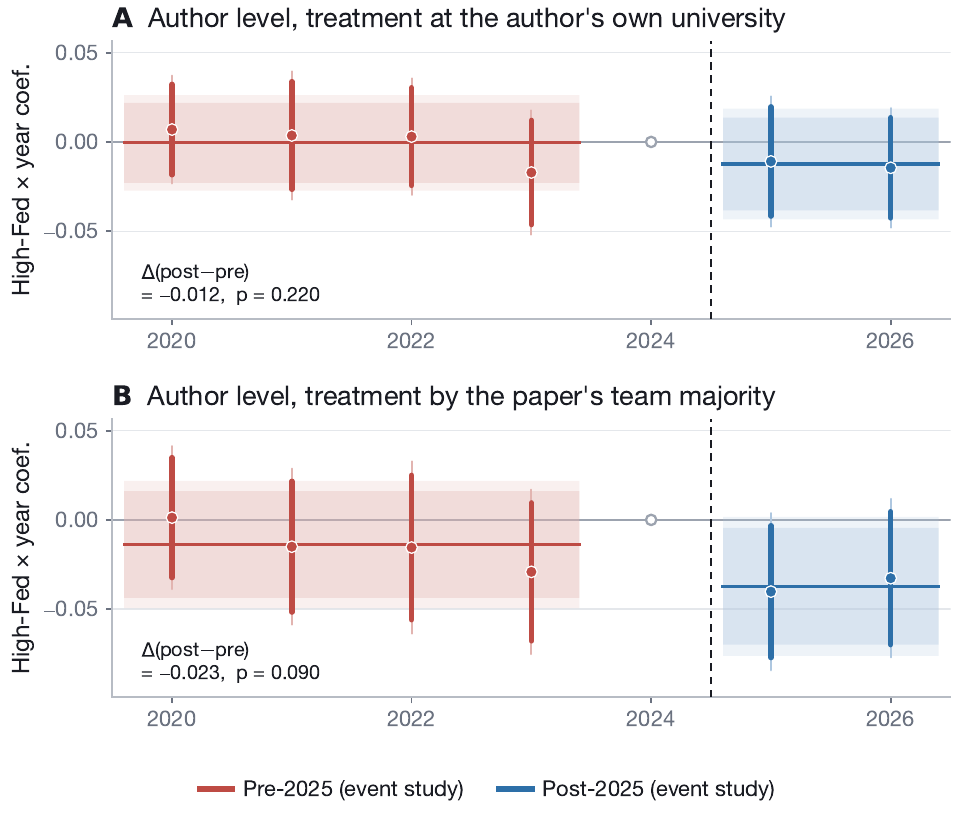}
\caption{Event Study at the Author Level: Gender, Race, and Environment Content}
\label{fig:es_author}
\begin{minipage}{0.95\textwidth}
\footnotesize
\textit{Notes:} This figure reports the author-level event studies of equations~(\ref{eq:eventstudy_author}) (Panel A) and~(\ref{eq:eventstudy_teammaj}) (Panel B), the author-level counterpart of Figure~\ref{fig:event_studies} (Panel B). The unit of observation is the author-paper pair. Each panel plots yearly coefficients on High-Fed $\times$ year, comparing US High-Fed to US Low-Fed authors, with 2024 (the last full year before the policy change) as the reference period. Panel A assigns each author the High-Fed status of their own university, held fixed at the last pre-2025 affiliation. Panel B assigns each author-paper observation the classification of the paper itself (team majority) and restricts the sample to US papers: a US paper is High-Fed if at least half of all its authors are at High-Fed universities (UK-affiliated and unmatched authors count as not High-Fed; ties assigned to High-Fed). High-Fed universities are those whose total-federal share of R\&D expenditure (HERD FY2024, all fields) exceeds the cross-university median; Low-Fed universities are at or below the median. The dependent variable is a binary indicator for whether the paper contains any targeted term used in a Gender, Race, or Environment (GRE) context in its abstract (GPT-OSS classification). All specifications include author fixed effects and source (NBER/CEPR/arXiv) fixed effects; standard errors are two-way clustered by university and paper. All specifications control for economics research capacity interacted with year (Section~\ref{sec:methods}). Red shading indicates pre-treatment periods; blue shading indicates post-treatment periods. Darker bands represent 90\% confidence intervals; lighter bands represent 95\% confidence intervals. Horizontal lines show pooled pre- and post-treatment averages. Linear probability model estimates. US authors only (affiliation fixed at the last pre-2025 paper).
\end{minipage}
\end{figure}
\clearpage
}{}
\IfFileExists{results_20may/fig_app_2024pre.pdf}{%
\begin{figure}[htbp]
\centering
\includegraphics[width=\textwidth]{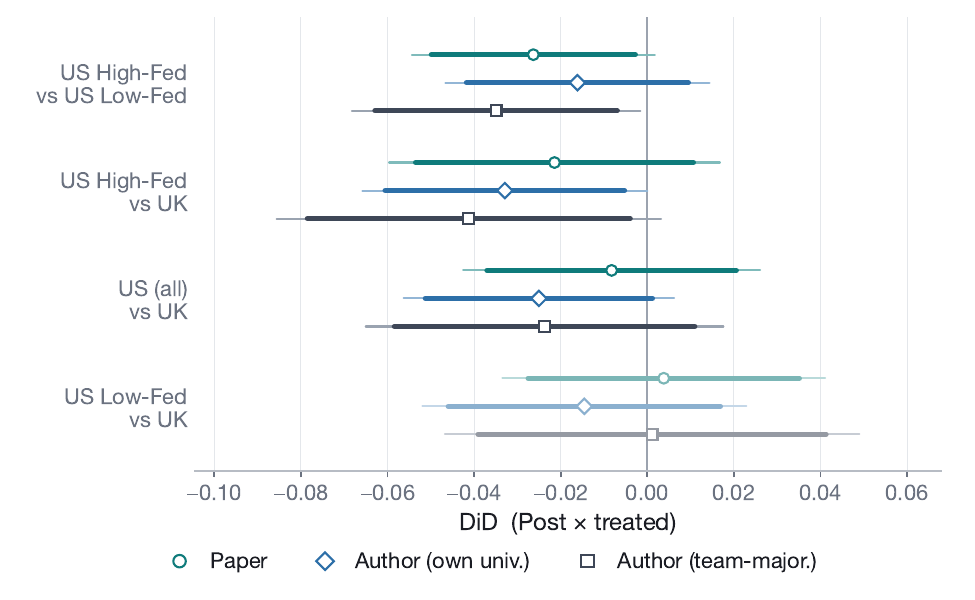}
\caption{Reference-Group Comparisons: Two-Year Pre-Period Window}
\label{fig:coefplot_2024pre}
\begin{minipage}{0.95\textwidth}
\footnotesize
\textit{Notes:} This figure replicates Figure~\ref{fig:coefplot_refgroups} with a two-year pre-period window: the pre-treatment period is restricted to 2023--2024, against the 2025--2026 post period (rather than the full 2020--2024 pre-treatment window), a check on robustness to potential pre-trends by restricting attention to the two years immediately preceding treatment. The dependent variable is a binary indicator for whether the abstract contains any targeted term used in a Gender, Race, or Environment context (GPT-OSS classification). With author fixed effects, the own-university estimates (blue diamonds) are identified only from authors who publish both in 2023--2024 and in the post period, a smaller sample than in the main specification. Team-majority estimates are shown as slate squares and paper-level estimates as teal circles. High-Fed universities are those whose total-federal share of R\&D expenditure (HERD FY2024, all fields) exceeds the cross-university median. Treatment assignment, including the tie rules, follows Figure~\ref{fig:coefplot_refgroups}. Thick bars represent 90\% confidence intervals; thin lines represent 95\% confidence intervals. Paper-level estimates use robust standard errors; author-level estimates include author fixed effects and cluster standard errors two-way by university and paper. All regressions control for economics research capacity interacted with Post (Section~\ref{sec:methods}). Linear probability model estimates. A tabular version is in Table~\ref{tab:app_2024pre}.
\end{minipage}
\end{figure}
\clearpage
}{}

\IfFileExists{results_20may/fig_app_contdose_fed.pdf}{%
\begin{figure}[htbp]
\centering
\includegraphics[width=\textwidth]{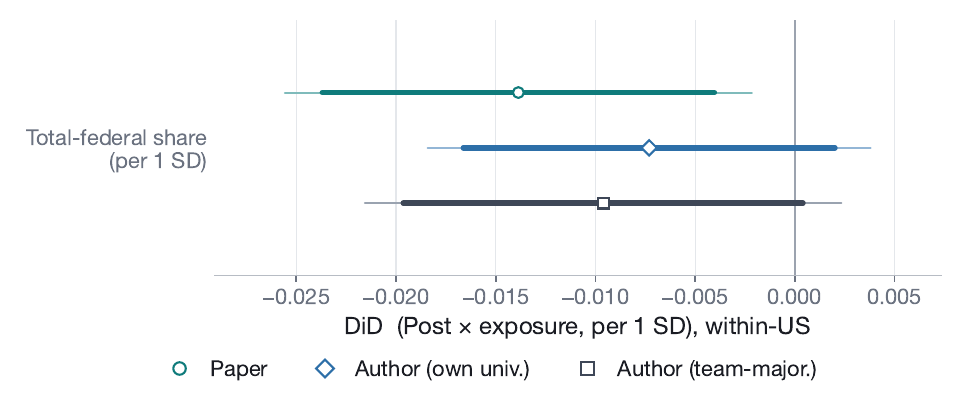}
\caption{Robustness: Continuous Federal-Share Exposure (Within US)}
\label{fig:coefplot_contdose}
\begin{minipage}{0.95\textwidth}
\footnotesize
\textit{Notes:} This figure examines the within-US treatment margin of Figure~\ref{fig:coefplot_refgroups} using continuous exposure instead of the binary median split: the total-federal share of a university's R\&D expenditure (HERD FY2024), standardized within the estimation sample, so coefficients are per one standard deviation of the share. The paper level (teal circles) uses the paper's average share across all its authors (UK-affiliated and unmatched authors count as zero) with one observation per paper and robust standard errors; the author level with own-university exposure (blue diamonds) assigns each author the basefixed share of their own university; the team-majority analog (slate squares) assigns the paper's average share; both author levels include author fixed effects and cluster standard errors two-way by university and paper. US papers have at least as many US-affiliated (HERD-matched) as UK-affiliated (HESA-matched) authors (ties assigned to US). The dependent variable is a binary indicator for whether the abstract contains any targeted term used in a Gender, Race, or Environment context (GPT-OSS classification). Thick bars represent 90\% confidence intervals; thin lines represent 95\% confidence intervals. All regressions control for economics research capacity interacted with Post (Section~\ref{sec:methods}). Linear probability model estimates. A tabular version is in Table~\ref{tab:app_contdose}.
\end{minipage}
\end{figure}
\clearpage
}{}

\IfFileExists{results_20may/fig_app_dense.pdf}{%
\begin{figure}[htbp]
\centering
\includegraphics[width=\textwidth]{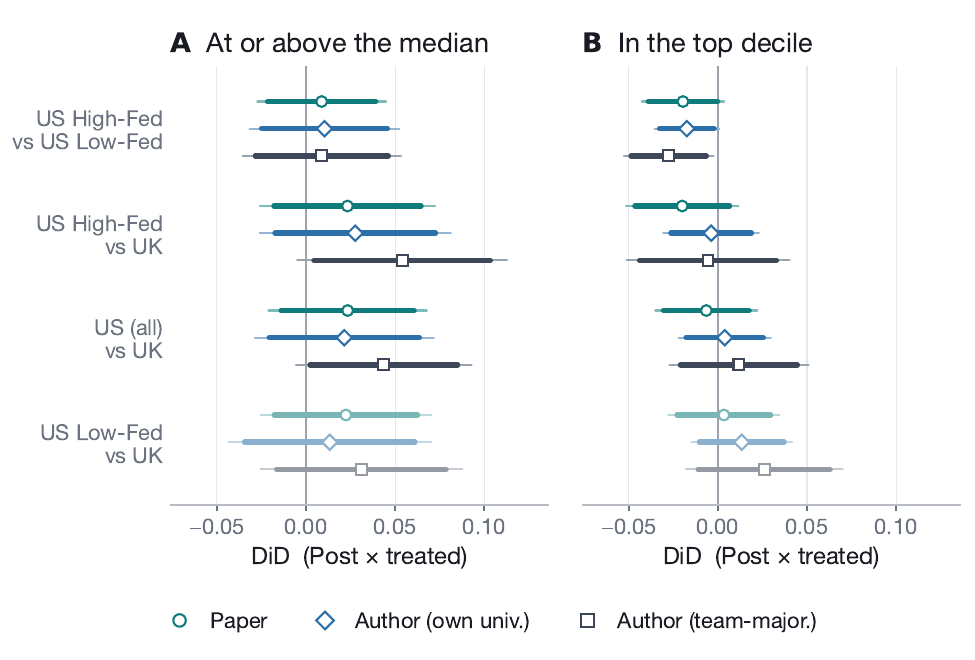}
\caption{Reference-Group Comparisons: Dense Use of Targeted Terms in the Full Text}
\label{fig:app_dense}
\begin{minipage}{0.95\textwidth}
\footnotesize
\textit{Notes:} This figure replicates Figure~\ref{fig:coefplot_refgroups} using, as the outcome, an indicator for dense use of targeted terms in the full paper text, with no context classification: a paper's term density is the number of targeted-term mentions in its full text per page, and the indicator equals one if the density is at or above the median (Panel A; 0.67 mentions per page) or the 90th percentile (Panel B; 2.87 mentions per page) of the density among all papers dated before 2025 (all three series pooled). In the within-US comparison, the share of papers at or above the median does not change significantly, while the share in the top decile declines. High-Fed universities are those whose total-federal share of R\&D expenditure (HERD FY2024, all fields) exceeds the cross-university median. Treatment assignment, including the tie rules, follows Figure~\ref{fig:coefplot_refgroups}. Thick bars represent 90\% confidence intervals; thin lines represent 95\% confidence intervals. Paper-level estimates (teal circles) use robust standard errors; author-level estimates (own university: blue diamonds; team majority: slate squares) include author fixed effects and cluster standard errors two-way by university and paper. All regressions control for economics research capacity interacted with Post (Section~\ref{sec:methods}). Linear probability model estimates. A tabular version of Panel B is in Table~\ref{tab:fig3} (``Dense full text (top decile)'' column).
\end{minipage}
\end{figure}
\clearpage
}{}

\IfFileExists{results_20may/fig_app_anyword.pdf}{%
\begin{figure}[htbp]
\centering
\includegraphics[width=\textwidth]{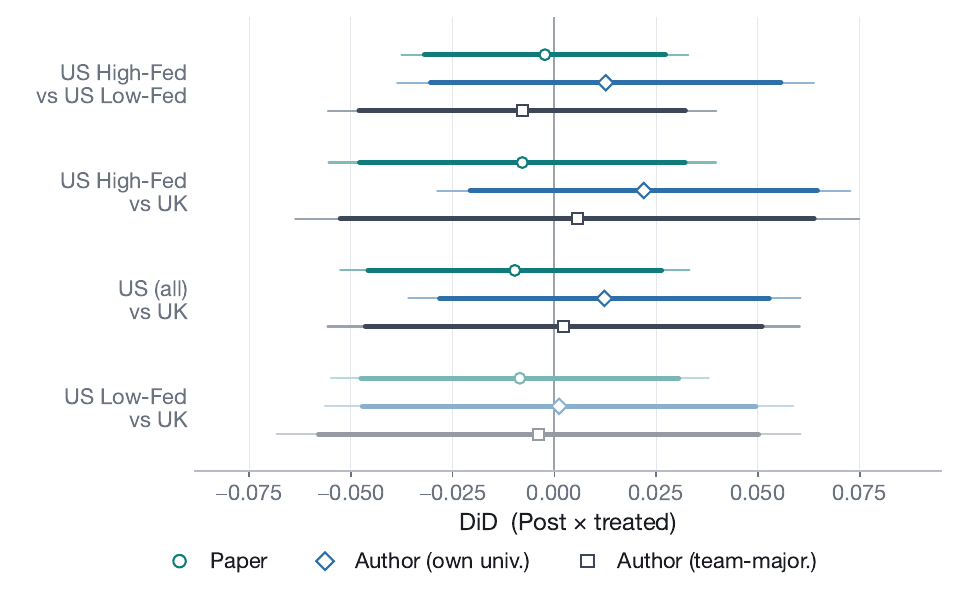}
\caption{Reference-Group Comparisons: Any Targeted Word in the Abstract}
\label{fig:coefplot_anyword}
\begin{minipage}{0.95\textwidth}
\footnotesize
\textit{Notes:} This figure replicates Figure~\ref{fig:coefplot_refgroups} using a measure that does not condition on the GPT-OSS context classification: a binary indicator for whether any targeted word appears anywhere in the abstract. This broader measure shows no differential change in any comparison, consistent with many targeted terms having incidental technical uses in economics that dilute the sensitive margin. High-Fed universities are those whose total-federal share of R\&D expenditure (HERD FY2024, all fields) exceeds the cross-university median. Treatment assignment, including the tie rules, follows Figure~\ref{fig:coefplot_refgroups}. Thick bars represent 90\% confidence intervals; thin lines represent 95\% confidence intervals. Paper-level estimates (teal circles) use robust standard errors; author-level estimates (own university: blue diamonds; team majority: slate squares) include author fixed effects and cluster standard errors two-way by university and paper. All regressions control for economics research capacity interacted with Post (Section~\ref{sec:methods}). Linear probability model estimates. A tabular version is in Table~\ref{tab:app_anyword} (``Any targeted word (abstract)'' column).
\end{minipage}
\end{figure}
\clearpage
}{}

\IfFileExists{results_20may/fig_app_anyword_fulltext.pdf}{%
\begin{figure}[htbp]
\centering
\includegraphics[width=\textwidth]{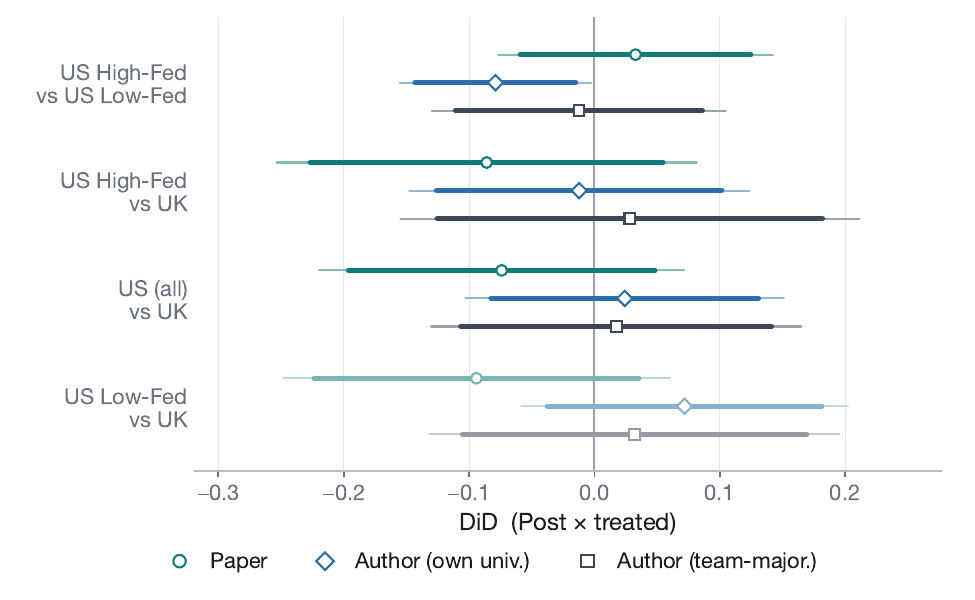}
\caption{Reference-Group Comparisons: Targeted-Word Count in the Full Text}
\label{fig:coefplot_anyword_fulltext}
\begin{minipage}{0.95\textwidth}
\footnotesize
\textit{Notes:} This figure replicates Figure~\ref{fig:coefplot_refgroups} using the raw count of targeted words in the full paper text, with no context classification. In the within-US comparison, the raw count declines at the own-university author level, while the paper-level and team-majority estimates are not statistically distinguishable from zero. High-Fed universities are those whose total-federal share of R\&D expenditure (HERD FY2024, all fields) exceeds the cross-university median. Treatment assignment, including the tie rules, follows Figure~\ref{fig:coefplot_refgroups}. The page count is included as a control. Thick bars represent 90\% confidence intervals; thin lines represent 95\% confidence intervals. Paper-level estimates (teal circles) use robust standard errors; author-level estimates (own university: blue diamonds; team majority: slate squares) include author fixed effects and cluster standard errors two-way by university and paper. All regressions control for economics research capacity interacted with Post (Section~\ref{sec:methods}). Poisson pseudo-maximum-likelihood (PPML) estimates. A tabular version is in Table~\ref{tab:app_anyword} (``Targeted words (full text)'' column).
\end{minipage}
\end{figure}
\clearpage
}{}

\IfFileExists{results_20may/fig4_content_union_addcomps_nov05.pdf}{%
\begin{figure}[htbp]
\centering
\includegraphics[width=\textwidth]{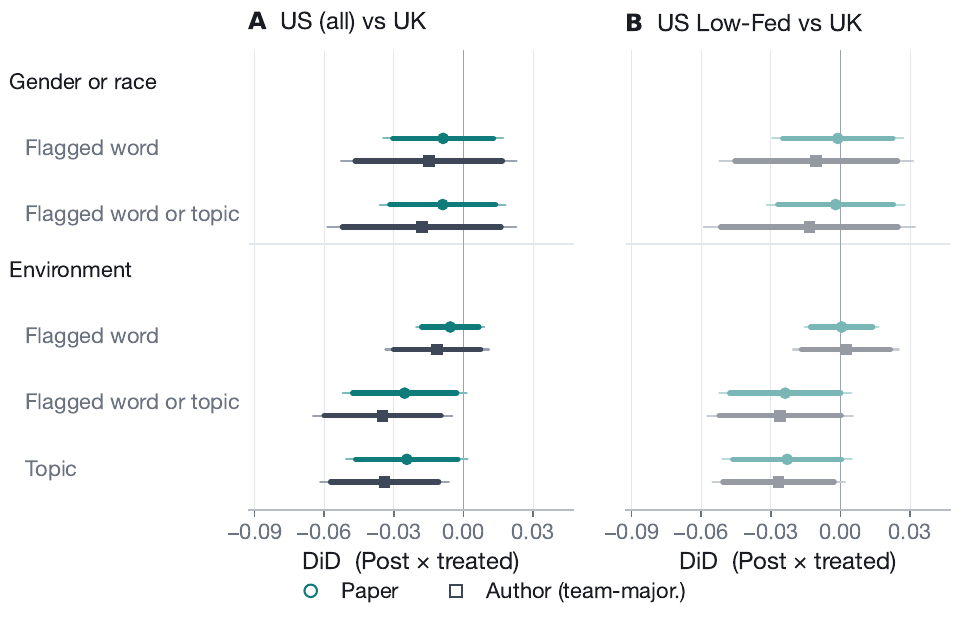}
\caption{Treatment Effects by Content: Remaining Comparisons}
\label{fig:topics_addcomps}
\begin{minipage}{0.95\textwidth}
\footnotesize
\textit{Notes:} This figure completes Figure~\ref{fig:topics}, showing the content rows of Figure~\ref{fig:topics} for the remaining comparisons: all US vs.\ UK (Panel A) and US Low-Fed vs.\ UK (Panel B; lighter shades, as neither group is highly exposed to federal funding). The gender-or-race rows do not change significantly in these two comparisons; the environment rows that include the topic are negative in both (topic row, paper level: $-0.024$ for all US and $-0.023$ for US Low-Fed vs.\ UK; team majority: $-0.034$, $p = 0.02$, and $-0.027$). Rows, series, samples, and treatment definitions follow Figure~\ref{fig:topics}. Thick bars represent 90\% confidence intervals; thin lines represent 95\% confidence intervals. All regressions control for economics research capacity interacted with Post (Section~\ref{sec:methods}). Linear probability model estimates. All rows are tabulated, with all three assignments, in Table~\ref{tab:content_union}.
\end{minipage}
\end{figure}
\clearpage
}{}

\IfFileExists{results_20may/fig_cruz_aug.pdf}{
\begin{figure}[htbp]
\centering
\includegraphics[width=\textwidth]{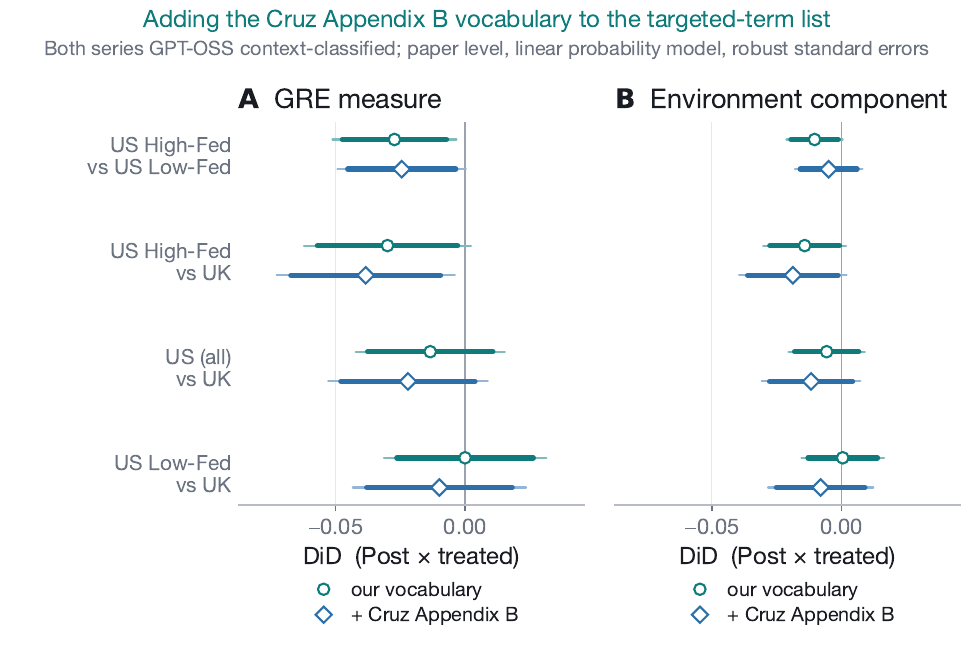}
\caption{Robustness: Adding the Senate-Appendix Vocabulary to the Targeted-Term List}
\label{fig:cruz_aug}
\begin{minipage}{0.95\textwidth}
\footnotesize
\textit{Notes:} This figure displays difference-in-differences coefficients (Post $\times$ Treated) comparing our published outcome with an augmented outcome that adds the Senate-appendix vocabulary (Section~\ref{sec_data}): the 563 Appendix B keywords that are not identical to one of our search patterns, including 34 environmental terms absent from the \citet{yourish2025words} compilation (``climate change,'' ``net zero,'' ``ESG''). Occurrences of the added keywords (1,455 word-paper pairs across 1,297 abstracts, of which 538 are classified as Environment) are classified with the identical GPT-OSS context prompt; occurrences of overlapping words retain their published classifications. Panel A shows the GRE indicator (any targeted term used in a Gender, Race, or Environment context; share of abstracts 10.7\% for our vocabulary and 12.2\% augmented); Panel B shows the Environment component alone (2.4\% and 3.8\%). Rows are the four treatment-control comparisons. All estimates are at the paper level with robust standard errors and source indicators. High-Fed universities are those whose total-federal share of R\&D expenditure (HERD FY2024, all fields) exceeds the cross-university median. US papers have at least as many US-affiliated (HERD-matched) as UK-affiliated (HESA-matched) authors (ties assigned to US); UK papers have strictly more UK-affiliated authors. A US paper is High-Fed if at least half of all its authors are at High-Fed universities (UK-affiliated and unmatched authors count as not High-Fed; ties assigned to High-Fed). Thick bars represent 90\% confidence intervals; thin lines represent 95\% confidence intervals. All regressions control for economics research capacity interacted with Post (Section~\ref{sec:methods}). Linear probability model estimates.
\end{minipage}
\end{figure}
\clearpage
}{}

\IfFileExists{results_20may/fig5_agency_addcomps.pdf}{%
\begin{figure}[htbp]
\centering
\includegraphics[width=\textwidth]{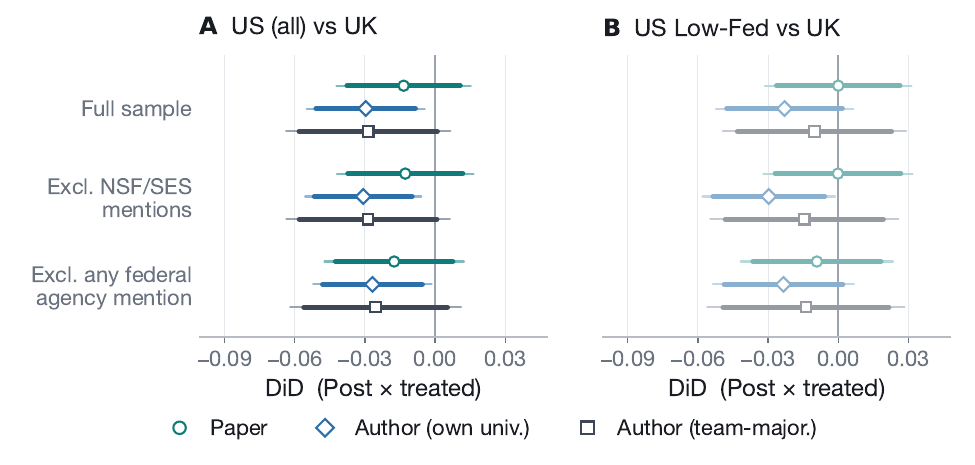}
\caption{Treatment Effects Excluding Federal-Agency Mentions: Remaining Comparisons}
\label{fig:nsf_exclude_addcomps}
\begin{minipage}{0.95\textwidth}
\footnotesize
\textit{Notes:} This figure completes Figure~\ref{fig:nsf_exclude}, showing the federal-acknowledgment exclusions for the remaining comparisons: all US vs.\ UK (Panel A) and US Low-Fed vs.\ UK (Panel B; lighter shades, as neither group is highly exposed to federal funding). Samples, series, and treatment definitions follow Figure~\ref{fig:nsf_exclude}; the x-axis matches Figures~\ref{fig:coefplot_refgroups} and~\ref{fig:nsf_exclude}. Thick bars represent 90\% confidence intervals; thin lines represent 95\% confidence intervals. All regressions control for economics research capacity interacted with Post (Section~\ref{sec:methods}). Linear probability model estimates. A tabular version is in Table~\ref{tab:fig5}.
\end{minipage}
\end{figure}
\clearpage
}{}

\IfFileExists{results_20may/fig_app_redblue.pdf}{%
\begin{figure}[htbp]
\centering
\includegraphics[width=\textwidth]{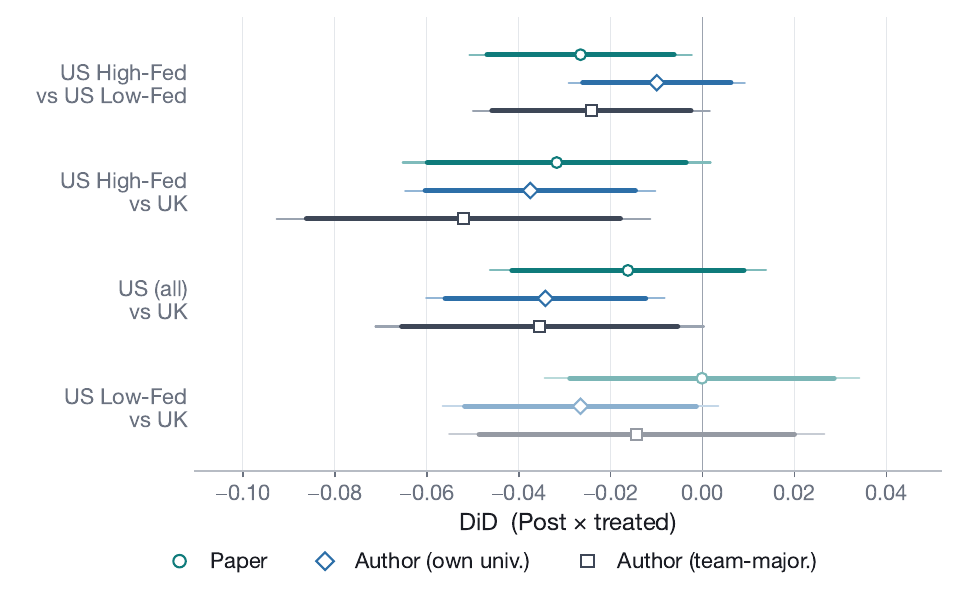}
\caption{Reference-Group Comparisons: Red/Blue State $\times$ Post Control}
\label{fig:coefplot_redblue}
\begin{minipage}{0.95\textwidth}
\footnotesize
\textit{Notes:} This figure replicates Figure~\ref{fig:coefplot_refgroups} while adding Red State $\times$ Post as an additional control, where Red State indicates universities located in states that voted Republican in the 2024 presidential election. This allows for a differential post-2024 shift between universities in Republican- and Democratic-leaning states. The dependent variable is a binary indicator for whether the abstract contains any targeted term used in a Gender, Race, or Environment context (GPT-OSS classification). High-Fed universities are those whose total-federal share of R\&D expenditure (HERD FY2024, all fields) exceeds the cross-university median. Treatment assignment, including the tie rules, follows Figure~\ref{fig:coefplot_refgroups}. Thick bars represent 90\% confidence intervals; thin lines represent 95\% confidence intervals. Paper-level estimates (teal circles) use robust standard errors; author-level estimates (own university: blue diamonds; team majority: slate squares) include author fixed effects and cluster standard errors two-way by university and paper. All regressions control for economics research capacity interacted with Post (Section~\ref{sec:methods}). Linear probability model estimates. A tabular version is in Table~\ref{tab:app_redblue}.
\end{minipage}
\end{figure}
\clearpage
}{}

\begin{figure}[htbp]
\centering
\includegraphics[width=\textwidth]{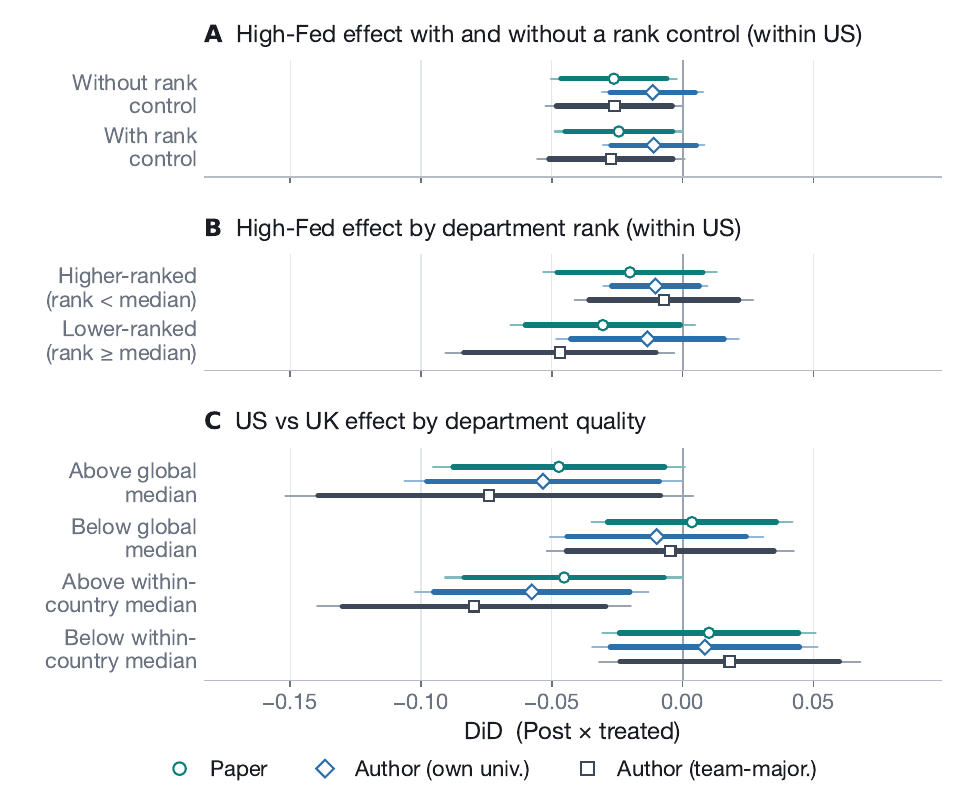}
\caption{Treatment Effects by Department Rank and Country}
\label{fig:rank}
\begin{minipage}{0.95\textwidth}
\footnotesize
\textit{Notes:} This figure assesses the role of institutional research standing, measured by the RePEc\slash IDEAS ranking of economics departments (a lower rank number denotes higher standing). Panel A reports the US High-Fed versus Low-Fed effect with and without a department-rank control (the rank level and its interaction with Post; sample restricted to papers with a nonmissing rank); Panel B reports the same effect estimated separately for papers at higher-ranked (rank below the median) and lower-ranked (rank at or above the median) departments; Panel C reports the US versus UK comparison estimated separately within above- and below-median quality halves, with the median split computed both globally (across the pooled sample) and within country. All panels show three series: paper-level estimates (teal circles), own-university author-level estimates (blue diamonds), and team-majority author-level estimates (slate squares); Panel A also reports the author-level estimates alongside the paper-level estimates. The dependent variable is a binary indicator for whether a paper contains any targeted term used in a Gender, Race, or Environment (GRE) context in its abstract (GPT-OSS classification). High-Fed universities are those whose total-federal share of R\&D expenditure (HERD FY2024, all fields) exceeds the cross-university median; US papers have at least as many US-affiliated (HERD-matched) as UK-affiliated (HESA-matched) authors (ties assigned to US), UK papers strictly more UK-affiliated authors. Thick bars represent 90\% confidence intervals; thin lines represent 95\% confidence intervals. Paper-level estimates use robust standard errors; author-level estimates include author fixed effects and cluster standard errors two-way by university and paper. All regressions control for economics research capacity interacted with Post (Section~\ref{sec:methods}). Linear probability model estimates. A tabular version is in Table~\ref{tab:fig6}.
\end{minipage}
\end{figure}
\clearpage

\IfFileExists{results_20may/fig_app_altinf.pdf}{%
\begin{figure}[htbp]
\centering
\includegraphics[width=\textwidth]{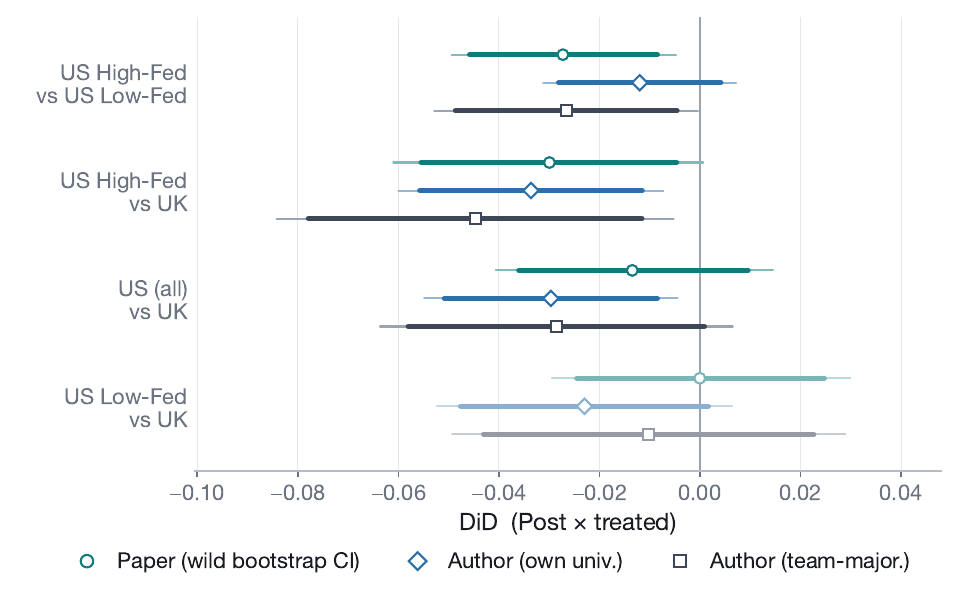}
\caption{Robustness: Alternative Inference (Wild Cluster Bootstrap)}
\label{fig:coefplot_altinf}
\begin{minipage}{0.95\textwidth}
\footnotesize
\textit{Notes:} This figure replicates Figure~\ref{fig:coefplot_refgroups} under the alternative inference scheme: for the paper-level estimates (teal circles), standard errors are clustered by the paper's modal university among its matched authors (UK universities identified by their HESA provider codes) and the confidence intervals are computed by wild cluster bootstrap test inversion (Rademacher weights, 99,999 replications, imposing the null); the sample is restricted to papers with an assignable modal university. The author-level series are unchanged from Figure~\ref{fig:coefplot_refgroups} -- own-university (blue diamonds) and team-majority (slate squares) estimates include author fixed effects and cluster standard errors two-way by university and paper -- and are shown for comparison. Treatment assignment, including the tie rules, follows Figure~\ref{fig:coefplot_refgroups}. Thick bars represent 90\% confidence intervals; thin lines represent 95\% confidence intervals. All regressions control for economics research capacity interacted with Post (Section~\ref{sec:methods}). Linear probability model estimates. A tabular version, including cluster-robust and wild bootstrap $p$-values, is in Table~\ref{tab:app_altinf}.
\end{minipage}
\end{figure}
\clearpage
}{}

\IfFileExists{results_20may/fig_app_paperlength.pdf}{%
\begin{figure}[htbp]
\centering
\includegraphics[width=\textwidth]{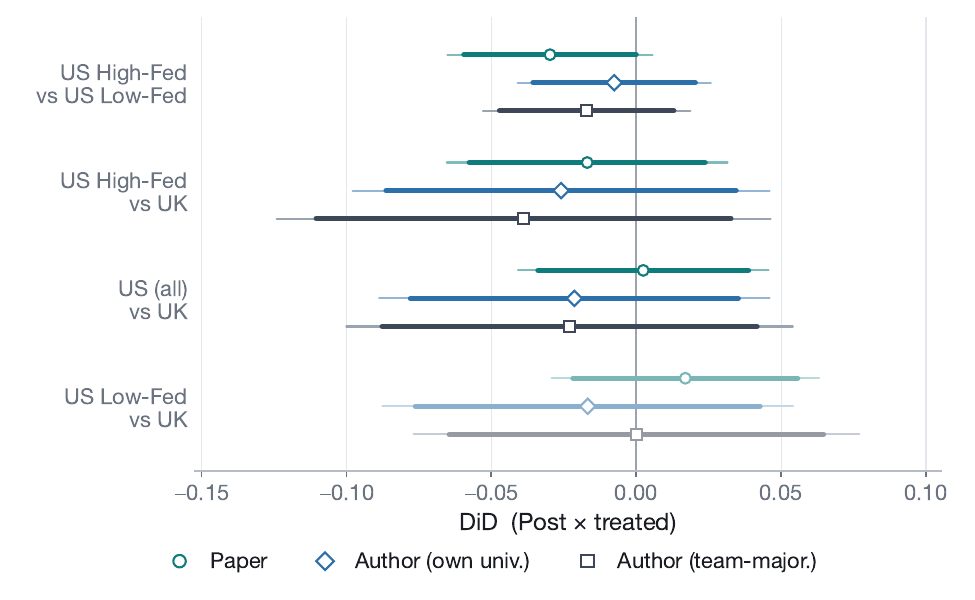}
\caption{Reference-Group Comparisons: Paper Length (Number of Pages)}
\label{fig:paperlength}
\begin{minipage}{0.95\textwidth}
\footnotesize
\textit{Notes:} This figure reports pooled difference-in-differences estimates with the number of pages as the outcome, across the reference-group comparisons of Figure~\ref{fig:coefplot_refgroups}. The within-US paper-level estimate is $-0.030$ ($p = 0.10$), about 3\% of pages; the other paper-level estimates are close to zero, and the author-level cross-country point estimates are reductions of at most about 4\% of pages. We find little evidence that authors shortened their papers; small changes in total pages cannot rule out that short sections (for example, demographic-heterogeneity sections) were dropped. High-Fed universities are those whose total-federal share of R\&D expenditure (HERD FY2024, all fields) exceeds the cross-university median. Treatment assignment, including the tie rules, follows Figure~\ref{fig:coefplot_refgroups}. Thick bars represent 90\% confidence intervals; thin lines represent 95\% confidence intervals. Paper-level estimates (teal circles) use robust standard errors; author-level estimates (own university: blue diamonds; team majority: slate squares) include author fixed effects and cluster standard errors two-way by university and paper. All regressions control for economics research capacity interacted with Post (Section~\ref{sec:methods}). Poisson pseudo-maximum-likelihood (PPML) estimates. A tabular version is in Table~\ref{tab:app_paperlength}.
\end{minipage}
\end{figure}
\clearpage
}{}

\IfFileExists{results_20may/fig_app_papercounts.pdf}{%
\begin{figure}[htbp]
\centering
\includegraphics[width=\textwidth]{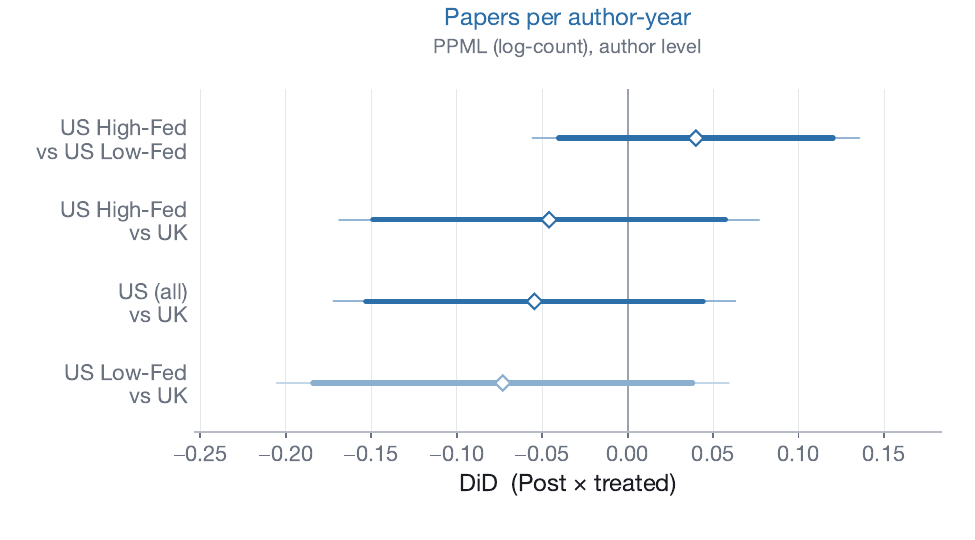}
\caption{Reference-Group Comparisons: Number of Papers per Author per Year}
\label{fig:papercounts}
\begin{minipage}{0.95\textwidth}
\footnotesize
\textit{Notes:} This figure reports pooled difference-in-differences estimates with the average number of papers per author per year as the outcome, across the reference-group comparisons of Figure~\ref{fig:coefplot_refgroups}. For each author, the paper count is computed separately for the pre-period (2020--2024, divided by 5) and the post-period (papers dated 2025 onward, through June 2026, divided by 1.5), yielding two observations per author; this averaging avoids having to impute zeros for years in which an author is absent from the data. The within-US High-Fed versus Low-Fed estimate is positive and not statistically significant ($0.040$, $p = 0.41$), and the negative cross-country estimates are, if anything, larger for US Low-Fed than for US High-Fed authors, providing no evidence of a High-Fed-specific reduction in output. Only author-level estimates are shown, as paper counts are undefined at the paper level; because the unit is the author-period, the team-majority classification does not apply here and treatment is the author's own (basefixed) university. High-Fed universities are those whose total-federal share of R\&D expenditure (HERD FY2024, all fields) exceeds the cross-university median. Thick bars represent 90\% confidence intervals; thin lines represent 95\% confidence intervals. Estimation is on an author-by-period panel with author fixed effects; standard errors are clustered by university. All regressions control for economics research capacity interacted with Post (Section~\ref{sec:methods}). Poisson pseudo-maximum-likelihood (PPML) estimates. A tabular version is in Table~\ref{tab:app_papercounts}.
\end{minipage}
\end{figure}
\clearpage
}{}

\IfFileExists{results_20may/fig_app_papercounts_split.pdf}{%
\begin{figure}[htbp]
\centering
\includegraphics[width=\textwidth]{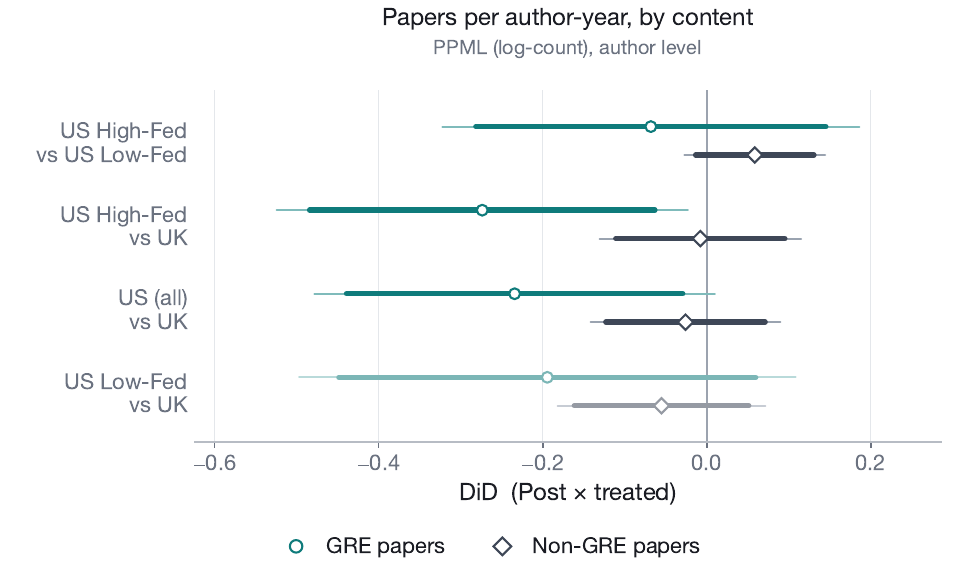}
\caption{Papers per Author per Year: GRE versus Non-GRE Content}
\label{fig:papercounts_split}
\begin{minipage}{0.95\textwidth}
\footnotesize
\textit{Notes:} This figure splits the author-productivity difference-in-differences of Figure~\ref{fig:papercounts} into two outcomes -- the average number of \textit{GRE} papers per author per year and the average number of \textit{non-GRE} papers per author per year -- to test whether any post-2025 decline in output is specific to Gender, Race, and Environment content or general. A paper is classified as GRE if it contains at least one targeted term used in a Gender, Race, or Environment context (GPT-OSS classification of the abstract). Each count is constructed exactly as in Figure~\ref{fig:papercounts} (pre-period 2020--2024 divided by 5; post-period 2025 onward, through June 2026, divided by 1.5), and the two outcomes partition the total paper count exactly. Across the cross-country comparisons the larger declines are for GRE papers. A cluster bootstrap over universities (re-estimating both Poisson models jointly on each of 300 draws, with UK universities identified by their HESA provider codes) distinguishes the GRE-versus-non-GRE gap at the 5\% level only for US High-Fed versus UK (difference $-0.27$, $p = 0.025$; US versus UK: $-0.21$, $p = 0.056$; within-US High-Fed versus Low-Fed: $p = 0.20$; US Low-Fed versus UK: $p = 0.34$). High-Fed universities are those whose total-federal share of R\&D expenditure (HERD FY2024, all fields) exceeds the cross-university median. Estimation is on an author-by-period panel with author fixed effects; standard errors are clustered by university; PPML estimates. All regressions control for economics research capacity interacted with Post (Section~\ref{sec:methods}). Thick bars represent 90\% confidence intervals; thin lines represent 95\% confidence intervals.
\end{minipage}
\end{figure}
\clearpage
}{}

\IfFileExists{results_20may/fig_randinf_treatment_teammaj_nov13.pdf}{%
\begin{figure}[htbp]
\centering
\includegraphics[width=\textwidth]{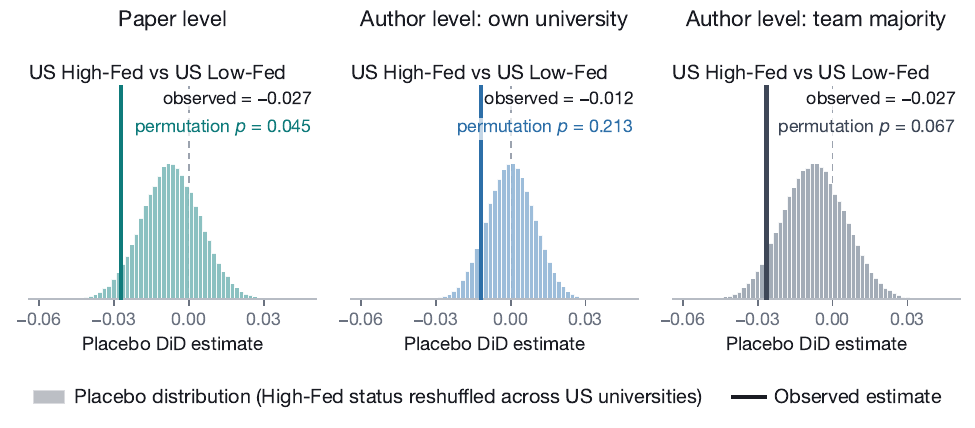}
\caption{Randomization Inference: High-Fed Status Reassigned across US Universities}
\label{fig:randinf}
\begin{minipage}{0.95\textwidth}
\footnotesize
\textit{Notes:} This figure reports a placebo exercise for the comparison of US High-Fed with US Low-Fed, at the paper level (left) and at the author level under the own-university (middle) and team-majority (right) assignments. Holding fixed the papers, authors, affiliations, outcomes, and dates, we randomly reassign High-Fed status across the 373 US universities with a High-Fed classification in our sample, keeping the number of High-Fed universities fixed at 169, re-classify every paper and author from the reassigned labels by the rules of Figure~\ref{fig:coefplot_refgroups}, and re-estimate the same difference-in-differences on each of $B = 100{,}000$ draws. The histograms show the placebo estimates; the vertical line marks the observed estimate, and the annotation reports the two-sided permutation $p$-value (the share of draws at least as extreme in absolute value as the observed estimate). The paper-level placebo distribution is centered slightly below zero (mean $-0.007$) because coauthors outside US universities count as not High-Fed: under any assignment, High-Fed papers are disproportionately papers written entirely at US universities, which reduced GRE content by more. Under the team-majority assignment (right), each author-paper observation takes the reassigned status of its paper. Its placebo distribution is centered at the same value, for the same reason. Reassigning High-Fed status only within quintiles of university size yields $p = 0.040$ at the paper level, $p = 0.21$ at the own-university author level, and $p = 0.062$ under the team-majority assignment. Estimation (including the capacity control), samples, and the treatment definition are identical to Figure~\ref{fig:coefplot_refgroups} (paper-level, own-university, and team-majority author series).
\end{minipage}
\end{figure}
\clearpage
}{}

\begin{table}[htbp]
\centering
\caption{Change in GRE Content After 2024, by Reference Group}
\label{tab:fig3}
\footnotesize
\setlength{\tabcolsep}{4.5pt}
\begin{tabular}{lccc}
\toprule
 & GRE in abstract & Dense full text (top decile) & $N$ \\
\midrule
\addlinespace[3pt]
\multicolumn{4}{l}{\textit{US High-Fed vs.\ US Low-Fed}} \\
\addlinespace[2pt]
  Paper & -0.027** & -0.020* & 15,022 \\
   & (0.012) & (0.012) &  \\
  Author (own univ.) & -0.012 & -0.017* & 25,506 \\
   & (0.010) & (0.009) &  \\
  Author (team-maj.) & -0.027** & -0.028** & 24,817 \\
   & (0.013) & (0.013) &  \\
\addlinespace[3pt]
\multicolumn{4}{l}{\textit{US High-Fed vs.\ UK}} \\
\addlinespace[2pt]
  Paper & -0.030* & -0.020 & 10,564 \\
   & (0.016) & (0.016) &  \\
  Author (own univ.) & -0.034** & -0.004 & 20,287 \\
   & (0.013) & (0.014) &  \\
  Author (team-maj.) & -0.045** & -0.006 & 16,823 \\
   & (0.020) & (0.023) &  \\
\addlinespace[3pt]
\multicolumn{4}{l}{\textit{US (all) vs.\ UK}} \\
\addlinespace[2pt]
  Paper & -0.013 & -0.006 & 17,965 \\
   & (0.015) & (0.015) &  \\
  Author (own univ.) & -0.030** & 0.004 & 30,034 \\
   & (0.013) & (0.013) &  \\
  Author (team-maj.) & -0.029 & 0.012 & 29,505 \\
   & (0.018) & (0.020) &  \\
\addlinespace[3pt]
\multicolumn{4}{l}{\textit{US Low-Fed vs.\ UK}} \\
\addlinespace[2pt]
  Paper & 0.000 & 0.003 & 10,344 \\
   & (0.016) & (0.016) &  \\
  Author (own univ.) & -0.023 & 0.013 & 14,275 \\
   & (0.015) & (0.014) &  \\
  Author (team-maj.) & -0.010 & 0.026 & 14,347 \\
   & (0.020) & (0.023) &  \\
\bottomrule
\end{tabular}
\begin{minipage}{0.95\textwidth}
\vspace{4pt}
\footnotesize
\textit{Notes:} Each cell is the difference-in-differences coefficient on Post $\times$ Treatment, where Post indicates papers dated 2025 or later. Standard errors in parentheses; paper-level standard errors are robust, author-level standard errors are two-way clustered by university and paper. US papers have at least as many US- as UK-affiliated matched authors (ties assigned to US); UK papers have strictly more UK-affiliated authors; High-Fed papers have at least half of all their authors at High-Fed universities (UK-affiliated and unmatched authors count as not High-Fed; ties assigned to High-Fed). ``Author (own univ.)'' assigns treatment by the author's own (basefixed) university with author fixed effects; ``Author (team-majority)'' assigns treatment by the paper's classification, with author fixed effects. The outcome ``GRE in abstract'' is an indicator for whether the abstract contains a targeted term used in a Gender, Race, or Environment context (GPT-OSS classification), estimated as a linear probability model; the coefficient is a change in probability. ``Dense full text (top decile)'' is a linear probability model on an indicator for whether the paper's targeted-term mentions per page are at or above the 90th percentile of pre-2025 papers (all series pooled; no context classification); the coefficient is a change in probability. High-Fed = above-median total-federal R\&D share (HERD FY2024, all fields). All regressions control for economics research capacity interacted with Post: the log of one plus the number of papers in the sample dated before 2025 with at least one author at the university, averaged over the paper's US- and UK-affiliated authors (the author's own university in the own-university specification). $^{*}\,p<0.10$, $^{**}\,p<0.05$, $^{***}\,p<0.01$. $N$ is for the abstract specification.
\end{minipage}
\end{table}

\clearpage

\begin{table}[htbp]
\centering
\caption{Change in Content After 2024: Flagged Words and Topics (Abstract)}
\label{tab:content_union}
\small
\setlength{\tabcolsep}{4pt}
\resizebox{\textwidth}{!}{%
\begin{tabular}{lcccccccc}
\toprule
 & \multicolumn{3}{c}{Gender or race} & & \multicolumn{3}{c}{Environment} & \\
\cmidrule(lr){2-4} \cmidrule(lr){6-8}
 & Flagged word & Flagged word or topic & Topic & & Flagged word & Flagged word or topic & Topic & $N$ \\
\midrule
\addlinespace[3pt]
\multicolumn{9}{l}{\textit{US High-Fed vs.\ US Low-Fed}} \\
\addlinespace[2pt]
  Paper & -0.018 & -0.017 & -0.010 & & -0.010* & -0.001 & -0.001 & 15,022 \\
   & (0.011) & (0.012) & (0.012) & & (0.006) & (0.010) & (0.010) & \\
  Author (own univ.) & -0.004 & 0.000 & -0.001 & & -0.010 & -0.002 & -0.001 & 25,506 \\
   & (0.009) & (0.010) & (0.009) & & (0.008) & (0.008) & (0.008) & \\
  Author (team-maj.) & -0.015 & -0.014 & -0.009 & & -0.011* & -0.007 & -0.009 & 24,817 \\
   & (0.013) & (0.013) & (0.013) & & (0.007) & (0.009) & (0.009) & \\
\addlinespace[3pt]
\multicolumn{9}{l}{\textit{US High-Fed vs.\ UK}} \\
\addlinespace[2pt]
  Paper & -0.017 & -0.015 & -0.008 & & -0.014* & -0.030** & -0.028* & 10,564 \\
   & (0.015) & (0.016) & (0.015) & & (0.008) & (0.015) & (0.015) & \\
  Author (own univ.) & -0.017 & -0.017 & -0.011 & & -0.015** & -0.022* & -0.018* & 20,287 \\
   & (0.014) & (0.014) & (0.012) & & (0.007) & (0.011) & (0.010) & \\
  Author (team-maj.) & -0.021 & -0.024 & -0.011 & & -0.021 & -0.044** & -0.043*** & 16,823 \\
   & (0.021) & (0.023) & (0.019) & & (0.014) & (0.017) & (0.016) & \\
\addlinespace[3pt]
\multicolumn{9}{l}{\textit{US (all) vs.\ UK}} \\
\addlinespace[2pt]
  Paper & -0.009 & -0.009 & -0.006 & & -0.006 & -0.025* & -0.024* & 17,965 \\
   & (0.013) & (0.014) & (0.013) & & (0.008) & (0.014) & (0.013) & \\
  Author (own univ.) & -0.019 & -0.019 & -0.012 & & -0.009 & -0.020* & -0.017* & 30,034 \\
   & (0.013) & (0.014) & (0.011) & & (0.007) & (0.011) & (0.009) & \\
  Author (team-maj.) & -0.015 & -0.018 & -0.007 & & -0.011 & -0.035** & -0.034** & 29,505 \\
   & (0.019) & (0.021) & (0.017) & & (0.011) & (0.015) & (0.014) & \\
\addlinespace[3pt]
\multicolumn{9}{l}{\textit{US Low-Fed vs.\ UK}} \\
\addlinespace[2pt]
  Paper & -0.001 & -0.002 & -0.002 & & 0.000 & -0.024 & -0.023 & 10,344 \\
   & (0.014) & (0.015) & (0.014) & & (0.008) & (0.015) & (0.014) & \\
  Author (own univ.) & -0.018 & -0.020 & -0.013 & & -0.001 & -0.018 & -0.016 & 14,275 \\
   & (0.015) & (0.016) & (0.014) & & (0.009) & (0.012) & (0.010) & \\
  Author (team-maj.) & -0.010 & -0.013 & -0.004 & & 0.002 & -0.026 & -0.027* & 14,347 \\
   & (0.021) & (0.023) & (0.019) & & (0.012) & (0.016) & (0.015) & \\
\bottomrule
\end{tabular}}
\begin{minipage}{0.95\textwidth}
\vspace{4pt}
\footnotesize
\textit{Notes:} Each cell is the difference-in-differences coefficient on Post $\times$ Treatment, where Post indicates papers dated 2025 or later, for a binary abstract-level outcome: ``Flagged word'' = the abstract uses a targeted term in a gender-or-race (left) or environment (right) context (GPT-OSS context classification); ``Topic'' = a separate GPT-OSS call, which does not use the list of targeted terms, says that the abstract discusses the topic (Appendix~\ref{app_subsec_topics}); ``Flagged or topic'' = either. Paper-level rows use robust standard errors; author-level rows include author fixed effects and cluster standard errors two-way by university and paper, with treatment at the author's own basefixed university (own univ.) or given by the paper's team classification (team-maj.); at the author level, $N$ counts author-paper observations. US papers have at least as many US-affiliated (HERD-matched) as UK-affiliated (HESA-matched) authors (ties assigned to US); UK papers have strictly more UK-affiliated authors. High-Fed universities are those whose total-federal share of R\&D expenditure (HERD FY2024, all fields) exceeds the cross-university median; a US paper is High-Fed if at least half of all its authors are at High-Fed universities (UK-affiliated and unmatched authors count as not High-Fed; ties assigned to High-Fed). All regressions include source fixed effects and control for economics research capacity interacted with Post (Section~\ref{sec:methods}). Linear probability model estimates; standard errors in parentheses. Stars: * $p<0.10$, ** $p<0.05$, *** $p<0.01$ (normal approximation).
\end{minipage}
\end{table}

\clearpage

\begin{table}[htbp]
\centering
\caption{Robustness to Dropping Federal-Agency Mentions (Abstract)}
\label{tab:fig5}
\small
\setlength{\tabcolsep}{4.5pt}
\begin{tabular}{lcccc}
\toprule
 & All papers & Excl.\ NSF/SES mentions & Excl.\ any agency mention & $N$ \\
\midrule
\addlinespace[3pt]
\multicolumn{5}{l}{\textit{US High-Fed vs.\ US Low-Fed}} \\
\addlinespace[2pt]
  Paper & -0.027** & -0.026** & -0.022 & 15,022 \\
   & (0.012) & (0.013) & (0.013) &  \\
  Author (own univ.) & -0.012 & -0.003 & -0.009 & 25,506 \\
   & (0.010) & (0.009) & (0.011) &  \\
  Author (team-maj.) & -0.027** & -0.020 & -0.025* & 24,817 \\
   & (0.013) & (0.014) & (0.015) &  \\
\addlinespace[3pt]
\multicolumn{5}{l}{\textit{US High-Fed vs.\ UK}} \\
\addlinespace[2pt]
  Paper & -0.030* & -0.029* & -0.027 & 10,564 \\
   & (0.016) & (0.017) & (0.017) &  \\
  Author (own univ.) & -0.034** & -0.030** & -0.026* & 20,287 \\
   & (0.013) & (0.013) & (0.014) &  \\
  Author (team-maj.) & -0.045** & -0.045** & -0.040** & 16,823 \\
   & (0.020) & (0.020) & (0.020) &  \\
\addlinespace[3pt]
\multicolumn{5}{l}{\textit{US (all) vs.\ UK}} \\
\addlinespace[2pt]
  Paper & -0.013 & -0.013 & -0.018 & 17,965 \\
   & (0.015) & (0.015) & (0.015) &  \\
  Author (own univ.) & -0.030** & -0.031** & -0.027** & 30,034 \\
   & (0.013) & (0.013) & (0.013) &  \\
  Author (team-maj.) & -0.029 & -0.029 & -0.025 & 29,505 \\
   & (0.018) & (0.018) & (0.019) &  \\
\addlinespace[3pt]
\multicolumn{5}{l}{\textit{US Low-Fed vs.\ UK}} \\
\addlinespace[2pt]
  Paper & 0.000 & 0.000 & -0.009 & 10,344 \\
   & (0.016) & (0.016) & (0.017) &  \\
  Author (own univ.) & -0.023 & -0.030** & -0.023 & 14,275 \\
   & (0.015) & (0.014) & (0.015) &  \\
  Author (team-maj.) & -0.010 & -0.015 & -0.014 & 14,347 \\
   & (0.020) & (0.020) & (0.021) &  \\
\bottomrule
\end{tabular}
\begin{minipage}{0.95\textwidth}
\vspace{4pt}
\footnotesize
\textit{Notes:} Each cell is the difference-in-differences coefficient on Post $\times$ Treatment, where Post indicates papers dated 2025 or later. Standard errors in parentheses; paper-level standard errors are robust, author-level standard errors are two-way clustered by university and paper. US papers have at least as many US- as UK-affiliated matched authors (ties assigned to US); UK papers have strictly more UK-affiliated authors; High-Fed papers have at least half of all their authors at High-Fed universities (UK-affiliated and unmatched authors count as not High-Fed; ties assigned to High-Fed). ``Author (own univ.)'' assigns treatment by the author's own (basefixed) university with author fixed effects; ``Author (team-majority)'' assigns treatment by the paper's classification, with author fixed effects. The outcome ``GRE in abstract'' is an indicator for whether the abstract contains a targeted term used in a Gender, Race, or Environment context (GPT-OSS classification), estimated as a linear probability model; the coefficient is a change in probability. High-Fed = above-median total-federal R\&D share (HERD FY2024, all fields). All regressions control for economics research capacity interacted with Post: the log of one plus the number of papers in the sample dated before 2025 with at least one author at the university, averaged over the paper's US- and UK-affiliated authors (the author's own university in the own-university specification). $^{*}\,p<0.10$, $^{**}\,p<0.05$, $^{***}\,p<0.01$. Columns progressively drop papers that mention a federal agency: the NSF/SES program, then any federal agency. $N$ is for the all-papers column.
\end{minipage}
\end{table}

\clearpage

\begin{table}[htbp]
\centering
\caption{University Rank: Controls, Heterogeneity, and US--UK Comparison by Quality}
\label{tab:fig6}
\scriptsize
\setlength{\tabcolsep}{3pt}\renewcommand{\arraystretch}{0.75}
\begin{tabular}{lcc}
\toprule
 & GRE in abstract & $N$ \\
\midrule
\addlinespace[3pt]
\multicolumn{3}{l}{\textit{US High-Fed vs.\ US Low-Fed: with and without rank control (level + rank $\times$ Post)}} \\
\addlinespace[2pt]
  No rank control -- Paper & -0.026** & 14,871 \\
   & (0.012) &  \\
  No rank control -- Author (own univ.) & -0.011 & 24,888 \\
   & (0.010) &  \\
  No rank control -- Author (team-maj.) & -0.026* & 24,671 \\
   & (0.013) &  \\
  With rank $\times$ Post control -- Paper & -0.024* & 14,871 \\
   & (0.012) &  \\
  With rank $\times$ Post control -- Author (own univ.) & -0.011 & 24,888 \\
   & (0.010) &  \\
  With rank $\times$ Post control -- Author (team-maj.) & -0.027* & 24,671 \\
   & (0.014) &  \\
\addlinespace[3pt]
\multicolumn{3}{l}{\textit{US High-Fed vs.\ US Low-Fed: by department rank}} \\
\addlinespace[2pt]
  Higher-ranked (rank $<$ median) -- Paper & -0.020 & 7,435 \\
   & (0.017) &  \\
  Higher-ranked (rank $<$ median) -- Author (own univ.) & -0.010 & 12,999 \\
   & (0.010) &  \\
  Higher-ranked (rank $<$ median) -- Author (team-maj.) & -0.007 & 12,332 \\
   & (0.017) &  \\
  Lower-ranked (rank $\geq$ median) -- Paper & -0.030* & 7,436 \\
   & (0.018) &  \\
  Lower-ranked (rank $\geq$ median) -- Author (own univ.) & -0.013 & 12,237 \\
   & (0.018) &  \\
  Lower-ranked (rank $\geq$ median) -- Author (team-maj.) & -0.047** & 11,125 \\
   & (0.022) &  \\
\addlinespace[3pt]
\multicolumn{3}{l}{\textit{US (all) vs.\ UK: by department quality (global median)}} \\
\addlinespace[2pt]
  Above global median -- Paper & -0.047* & 8,734 \\
   & (0.025) &  \\
  Above global median -- Author (own univ.) & -0.053** & 14,533 \\
   & (0.027) &  \\
  Above global median -- Author (team-maj.) & -0.074* & 14,479 \\
   & (0.040) &  \\
  Below global median -- Paper & 0.004 & 8,878 \\
   & (0.020) &  \\
  Below global median -- Author (own univ.) & -0.010 & 13,190 \\
   & (0.021) &  \\
  Below global median -- Author (team-maj.) & -0.005 & 13,095 \\
   & (0.024) &  \\
\addlinespace[3pt]
\multicolumn{3}{l}{\textit{US (all) vs.\ UK: by department quality (within-country median)}} \\
\addlinespace[2pt]
  Above within-country median -- Paper & -0.045* & 8,732 \\
   & (0.023) &  \\
  Above within-country median -- Author (own univ.) & -0.058** & 14,486 \\
   & (0.023) &  \\
  Above within-country median -- Author (team-maj.) & -0.080*** & 14,373 \\
   & (0.031) &  \\
  Below within-country median -- Paper & 0.010 & 8,880 \\
   & (0.021) &  \\
  Below within-country median -- Author (own univ.) & 0.009 & 13,217 \\
   & (0.022) &  \\
  Below within-country median -- Author (team-maj.) & 0.018 & 13,162 \\
   & (0.025) &  \\
\addlinespace[1pt]
\bottomrule
\end{tabular}
\begin{minipage}{0.95\textwidth}
\vspace{4pt}
\scriptsize
\textit{Notes:} Each cell is the difference-in-differences coefficient on Post $\times$ Treatment, where Post indicates papers dated 2025 or later. Standard errors in parentheses; paper-level standard errors are robust, author-level standard errors are two-way clustered by university and paper. US papers have at least as many US- as UK-affiliated matched authors (ties assigned to US); UK papers have strictly more UK-affiliated authors; High-Fed papers have at least half of all their authors at High-Fed universities (UK-affiliated and unmatched authors count as not High-Fed; ties assigned to High-Fed). ``Author (own univ.)'' assigns treatment by the author's own (basefixed) university with author fixed effects; ``Author (team-majority)'' assigns treatment by the paper's classification, with author fixed effects. The outcome ``GRE in abstract'' is an indicator for whether the abstract contains a targeted term used in a Gender, Race, or Environment context (GPT-OSS classification), estimated as a linear probability model; the coefficient is a change in probability. High-Fed = above-median total-federal R\&D share (HERD FY2024, all fields). All regressions control for economics research capacity interacted with Post: the log of one plus the number of papers in the sample dated before 2025 with at least one author at the university, averaged over the paper's US- and UK-affiliated authors (the author's own university in the own-university specification). $^{*}\,p<0.10$, $^{**}\,p<0.05$, $^{***}\,p<0.01$. University rank/quality is the RePEc economics-department rank; the median split is computed across the relevant sample. The rank control uses the paper's average department rank at every level; in the split by rank, the own-university rows split authors by the rank of their own university, so their samples differ from the rank-control block. $N$ is for the abstract specification.
\end{minipage}
\end{table}

\clearpage

\begin{table}[htbp]
\centering
\caption{Change in GRE Content in the Abstract After 2024, by Author Characteristics}
\label{tab:fig7}
\scriptsize
\setlength{\tabcolsep}{3pt}\renewcommand{\arraystretch}{0.75}
\begin{tabular}{lcccc}
\toprule
 & US High-Fed vs.\ US Low-Fed & US High-Fed vs.\ UK & US (all) vs.\ UK & $N$ \\
\midrule
\addlinespace[3pt]
\multicolumn{5}{l}{\textit{Paper level}} \\
\addlinespace[2pt]
  Female & -0.032 & -0.010 & 0.012 & 6,366 \\
   & (0.021) & (0.030) & (0.026) &  \\
  Male & -0.028** & -0.039** & -0.021 & 13,982 \\
   & (0.012) & (0.016) & (0.015) &  \\
  White & -0.032** & -0.044** & -0.025 & 10,694 \\
   & (0.016) & (0.021) & (0.018) &  \\
  Asian & -0.043** & -0.032 & -0.003 & 6,559 \\
   & (0.017) & (0.027) & (0.025) &  \\
  Hispanic & -0.075* & -0.101* & -0.017 & 1,607 \\
   & (0.040) & (0.054) & (0.048) &  \\
\addlinespace[3pt]
\multicolumn{5}{l}{\textit{Author (own univ.) level}} \\
\addlinespace[2pt]
  Female & -0.017 & -0.015 & -0.005 & 4,843 \\
   & (0.023) & (0.033) & (0.032) &  \\
  Male & -0.020 & -0.040*** & -0.033*** & 19,320 \\
   & (0.013) & (0.013) & (0.013) &  \\
  White & -0.020 & -0.030* & -0.027 & 11,812 \\
   & (0.017) & (0.018) & (0.017) &  \\
  Asian & -0.020 & -0.046 & -0.029 & 5,592 \\
   & (0.022) & (0.030) & (0.028) &  \\
  Hispanic & -0.044 & -0.099 & -0.079 & 951 \\
   & (0.075) & (0.066) & (0.064) &  \\
\addlinespace[3pt]
\multicolumn{5}{l}{\textit{Author (team-maj.) level}} \\
\addlinespace[2pt]
  Female & -0.054 & -0.044 & -0.019 & 4,744 \\
   & (0.033) & (0.050) & (0.043) &  \\
  Male & -0.025 & -0.047** & -0.028 & 18,789 \\
   & (0.015) & (0.021) & (0.018) &  \\
  White & -0.047** & -0.044* & -0.010 & 11,520 \\
   & (0.020) & (0.025) & (0.022) &  \\
  Asian & -0.018 & -0.063 & -0.033 & 5,412 \\
   & (0.023) & (0.039) & (0.041) &  \\
  Hispanic & -0.054 & -0.029 & -0.025 & 936 \\
   & (0.053) & (0.065) & (0.063) &  \\
\addlinespace[1pt]
\bottomrule
\end{tabular}
\begin{minipage}{0.95\textwidth}
\vspace{4pt}
\footnotesize
\textit{Notes:} Each cell is the difference-in-differences coefficient on Post $\times$ Treatment, where Post indicates papers dated 2025 or later. Standard errors in parentheses; paper-level standard errors are robust, author-level standard errors are two-way clustered by university and paper. US papers have at least as many US- as UK-affiliated matched authors (ties assigned to US); UK papers have strictly more UK-affiliated authors; High-Fed papers have at least half of all their authors at High-Fed universities (UK-affiliated and unmatched authors count as not High-Fed; ties assigned to High-Fed). ``Author (own univ.)'' assigns treatment by the author's own (basefixed) university with author fixed effects; ``Author (team-majority)'' assigns treatment by the paper's classification, with author fixed effects. The outcome ``GRE in abstract'' is an indicator for whether the abstract contains a targeted term used in a Gender, Race, or Environment context (GPT-OSS classification), estimated as a linear probability model; the coefficient is a change in probability. High-Fed = above-median total-federal R\&D share (HERD FY2024, all fields). All regressions control for economics research capacity interacted with Post: the log of one plus the number of papers in the sample dated before 2025 with at least one author at the university, averaged over the paper's US- and UK-affiliated authors (the author's own university in the own-university specification). $^{*}\,p<0.10$, $^{**}\,p<0.05$, $^{***}\,p<0.01$. Outcome is the GRE-in-abstract indicator; samples are split by the author's predicted gender and Census-predicted ethnicity. Groups below the 600-paper display threshold of the figure (predicted-Black) are omitted. $N$ is for the first available comparison in each row.
\end{minipage}
\end{table}

\clearpage

\IfFileExists{results_20may/table_app_2024pre.tex}{%
  \begin{table}[htbp]
\centering
\caption{Robustness: Two-Year Pre-Period Window (2023--2024 vs.\ 2025--2026)}
\label{tab:app_2024pre}
\small
\setlength{\tabcolsep}{4.5pt}
\begin{tabular}{lcc}
\toprule
 & GRE in abstract & $N$ \\
\midrule
  US High-Fed vs.\ US Low-Fed -- Paper & -0.026* & 8,418 \\
   & (0.014) &  \\
  US High-Fed vs.\ US Low-Fed -- Author (own univ.) & -0.016 & 12,968 \\
   & (0.016) &  \\
  US High-Fed vs.\ US Low-Fed -- Author (team-maj.) & -0.035** & 12,652 \\
   & (0.017) &  \\
  US High-Fed vs.\ UK -- Paper & -0.021 & 5,922 \\
   & (0.019) &  \\
  US High-Fed vs.\ UK -- Author (own univ.) & -0.033* & 10,347 \\
   & (0.017) &  \\
  US High-Fed vs.\ UK -- Author (team-maj.) & -0.041* & 8,368 \\
   & (0.023) &  \\
  US (all) vs.\ UK -- Paper & -0.008 & 10,072 \\
   & (0.017) &  \\
  US (all) vs.\ UK -- Author (own univ.) & -0.025 & 15,229 \\
   & (0.016) &  \\
  US (all) vs.\ UK -- Author (team-maj.) & -0.024 & 14,995 \\
   & (0.021) &  \\
  US Low-Fed vs.\ UK -- Paper & 0.004 & 5,804 \\
   & (0.019) &  \\
  US Low-Fed vs.\ UK -- Author (own univ.) & -0.015 & 7,143 \\
   & (0.019) &  \\
  US Low-Fed vs.\ UK -- Author (team-maj.) & 0.001 & 7,062 \\
   & (0.024) &  \\
\bottomrule
\end{tabular}
\begin{minipage}{0.95\textwidth}
\vspace{4pt}
\footnotesize
\textit{Notes:} Each cell is the difference-in-differences coefficient on Post $\times$ Treatment, where Post indicates papers dated 2025 or later. Standard errors in parentheses; paper-level standard errors are robust, author-level standard errors are two-way clustered by university and paper. US papers have at least as many US- as UK-affiliated matched authors (ties assigned to US); UK papers have strictly more UK-affiliated authors; High-Fed papers have at least half of all their authors at High-Fed universities (UK-affiliated and unmatched authors count as not High-Fed; ties assigned to High-Fed). ``Author (own univ.)'' assigns treatment by the author's own (basefixed) university with author fixed effects; ``Author (team-majority)'' assigns treatment by the paper's classification, with author fixed effects. The outcome ``GRE in abstract'' is an indicator for whether the abstract contains a targeted term used in a Gender, Race, or Environment context (GPT-OSS classification), estimated as a linear probability model; the coefficient is a change in probability. High-Fed = above-median total-federal R\&D share (HERD FY2024, all fields). All regressions control for economics research capacity interacted with Post: the log of one plus the number of papers in the sample dated before 2025 with at least one author at the university, averaged over the paper's US- and UK-affiliated authors (the author's own university in the own-university specification). $^{*}\,p<0.10$, $^{**}\,p<0.05$, $^{***}\,p<0.01$.
\end{minipage}
\end{table}
\clearpage}{}
\IfFileExists{results_20may/table_app_contdose_fed.tex}{%
  \begin{table}[htbp]
\centering
\caption{Robustness: Continuous Federal-Share Exposure (Within US)}
\label{tab:app_contdose}
\small
\setlength{\tabcolsep}{4.5pt}
\begin{tabular}{lcc}
\toprule
 & Estimate (per 1 SD) & $N$ \\
\midrule
  Total-federal share -- Paper & -0.014** & 15,022 \\
   & (0.006) &  \\
  Total-federal share -- Author (own univ.) & -0.007 & 25,506 \\
   & (0.006) &  \\
  Total-federal share -- Author (team-maj.) & -0.010 & 24,827 \\
   & (0.006) &  \\
\bottomrule
\end{tabular}
\begin{minipage}{0.95\textwidth}
\vspace{4pt}
\footnotesize
\textit{Notes:} Each cell is the difference-in-differences coefficient on Post $\times$ exposure, where Post indicates papers dated 2025 or later and exposure is the continuous funding share of the university (FY2024 HERD), standardized to mean zero and unit standard deviation within the estimation sample; coefficients are per one standard deviation of the share. Samples are the within-US samples of Figure~\ref{fig:coefplot_refgroups} (US papers: at least as many US- as UK-affiliated matched authors, ties assigned to US). ``Paper'' uses the paper's average share across all its authors (UK-affiliated and unmatched authors count as zero), one observation per paper, and robust standard errors; ``Author (own univ.)'' assigns each author the (basefixed) share of their own university; ``Author (team-majority)'' assigns the paper's average share across all its authors (UK-affiliated and unmatched authors count as zero); both author levels include author fixed effects and two-way clustering by university and paper. The dependent variable is the GRE-in-abstract indicator (GPT-OSS classification); linear probability models. All regressions control for economics research capacity interacted with Post: the log of one plus the number of papers in the sample dated before 2025 with at least one author at the university, averaged over the paper's US- and UK-affiliated authors (the author's own university in the own-university specification). $^{*}\,p<0.10$, $^{**}\,p<0.05$, $^{***}\,p<0.01$.
\end{minipage}
\end{table}
\clearpage}{}
\IfFileExists{results_20may/table_app_size.tex}{%
  \begin{table}[htbp]
\centering
\caption{University Size versus Federal Dependence (Within US)}
\label{tab:app_size}
\small
\setlength{\tabcolsep}{4.5pt}
\begin{tabular}{lccc}
\toprule
 & Paper & Author (own univ.) & Author (team-maj.) \\
\midrule
\addlinespace[3pt]
\multicolumn{4}{l}{\textit{A. High-Fed versus Low-Fed, with a control for federal dollars}} \\
\addlinespace[2pt]
  High-Fed $\times$ Post & -0.027** & -0.012 & -0.027** \\
   & (0.012) & (0.010) & (0.013) \\
  High-Fed $\times$ Post, with log dollars $\times$ Post & -0.032** & -0.017 & -0.036** \\
   & (0.013) & (0.011) & (0.014) \\
  Log federal dollars $\times$ Post & 0.007 & 0.012 & 0.016* \\
   & (0.008) & (0.009) & (0.009) \\
  Observations & 15,022 & 25,506 & 24,817 \\
\addlinespace[3pt]
\multicolumn{4}{l}{\textit{B. Continuous exposure, per one standard deviation}} \\
\addlinespace[2pt]
  Federal share $\times$ Post, alone & -0.014** & -0.007 & -0.010 \\
   & (0.006) & (0.006) & (0.006) \\
  Log federal dollars $\times$ Post, alone & -0.001 & 0.008 & 0.006 \\
   & (0.008) & (0.008) & (0.009) \\
  Federal share $\times$ Post, jointly & -0.021*** & -0.011 & -0.020*** \\
   & (0.007) & (0.007) & (0.007) \\
  Log federal dollars $\times$ Post, jointly & 0.015 & 0.014 & 0.023** \\
   & (0.010) & (0.009) & (0.010) \\
  Observations & 15,022 & 25,506 & 24,827 \\
\addlinespace[3pt]
\multicolumn{4}{l}{\textit{C. High-Fed versus Low-Fed, by university size}} \\
\addlinespace[2pt]
  Smaller recipients (below median dollars) & -0.017 & -0.024 & -0.031 \\
   & (0.020) & (0.015) & (0.020) \\
  Larger recipients (above median dollars) & -0.027 & -0.014 & -0.026 \\
   & (0.018) & (0.011) & (0.024) \\
  Difference, larger minus smaller & -0.009 & 0.011 & 0.005 \\
   & (0.027) & (0.019) & (0.033) \\
  Observations & 15,022 & 25,506 & 23,058 \\
\bottomrule
\end{tabular}
\begin{minipage}{0.95\textwidth}
\vspace{4pt}
\footnotesize
\textit{Notes:} Each cell is a difference-in-differences coefficient, where Post indicates papers dated 2025 or later. The dependent variable is the GRE-in-abstract indicator (GPT-OSS classification); linear probability models. Samples are the within-US samples of Figure~\ref{fig:coefplot_refgroups}. Federal dollars are the university's total federal R\&D expenditure (HERD FY2024, all fields), in logs; the federal share is the same expenditure as a share of total R\&D. ``Paper'' uses the paper's average over its authors, one observation per paper, and robust standard errors; at the paper level and under the team-majority assignment, the federal share and federal dollars are averaged over all authors of the paper, with UK-affiliated and unmatched authors counted as zero, and dollars are logged after averaging. ``Author (own univ.)'' assigns each author the values of their own (basefixed) university; ``Author (team-maj.)'' assigns the paper's values; both author levels include author fixed effects and two-way clustering by university and paper. Panel A adds log federal dollars interacted with Post to the High-Fed specification of Figure~\ref{fig:coefplot_refgroups}. Panel B replaces the binary High-Fed indicator with the continuous federal share and log federal dollars, each standardized within the estimation sample, entered alone and jointly. Panel C estimates the High-Fed effect separately for universities below and above the sample median of log federal dollars; the difference is the triple interaction in the pooled regression. All regressions control for economics research capacity interacted with Post: the log of one plus the number of papers in the sample dated before 2025 with at least one author at the university, averaged over the paper's US- and UK-affiliated authors (the author's own university in the own-university specification). $^{*}\,p<0.10$, $^{**}\,p<0.05$, $^{***}\,p<0.01$.
\end{minipage}
\end{table}
\clearpage}{}
\IfFileExists{results_20may/table_app_teamcomp.tex}{%
  \begin{table}[htbp]
\centering
\caption{Robustness: Papers with All Authors Matched to HERD Universities (Within US)}
\label{tab:app_teamcomp}
\small
\setlength{\tabcolsep}{4.5pt}
\begin{tabular}{lccc}
\toprule
 & Paper & Author (own univ.) & Author (team-maj.) \\
\midrule
  Full within-US sample & -0.027** & -0.012 & -0.027** \\
   & (0.012) & (0.010) & (0.013) \\
  Observations & 15,022 & 25,506 & 24,817 \\
\addlinespace[3pt]
  All authors HERD-matched & -0.037** & -0.015 & -0.029 \\
   & (0.018) & (0.016) & (0.022) \\
  Observations & 7,465 & 12,790 & 12,790 \\
\addlinespace[3pt]
\bottomrule
\end{tabular}
\begin{minipage}{0.95\textwidth}
\vspace{4pt}
\footnotesize
\textit{Notes:} Each cell is the difference-in-differences coefficient on Post $\times$ High-Fed for the US High-Fed versus US Low-Fed comparison of Figure~\ref{fig:coefplot_refgroups}, where Post indicates papers dated 2025 or later. The dependent variable is the GRE-in-abstract indicator (GPT-OSS classification); linear probability models. The first row is the full within-US sample. The second row keeps only papers whose authors are all matched to HERD universities, so that no UK-affiliated or unmatched coauthor enters the High-Fed rule. ``Paper'' is one observation per paper with robust standard errors; ``Author (own univ.)'' assigns treatment by the author's own (basefixed) university; ``Author (team-maj.)'' assigns the paper's classification; both author levels include author fixed effects and two-way clustering by university and paper. All regressions control for economics research capacity interacted with Post: the log of one plus the number of papers in the sample dated before 2025 with at least one author at the university, averaged over the paper's US- and UK-affiliated authors (the author's own university in the own-university specification). $^{*}\,p<0.10$, $^{**}\,p<0.05$, $^{***}\,p<0.01$.
\end{minipage}
\end{table}
\clearpage}{}
\IfFileExists{results_20may/table_app_redblue.tex}{%
  \begin{table}[htbp]
\centering
\caption{Robustness: Red/Blue State $\times$ Post Control}
\label{tab:app_redblue}
\small
\setlength{\tabcolsep}{4.5pt}
\begin{tabular}{lcc}
\toprule
 & GRE in abstract & $N$ \\
\midrule
  US High-Fed vs.\ US Low-Fed -- Paper & -0.027** & 15,022 \\
   & (0.012) &  \\
  US High-Fed vs.\ US Low-Fed -- Author (own univ.) & -0.010 & 25,506 \\
   & (0.010) &  \\
  US High-Fed vs.\ US Low-Fed -- Author (team-maj.) & -0.024* & 24,817 \\
   & (0.013) &  \\
  US High-Fed vs.\ UK -- Paper & -0.032* & 10,564 \\
   & (0.017) &  \\
  US High-Fed vs.\ UK -- Author (own univ.) & -0.037*** & 20,287 \\
   & (0.014) &  \\
  US High-Fed vs.\ UK -- Author (team-maj.) & -0.052** & 16,823 \\
   & (0.021) &  \\
  US (all) vs.\ UK -- Paper & -0.016 & 17,965 \\
   & (0.015) &  \\
  US (all) vs.\ UK -- Author (own univ.) & -0.034*** & 30,034 \\
   & (0.013) &  \\
  US (all) vs.\ UK -- Author (team-maj.) & -0.035* & 29,505 \\
   & (0.018) &  \\
  US Low-Fed vs.\ UK -- Paper & 0.000 & 10,344 \\
   & (0.017) &  \\
  US Low-Fed vs.\ UK -- Author (own univ.) & -0.027* & 14,275 \\
   & (0.015) &  \\
  US Low-Fed vs.\ UK -- Author (team-maj.) & -0.014 & 14,347 \\
   & (0.021) &  \\
\bottomrule
\end{tabular}
\begin{minipage}{0.95\textwidth}
\vspace{4pt}
\footnotesize
\textit{Notes:} Each cell is the difference-in-differences coefficient on Post $\times$ Treatment, where Post indicates papers dated 2025 or later. Standard errors in parentheses; paper-level standard errors are robust, author-level standard errors are two-way clustered by university and paper. US papers have at least as many US- as UK-affiliated matched authors (ties assigned to US); UK papers have strictly more UK-affiliated authors; High-Fed papers have at least half of all their authors at High-Fed universities (UK-affiliated and unmatched authors count as not High-Fed; ties assigned to High-Fed). ``Author (own univ.)'' assigns treatment by the author's own (basefixed) university with author fixed effects; ``Author (team-majority)'' assigns treatment by the paper's classification, with author fixed effects. The outcome ``GRE in abstract'' is an indicator for whether the abstract contains a targeted term used in a Gender, Race, or Environment context (GPT-OSS classification), estimated as a linear probability model; the coefficient is a change in probability. High-Fed = above-median total-federal R\&D share (HERD FY2024, all fields). All regressions control for economics research capacity interacted with Post: the log of one plus the number of papers in the sample dated before 2025 with at least one author at the university, averaged over the paper's US- and UK-affiliated authors (the author's own university in the own-university specification). $^{*}\,p<0.10$, $^{**}\,p<0.05$, $^{***}\,p<0.01$.
\end{minipage}
\end{table}
\clearpage}{}
\IfFileExists{results_20may/table_app_anyword.tex}{%
  \begin{table}[htbp]
\centering
\caption{Robustness: Any Targeted Word (Abstract and Full Text)}
\label{tab:app_anyword}
\scriptsize
\setlength{\tabcolsep}{3pt}\renewcommand{\arraystretch}{0.75}
\begin{tabular}{lccc}
\toprule
 & Any targeted word (abstract) & Targeted words (full text) & $N$ \\
\midrule
  US High-Fed vs.\ US Low-Fed -- Paper & -0.002 & 0.033 & 15,022 \\
   & (0.018) & (0.056) &  \\
  US High-Fed vs.\ US Low-Fed -- Author (own univ.) & 0.013 & -0.079** & 25,506 \\
   & (0.026) & (0.039) &  \\
  US High-Fed vs.\ US Low-Fed -- Author (team-maj.) & -0.008 & -0.012 & 24,817 \\
   & (0.024) & (0.060) &  \\
  US High-Fed vs.\ UK -- Paper & -0.008 & -0.086 & 10,564 \\
   & (0.024) & (0.085) &  \\
  US High-Fed vs.\ UK -- Author (own univ.) & 0.022 & -0.012 & 20,287 \\
   & (0.026) & (0.069) &  \\
  US High-Fed vs.\ UK -- Author (team-maj.) & 0.006 & 0.028 & 16,823 \\
   & (0.035) & (0.093) &  \\
  US (all) vs.\ UK -- Paper & -0.010 & -0.074 & 17,965 \\
   & (0.022) & (0.074) &  \\
  US (all) vs.\ UK -- Author (own univ.) & 0.012 & 0.024 & 30,034 \\
   & (0.025) & (0.065) &  \\
  US (all) vs.\ UK -- Author (team-maj.) & 0.002 & 0.017 & 29,505 \\
   & (0.030) & (0.075) &  \\
  US Low-Fed vs.\ UK -- Paper & -0.008 & -0.094 & 10,344 \\
   & (0.024) & (0.079) &  \\
  US Low-Fed vs.\ UK -- Author (own univ.) & 0.001 & 0.072 & 14,275 \\
   & (0.029) & (0.066) &  \\
  US Low-Fed vs.\ UK -- Author (team-maj.) & -0.004 & 0.032 & 14,347 \\
   & (0.033) & (0.083) &  \\
\addlinespace[1pt]
\bottomrule
\end{tabular}
\begin{minipage}{0.95\textwidth}
\vspace{4pt}
\footnotesize
\textit{Notes:} Each cell is the difference-in-differences coefficient on Post $\times$ Treatment, where Post indicates papers dated 2025 or later. Standard errors in parentheses; paper-level standard errors are robust, author-level standard errors are two-way clustered by university and paper. US papers have at least as many US- as UK-affiliated matched authors (ties assigned to US); UK papers have strictly more UK-affiliated authors; High-Fed papers have at least half of all their authors at High-Fed universities (UK-affiliated and unmatched authors count as not High-Fed; ties assigned to High-Fed). ``Author (own univ.)'' assigns treatment by the author's own (basefixed) university with author fixed effects; ``Author (team-maj.)'' assigns treatment by the paper's classification, with author fixed effects. Outcomes use the raw targeted-term dictionary with no context classification: ``Any targeted word (abstract)'' is a linear probability model on an indicator for whether any targeted term appears in the abstract; ``Targeted words (full text)'' is a Poisson regression on the raw count of targeted terms in the full paper text with a page-count control. High-Fed = above-median total-federal R\&D share (HERD FY2024, all fields). All regressions control for economics research capacity interacted with Post: the log of one plus the number of papers in the sample dated before 2025 with at least one author at the university, averaged over the paper's US- and UK-affiliated authors (the author's own university in the own-university specification). Stars: * $p<0.10$, ** $p<0.05$, *** $p<0.01$. $N$ is for the abstract specification.
\end{minipage}
\end{table}
\clearpage}{}
\IfFileExists{results_20may/table_app_altinf.tex}{%
  \begin{table}[htbp]
\centering
\caption{Robustness: Alternative Inference (Paper Level)}
\label{tab:app_altinf}
\scriptsize
\setlength{\tabcolsep}{3pt}\renewcommand{\arraystretch}{0.75}
\begin{tabular}{lcccccc}
\toprule
 & Coefficient & Cluster $p$ & Wild bootstrap $p$ & Wild bootstrap 95\% CI & Clusters & $N$ \\
\midrule
  US High-Fed vs.\ US Low-Fed & -0.027 & 0.015 & 0.020 & [-0.049, -0.005] & 321 & 15,022 \\
   & (0.011) & & & & & \\
  US High-Fed vs.\ UK & -0.030 & 0.052 & 0.054 & [-0.061, 0.001] & 306 & 10,564 \\
   & (0.015) & & & & & \\
  US (all) vs.\ UK & -0.013 & 0.326 & 0.330 & [-0.041, 0.015] & 403 & 17,965 \\
   & (0.014) & & & & & \\
  US Low-Fed vs.\ UK & 0.000 & 0.998 & 0.998 & [-0.029, 0.030] & 341 & 10,344 \\
   & (0.015) & & & & & \\
\addlinespace[1pt]
\bottomrule
\end{tabular}
\begin{minipage}{0.95\textwidth}
\vspace{4pt}
\footnotesize
\textit{Notes:} Alternative inference for the paper-level reference-group estimates of Figure~\ref{fig:coefplot_refgroups} (GRE in abstract, linear probability model; same specification as the paper rows of Table~\ref{tab:fig3}, restricted to papers with an assignable modal university). Each coefficient is the difference-in-differences estimate on Post $\times$ Treatment, with the modal-university-clustered standard error in parentheses; the baseline robust standard errors are in Table~\ref{tab:fig3}. The cluster is the paper's modal university among its matched authors (UK universities identified by their HESA provider codes). ``Wild bootstrap $p$'' and the 95\% confidence interval are from a wild cluster bootstrap over the same clusters (Rademacher weights, 99,999 replications, imposing the null; confidence interval by test inversion). All regressions control for economics research capacity interacted with Post: the log of one plus the number of papers in the sample dated before 2025 with at least one author at the university, averaged over the paper's US- and UK-affiliated authors (the author's own university in the own-university specification). No significance stars are shown, as each row reports several inference schemes. The author-level estimates of Figure~\ref{fig:coefplot_refgroups} are two-way clustered by university and paper and are unchanged by this exercise (Appendix Figure~\ref{fig:coefplot_altinf}). US papers have at least as many US- as UK-affiliated matched authors (ties assigned to US); UK papers have strictly more UK-affiliated authors; High-Fed papers have at least half of all their authors at High-Fed universities (UK-affiliated and unmatched authors count as not High-Fed; ties assigned to High-Fed). High-Fed = above-median total-federal R\&D share (HERD FY2024, all fields).
\end{minipage}
\end{table}
\clearpage}{}

\IfFileExists{results_20may/table_app_paperlength.tex}{%
  \begin{table}[htbp]
\centering
\caption{Reference-Group Comparisons: Paper Length}
\label{tab:app_paperlength}
\small
\setlength{\tabcolsep}{4.5pt}
\begin{tabular}{lcc}
\toprule
 & Number of pages & $N$ \\
\midrule
  US High-Fed vs.\ US Low-Fed -- Paper & -0.030* & 14,965 \\
   & (0.018) &  \\
  US High-Fed vs.\ US Low-Fed -- Author (own univ.) & -0.008 & 25,405 \\
   & (0.017) &  \\
  US High-Fed vs.\ US Low-Fed -- Author (team-maj.) & -0.017 & 24,723 \\
   & (0.018) &  \\
  US High-Fed vs.\ UK -- Paper & -0.017 & 10,513 \\
   & (0.025) &  \\
  US High-Fed vs.\ UK -- Author (own univ.) & -0.026 & 20,181 \\
   & (0.037) &  \\
  US High-Fed vs.\ UK -- Author (team-maj.) & -0.039 & 16,729 \\
   & (0.043) &  \\
  US (all) vs.\ UK -- Paper & 0.002 & 17,883 \\
   & (0.022) &  \\
  US (all) vs.\ UK -- Author (own univ.) & -0.021 & 29,894 \\
   & (0.034) &  \\
  US (all) vs.\ UK -- Author (team-maj.) & -0.023 & 29,373 \\
   & (0.039) &  \\
  US Low-Fed vs.\ UK -- Paper & 0.017 & 10,288 \\
   & (0.024) &  \\
  US Low-Fed vs.\ UK -- Author (own univ.) & -0.017 & 14,202 \\
   & (0.036) &  \\
  US Low-Fed vs.\ UK -- Author (team-maj.) & 0.000 & 14,278 \\
   & (0.039) &  \\
\bottomrule
\end{tabular}
\begin{minipage}{0.95\textwidth}
\vspace{4pt}
\footnotesize
\textit{Notes:} Each cell is the difference-in-differences coefficient on Post $\times$ Treatment, where Post indicates papers dated 2025 or later. Standard errors in parentheses; paper-level standard errors are robust, author-level standard errors are two-way clustered by university and paper. US papers have at least as many US- as UK-affiliated matched authors (ties assigned to US); UK papers have strictly more UK-affiliated authors; High-Fed papers have at least half of all their authors at High-Fed universities (UK-affiliated and unmatched authors count as not High-Fed; ties assigned to High-Fed). ``Author (own univ.)'' assigns treatment by the author's own (basefixed) university with author fixed effects; ``Author (team-maj.)'' assigns treatment by the paper's classification, with author fixed effects. The outcome is the number of pages, estimated by Poisson pseudo-maximum likelihood; the coefficient is an approximate proportional change. High-Fed = above-median total-federal R\&D share (HERD FY2024, all fields). All regressions control for economics research capacity interacted with Post: the log of one plus the number of papers in the sample dated before 2025 with at least one author at the university, averaged over the paper's US- and UK-affiliated authors (the author's own university in the own-university specification). Stars: * $p<0.10$, ** $p<0.05$, *** $p<0.01$.
\end{minipage}
\end{table}
\clearpage}{}

\IfFileExists{results_20may/table_app_papercounts.tex}{%
  \begin{table}[htbp]
\centering
\caption{Reference-Group Comparisons: Papers per Author per Year}
\label{tab:app_papercounts}
\small
\setlength{\tabcolsep}{4.5pt}
\begin{tabular}{lcc}
\toprule
 & Papers per author-year & $N$ \\
\midrule
  US High-Fed vs.\ US Low-Fed -- Author (own univ.) & 0.040 & 21,878 \\
   & (0.049) &  \\
  US High-Fed vs.\ UK -- Author (own univ.) & -0.046 & 17,814 \\
   & (0.063) &  \\
  US (all) vs.\ UK -- Author (own univ.) & -0.055 & 26,740 \\
   & (0.060) &  \\
  US Low-Fed vs.\ UK -- Author (own univ.) & -0.073 & 13,788 \\
   & (0.067) &  \\
\bottomrule
\end{tabular}
\begin{minipage}{0.95\textwidth}
\vspace{4pt}
\footnotesize
\textit{Notes:} Each cell is the difference-in-differences coefficient on Post $\times$ Treatment, where Post indicates papers dated 2025 or later. The outcome is the average number of papers per author per year, estimated by Poisson pseudo-maximum likelihood with author fixed effects on an author-by-period panel; standard errors in parentheses are clustered by university. All regressions control for economics research capacity interacted with Post: the log of one plus the number of papers in the sample dated before 2025 with at least one author at the university, averaged over the paper's US- and UK-affiliated authors (the author's own university in the own-university specification). Stars: * $p<0.10$, ** $p<0.05$, *** $p<0.01$.
\end{minipage}
\end{table}
\clearpage}{}

\IfFileExists{results_20may/table_app_fedfunding.tex}{%
  \begin{table}[htbp]
\centering
\caption{Direct Federal Funding of Economics Research}
\label{tab:app_fedfunding}
\small
\begin{tabular}{lcccc}
\toprule
\multicolumn{5}{l}{\textit{Panel A: Papers acknowledging federal funding}} \\[2pt]
 & Papers & NSF & SES & Any federal agency \\
\midrule
US High-Fed & 7,621 & 7.8\% & 3.7\% & 19.9\% \\
US Low-Fed & 7,401 & 5.3\% & 2.6\% & 15.5\% \\
UK & 2,943 & 1.0\% & 0.2\% & 5.5\% \\
\midrule
\multicolumn{5}{l}{\textit{Panel B: Federally financed economics R\&D, in-sample US universities}} \\[2pt]
\multicolumn{4}{l}{In-sample HERD universities} & 369 \\
\multicolumn{4}{l}{~~with no federally financed economics R\&D} & 197 (53.4\%) \\
\multicolumn{4}{l}{~~with NSF-funded economics R\&D} & 108 (29.3\%) \\
\multicolumn{4}{l}{Median economics share of total federal R\&D} & 0.000\% \\
\multicolumn{4}{l}{Universities with economics share below 1\%} & 95.4\% \\
\bottomrule
\multicolumn{5}{p{0.93\textwidth}}{\footnotesize \textit{Notes:} Table~\ref{tab:summary_statistics} (Panel C) reports agencies' shares of institution-wide federal R\&D; this table instead measures direct federal funding of the economics research itself. Panel A reports the share of papers that mention the NSF, its Division of Social and Economic Sciences (SES), or any federal agency in their acknowledgment sections, by treatment group (paper-level definitions as in Figure~\ref{fig:coefplot_refgroups}: US papers have at least as many US- as UK-affiliated matched authors, ties assigned to US; UK papers have strictly more UK-affiliated authors; a US paper is High-Fed if at least half of all its authors are at High-Fed universities (UK-affiliated and unmatched authors count as not High-Fed), ties assigned to High-Fed). Acknowledgment flags are unavailable for 145 arXiv papers, which are treated as having no mention. Panel B uses the HERD FY2023 field-level survey (``Social sciences, economics'' expenditure line; the FY2024 vintage used for the treatment reports all fields only) for the 369 of the 376 in-sample HERD universities (those with at least one matched author) that appear in the FY2023 field-level file; the other 7 are absent from it. The economics share divides federally financed economics R\&D by the university's total federal R\&D expenditure across all fields (FY2023). Amounts and shares treat universities without an economics field line as zero.} \\
\end{tabular}
\end{table}
\clearpage}{}

\end{document}